\documentclass[useAMS,usenatbib]{mn2e}
\usepackage{graphicx}
\usepackage{color}
\usepackage{epsf}
\usepackage{ulem}
\usepackage{amsmath}
\usepackage{amssymb}
\usepackage{times}
\usepackage{fixltx2e}
\usepackage{astrobib_mnras2e}

\newcommand{\hMpc}{h^{-1}\rm{Mpc}}

\newcommand{\rsf}{r_{s}^{\rm fid}}

\begin{document}

\title[Measuring $D_A$ and $H$ using BAO]
{The clustering of galaxies in the SDSS-III Baryon Oscillation
Spectroscopic Survey:  Measuring $D_A$ and $H$ at $z=0.57$
from the Baryon Acoustic Peak in the Data Release 9 Spectroscopic
Galaxy Sample}

\author[Anderson  et al.]{\parbox{\textwidth}{\Large
Lauren Anderson$^{1}$,
Eric Aubourg$^{2}$,
Stephen Bailey$^{3}$,
Florian Beutler$^{3}$,
Adam S. Bolton$^{4}$,
J. Brinkmann$^{5}$,
Joel R. Brownstein$^{4}$,
Chia-Hsun Chuang$^{6}$,
Antonio J. Cuesta$^{7}$,
Kyle S. Dawson$^{4}$,
Daniel J. Eisenstein$^{8}$,
Klaus Honscheid$^{9}$,
Eyal A. Kazin$^{10}$,
David Kirkby$^{11}$,
Marc Manera$^{12}$,
Cameron K. McBride$^{8}$,
O. Mena$^{13}$,
Robert C. Nichol$^{12}$,
Matthew D. Olmstead$^{4}$,
Nikhil Padmanabhan$^{7}$,
N. Palanque-Delabrouille$^{14}$,
Will J. Percival$^{12}$,
Francisco Prada$^{6,15,16}$,
Ashley J. Ross$^{12}$,
Nicolas P. Ross$^{3}$,
Ariel G. S\'anchez$^{17}$,
Lado Samushia$^{12,18}$,
David J. Schlegel$^{3}$,
Donald P. Schneider$^{19,20}$,
Hee-Jong Seo$^{3}$,
Michael A. Strauss$^{21}$,
Daniel Thomas$^{12}$,
Jeremy L. Tinker$^{22}$,
Rita Tojeiro$^{12}$,
Licia Verde$^{23}$,
David H. Weinberg$^{9}$,
Xiaoying Xu$^{24}$,
Christophe Y\`{e}che$^{14}$
} \vspace*{4pt} \\ 
\scriptsize $^{1}$Department of Astronomy, University of Washington, Box 351580 Seattle, WA 98195 USA \vspace*{-2pt} \\
\scriptsize $^{2}$APC, University Paris Diderot, CNRS/IN2P3, CEA/Irfu, Obs de Paris, Sorbonne Paris, France. \vspace*{-2pt} \\
\scriptsize $^{3}$Lawrence Berkeley National Lab, 1 Cyclotron Rd, Berkeley CA 94720, USA. \vspace*{-2pt} \\
\scriptsize $^{4}$Department of Physics and Astronomy, University of Utah, 115 S 1400 E, Salt Lake City, UT 84112, USA. \vspace*{-2pt} \\
\scriptsize $^{5}$Apache Point Observatory, PO Box 59 Sunspot, NM 88349-0059 USA. \vspace*{-2pt} \\
\scriptsize $^{6}$Instituto de F\'{\i}sica Te\'orica, (UAM/CSIC), Universidad Aut\'onoma de Madrid,  Cantoblanco, E-28049 Madrid, Spain. \vspace*{-2pt} \\
\scriptsize $^{7}$Department of Physics, Yale University, 260 Whitney Ave, New Haven, CT 06520, USA. \vspace*{-2pt} \\
\scriptsize $^{8}$Harvard-Smithsonian Center for Astrophysics, 60 Garden St., Cambridge, MA 02138, USA. \vspace*{-2pt} \\
\scriptsize $^{9}$Dept. of Astronomy and CCAPP, Ohio State University, Columbus, OH \vspace*{-2pt} \\
\scriptsize $^{10}$Centre for Astrophysics \& Supercomputing, Swinburne University of Technology, PO Box 218, Hawthorn, VIC 3122, Australia. \vspace*{-2pt} \\
\scriptsize $^{11}$Department of Physics and Astronomy, UC Irvine, 4129 Frederick Reines Hall Irvine, CA 92697 USA \vspace*{-2pt} \\
\scriptsize $^{12}$Institute of Cosmology \& Gravitation, University of Portsmouth, Dennis Sciama Building, Portsmouth PO1 3FX, UK. \vspace*{-2pt} \\
\scriptsize $^{13}$IFIC (CSIC-UV), Paterna, Valencia, Spain \vspace*{-2pt} \\
\scriptsize $^{14}$CEA, Centre de Saclay, Irfu/SPP,  F-91191 Gif-sur-Yvette, France. \vspace*{-2pt} \\
\scriptsize $^{15}$Instituto de Astrof\'{\i}sica de Andaluc\'{\i}a (CSIC), Glorieta de la Astronom\'{\i}a, E-18080 Granada, Spain. \vspace*{-2pt} \\
\scriptsize $^{16}$Campus of International Excellence UAM+CSIC, Cantoblanco, E-28049 Madrid, Spain. \vspace*{-2pt} \\
\scriptsize $^{17}$Max-Planck-Institut f\"ur Extraterrestrische Physik, Giessenbachstra\ss e, 85748 Garching, Germany. \vspace*{-2pt} \\
\scriptsize $^{18}$National Abastumani Astrophysical Observatory, Ilia State University, 2A Kazbegi Ave., GE-1060 Tbilisi, Georgia \vspace*{-2pt} \\
\scriptsize $^{19}$Department of Astronomy and Astrophysics, The Pennsylvania State University, University Park, PA 16802, USA. \vspace*{-2pt} \\
\scriptsize $^{20}$Institute for Gravitation and the Cosmos, The Pennsylvania State University, University Park, PA 16802, USA. \vspace*{-2pt} \\
\scriptsize $^{21}$Department of Astrophysical Sciences, Princeton University, Princeton NJ 08544 USA. \vspace*{-2pt} \\
\scriptsize $^{22}$Center for Cosmology and Particle Physics, New York University, New York, NY 10003, USA. \vspace*{-2pt} \\
\scriptsize $^{23}$ICREA \& ICC University of Barcelona (IEEC-UB), Marti i Franques 1, Barcelona 08028, Spain. \vspace*{-2pt} \\
\scriptsize $^{24}$Department of Physics, Carnegie Mellon University, 5000 Forbes Ave., Pittsburgh, PA 15213, USA. \vspace*{-2pt} \\
}

%\linenumbers

\topmargin-1cm
\maketitle

\label{firstpage}

%\clearpage

\begin{abstract}

We present measurements of the angular diameter distance to and
Hubble parameter at $z=0.57$ from the measurement of the baryon
acoustic peak in the correlation of galaxies from the Sloan Digital
Sky Survey III Baryon Oscillation Spectroscopic Survey.  Our
analysis is based on a sample from Data Release 9 of 264,283 galaxies
over 3275 square degrees in the redshift range $0.43<z<0.70$.
We use two different methods to provide robust measurement of the
acoustic peak position across and along the line of sight in order
to measure the cosmological distance scale.  We find $D_A(0.57) =
1408 \pm 45$~Mpc and $H(0.57) = 92.9\pm7.8$~km/s/Mpc for our fiducial
value of the sound horizon.  These results from the anisotropic fitting
are fully consistent with the analysis of the spherically averaged
acoustic peak position presented in \citet{2012MNRAS.427.3435A}.  Our distance
measurements are a close match to the predictions of the standard 
cosmological model featuring a cosmological constant and zero spatial
curvature.

\end{abstract}

\section{Introduction} \label{sec:intro}
The expansion history of the Universe is one of the most fundamental
measurements in cosmology.  Its importance has been magnified in
the last 15 years because of the discovery of the late-time
acceleration of the expansion rate \citep{1998AJ....116.1009R,1999ApJ...517..565P}.
Precision measurements of the cosmic distance scale are crucial for
probing the behavior of the acceleration and the nature of the dark
energy that might cause it \citep{2012arXiv1201.2434W}.

The baryon acoustic oscillation (BAO) method provides a powerful opportunity
to measure the cosmic expansion history in a manner that is both
precise and robust.  Sound waves propagating in the first 400,000
years after the Big Bang create an excess of clustering at 150 comoving Mpc 
in the late-time distribution of matter
\citep{1970Ap&SS...7....3S,1970ApJ...162..815P,1987MNRAS.226..655B,1996ApJ...471..542H}.  This length scale, known
as the acoustic scale, results from simple physics: it is the
distance that the sound waves travel prior to recombination.
Because the acoustic scale is large, the measurement is altered
only modestly by subsequent non-linear structure formation and galaxy clustering
bias \citep{1999MNRAS.304..851M}.  Simulations and analytic theory predict shifts 
below 1\% in conventional models 
\citep{2003ApJ...598..720S,2005Natur.435..629S,2007APh....26..351H,2007ApJ...665...14S,
2008MNRAS.383..755A,2009PhRvD..80f3508P,2010ApJ...720.1650S,2011ApJ...734...94M}.

The robustness of the scale of this distinctive clustering signature 
allows it to be used as a standard ruler to measure the cosmic distance
scale.  By observing an feature of known size in the Hubble flow,
one can use the redshift spread along the line of sight to measure
the Hubble parameter $H(z)$ and one can use the angular spread in
the transverse direction to measure the angular diameter distance
$D_A(z)$.  By repeating this at a variety of redshifts, one can map 
out the cosmic expansion history and constrain the properties of 
dark energy \citep{2002ASPC..280...35E,2003ApJ...594..665B,2003PhRvD..68f3004H,2003PhRvD..68h3504L,2003ApJ...598..720S}.

The imprint of the baryon acoustic oscillations has been detected in
a variety of low-redshift data sets.  The strongest signals have been
in galaxy redshift surveys, including the 
Sloan Digital Sky Survey 
\citep[SDSS][]{2005ApJ...633..560E,2006A&A...449..891H,2006PhRvD..74l3507T,2007MNRAS.381.1053P,2010MNRAS.401.2148P,2010ApJ...710.1444K,2012MNRAS.423.1474C,2012MNRAS.426..226C,2012MNRAS.427.2132P,2012MNRAS.427.2146X},
2dF Galaxy Redshift Survey \citep{2005MNRAS.362..505C},
WiggleZ survey \citep{2011MNRAS.415.2892B,2011MNRAS.418.1707B},
6dF Galaxy Survey \citep{2011MNRAS.416.3017B},
and the SDSS-III Baryon Oscillation Spectroscopic Survey \citep[BOSS][]{2012MNRAS.427.3435A}.
The BAO feature has also been detected in imaging data sets using photometric redshifts
\citep{2007MNRAS.378..852P,2007MNRAS.374.1527B,2012ApJ...761...13S} and 
in galaxy cluster samples\footnote{For early work on cluster samples, see also \citep{2001ApJ...555...68M}.} \citep{2010MNRAS.401.2477H}.
Most recently, the acoustic peak has been detected in the Lyman $\alpha$
forest \citep{2012arXiv1211.2616B,2013arXiv1301.3459S,2013arXiv1301.3456K}, thereby extending the measurement
of cosmic distance to $z\approx2.3$.

Most of these detections of the BAO have used spherically averaged
clustering statistics, yielding a measurement of $D_V = D_A^{2/3}
(cz/H(z))^{1/3}$.  However, it is important to separate the line-of-sight
and transverse information for several reasons.  First, measuring $H(z)$ and 
$D_A(z)$ separately can give additional cosmological constraints at 
high redshift \citep{1979Natur.281..358A}.  Second, the interplay of shot noise
and sample variance varies with the angle of a pair to the line of
sight, so one can weight the data more optimally.  Third, the acoustic
peak is degraded in the line of sight direction by redshift-space 
distortions both from large scales \citep{1987MNRAS.227....1K} and small-scale
fingers of god \citep{1972MNRAS.156P...1J}.  Fully tracking all of the BAO 
information requires a non-spherical analysis of the clustering signal.

Such anisotropic analyses have been performed on SDSS-II data
\citep{2008ApJ...676..889O,2009MNRAS.399.1663G,2012MNRAS.426..226C,2012arXiv1206.6732X}.
Because of the moderate redshift of this data, $z\approx0.35$, the split of 
$H(z)$ and $D_A(z)$ does not improve the cosmological constraints
\footnote{As $z \rightarrow 0$, the different cosmological distances become
degenerate.} above
those of the $D_V(z)$ measurements.
But these papers have been important for developing analysis methods to be applied to higher
redshift samples.  Of particular relevance to this paper, \citet{2012MNRAS.419.3223K} present a method
that uses a split of the full correlation function based on the angle
of the pair to the line of sight, resulting in a correlation function
in each of two angular wedges.  
\citet{2012arXiv1206.6732X} present a method based on the monopole
and quadrupole of the correlation function that includes the effects
of density-field reconstruction \citep{2007ApJ...664..675E,2012MNRAS.427.2132P}.
 \citet{2012MNRAS.426..226C} extract the anisotropic signal from direct fits to
the redshift-space correlation function $\xi(r_p,\pi)$, where $\pi$ is the
separation of the pairs along the line of sight
and $r_p$ is the transverse separation.

In this paper, we extend the analysis of the SDSS-III BOSS
Data Release 9 (DR9) galaxy sample presented in \citet{2012MNRAS.427.3435A}
to include the anisotropic BAO information.
This sample has already yielded a 5~$\sigma$ detection of the
acoustic peak in a spherically averaged analysis \citep{2012MNRAS.427.3435A},
the most significant single detection of the acoustic peak yet.
\citet{2012MNRAS.427.3435A} uses this detection to measure $D_V$ at $z=0.57$
to 1.7\%.  In this paper and its companion papers 
\citep{kazin13,sanchez13,chuang13}, we will decompose the acoustic
peak detection to measure $H(z)$ and $D_A(z)$.

This paper will focus solely on the acoustic peak information.
Other cosmological information is present in the anisotropic
clustering data, particularly the large-scale redshift distortion
that results from the growth of cosmological structure and the
measurement of the Alcock-Paczynski signal from the broadband shape
of the correlation function.  This additional information has been
studied in \citet{2012MNRAS.426.2719R}, \citet{2013MNRAS.429.1514S}, 
and \citet{2012MNRAS.424.2339T}.
\citet{sanchez13} and \citet{chuang13} continue this analysis.  In
this paper as well as in \citet{kazin13}, we remove this additional
information by including flexible broadband clustering terms in our
fits.  After marginalizing over these terms, the distance measurements
are dominated by the sharp acoustic peak.  \citet{kazin13} presents
an analysis using the clustering wedges method of \citet{2012MNRAS.419.3223K}, whereas this
paper performs a monopole-quadrupole analysis following \citet{2012arXiv1206.6732X}
and presents the consensus of the two methods and a short cosmological
interpretation.

We also use this analysis of the DR9 data and mock catalogs as an
opportunity to further improve and test the methods for extraction
of the anisotropic BAO signal.  As the detection of the BAO improves
in the BOSS survey and future higher redshift surveys, such anisotropic
analyses will become the preferred route to cosmology.  
Extraction of the BAO to sub-percent accuracy is challenging because
of the strongly anisotropic and imperfectly predicted effect imposed
by redshift distortions and the partial removal of this anisotropy
by density-field reconstruction.  However, we will argue
that the extraction methods have been tested enough that the measurements
presented are limited by statistical rather than systematic errors.

The outline of the paper is as follows : \S~\ref{sec:fidcosmo} defines our
fiducial cosmology and conventions. \S~\ref{sec:analysis} describes the data and
mock catalogues and outlines the correlation function analysis methodology.
\S~\ref{sec:method} then describes how we constrain the angular diameter
distances and Hubble parameters from the data. \S~\ref{sec:mock_results} and
\S~\ref{sec:data_results} summarize our results from the mocks and data
respectively, while \S~\ref{sec:compare} compares results with previous
analyses. \S~\ref{sec:cosmology} presents the cosmological implications of these
results. We present our conclusions in \S~\ref{sec:conclusions}.

\section{Fiducial Cosmology} \label{sec:fidcosmo}
We assume a fiducial $\Lambda$CDM cosmology with $\Omega_{M}=0.274$,
$\Omega_{b}=0.0457$, $h=0.7$ and $n_s = 0.95$. We report physical angular
diameter distances defined by \citep[eg.][]{1999astro.ph..5116H}
\begin{align}
	 D_A(z) = \frac{1}{1+z} \frac{c}{H_0} & \left\{  \begin{array}{lr}
	 \frac{1}{\sqrt{\Omega_k}} \sinh [\sqrt{\Omega_k} E(z)] & \text{\rm for } \Omega_{k} > 0 \\
	 E(z) & \text{\rm for } \Omega_{k} = 0 \\ 
	 \frac{1}{\sqrt{-\Omega_k}} \sin [\sqrt{-\Omega_k} E(z)] & \text{\rm for } \Omega_{k} < 0 \\
	\end{array} \right. 
\end{align}
where
\begin{equation}
E(z)  = \int_{0}^{z} \frac{H_0 dz'}{H(z')}
\end{equation}
The angular diameter distance to $z=0.57$ for our fiducial cosmology is
$D_A(0.57)=1359.72\,{\rm Mpc}$, while the Hubble parameter is 
$H(0.57) = 93.56\,{\rm km/s/Mpc}$.
The sound horizon for this cosmology
is $r_s = 153.19\,{\rm Mpc}$, where we adopt the conventions in
\citet{1998ApJ...496..605E}.
These distances are all in $\rm Mpc$, not $h^{-1} {\rm Mpc}$.
We note that slightly different definitions of
the sound horizon are in use; for example, the sound horizon quoted
by CAMB \citep{2000ApJ...538..473L} differs from our choice by 2\%.  For further
discussion, see \citet{2012MNRAS.427.2168M}.

\section{Analysis} \label{sec:analysis}
\subsection{Data}\label{sec:data}
SDSS-III BOSS \citep{2013AJ....145...10D} is a spectroscopic survey designed to obtain spectra and redshifts
for 1.35 million galaxies over 10,000 square degrees of sky and the course of
five years (2009-2014). BOSS galaxies are targeted from SDSS imaging, which was
obtained using the 2.5m Sloan Foundation Telescope \citep{2006AJ....131.2332G}
at Apache Point Observatory in New Mexico.  The five-band imaging
\citep{1996AJ....111.1748F,2002AJ....123.2121S,2010AJ....139.1628D} was taken
using a drift-scan mosaic CCD camera \citep{1998AJ....116.3040G} to a limiting
magnitude of  $r\simeq 22.5$; all magnitudes were corrected for Galactic
extinction using the maps of \citet{1998ApJ...500..525S}. A 1000 object
fiber-fed spectrograph \citep{2012arXiv1208.2233S} measures spectra for targeted objects. 
We refer the reader to the following
publications for details on astrometric calibration
\citep{2003AJ....125.1559P}, photometric reduction
\citep{2001ASPC..238..269L}, photometric calibration
\citep{2008ApJ...674.1217P} and spectral classification and redshift
measurements \citep{2012AJ....144..144B}. All of the BOSS targeting
is done on Data Release 8 \citep[DR8][]{2011ApJS..193...29A}
photometry, and all
spectroscopic data used in this paper has been released as part of
Data Release 9 \citep[DR9][]{2012ApJS..203...21A}.

BOSS targets two populations of galaxies, using two combinations
of colour-magnitude cuts to achieve a number density of
$3\times10^{-4}\,h^{3}{\rm Mpc}^{-3}$ at  $0.2 < z < 0.43$ (the
LOWZ sample) and $0.43 < z < 0.7$ (the CMASS sample). A description
of both target selection algorithms can be found in
\citet{2013AJ....145...10D}.  This paper focuses exclusively on the
CMASS sample.

\subsection{Catalogue creation}

The treatment of the sample is in every way identical to that presented in
\citet{2012MNRAS.427.3435A}, to which we refer the reader for full details on
the angular mask and catalogue creation.
We use the {\sc Mangle\/} software \citep{2008MNRAS.387.1391S} to trace the
areas covered by the survey, and to define the angular completeness in each
region. The final mask combines the outline of the survey regions and position
of the spectroscopic plates with a series of ``veto'' masks used to exclude
regions of poor photometric quality, regions around the centre posts of the
plates where fibers cannot be placed, regions around bright stars and regions
around higher-priority targets (mostly high-redshift quasars). In total, the
``veto'' mask excludes $\sim$ 5\% of the observed area.

We define weights to deal with the issues of close-pair
corrections, redshift-failure corrections, systematic targeting
effects and effective volume (again we refer to \citealt{2012MNRAS.427.3435A} for
full details, successes and caveats related to each weighting
scheme):

\begin{enumerate}
\item Close-pair correction ($w_{\rm cp}$) : We assign a weight of $w_{\rm
cp}=1$ to each galaxy by default, and we add one to this for each CMASS target
within $62^{\prime\prime}$ that failed to get a fiber allocated due to
collisions.
\item Redshift-failure correction ($w_{\rm rf}$) : For each target with a
failed redshift measurement we upweight the nearest target object for which a galaxy
redshift (or stellar classification) has been successfully achieved.
\item Systematic weights ($w_{\rm sys}$) : We correct for an observed
dependence of the angular targeting density on stellar density \citep{2012MNRAS.424..564R}
by computing a set of angular weights that depend on stellar density and fiber
magnitude in the $i$-band and that minimise this dependency.
\item FKP weights ($w_{\rm FKP}$) : We implement the weighting scheme of
\citet{1994ApJ...426...23F} in order to optimally balance the effect of
shot-noise and sample variance in our measurements.
\end{enumerate}

These weights are combined to give a total weight to each galaxy in the
catalogue as $w_{\rm tot} = w_{\rm FKP} w_{\rm sys}(w_{\rm rf} + w_{\rm cp}
-1)$. Both $w_{\rm rf}$ and $w_{\rm cp}$ are one in the absence of any correction,
and we therefore need subtract one (in general, one less than the number of additive
weights) from their sum. 

We use 264283 galaxies in the redshift range $0.43 < z <
0.7$, covering an effective area of 3275 sq. degrees (see Table 1
of \citealt{2012MNRAS.427.3435A} for more details). Random catalogues
with 70 times the density of the
corresponding galaxy catalogues and the same redshift and angular
window functions are computed for the Northern and Southern Galactic
Caps separately, using the ``shuffled" redshifts method defined in
\citet{2012MNRAS.424..564R}. Angular completeness, redshift failures and close
pairs are implemented exactly as in \citet{2012MNRAS.427.3435A}.

\subsection{Measuring the correlation function}
The two-dimensional  correlation function is computed using the
Landy-Szalay estimator \citep{1993ApJ...412...64L} as:
\begin{equation}
\xi(r,\mu) = \frac{ DD(r,\mu) -2DR(r,\mu)+RR(r,\mu)}{RR(r,\mu)}
\end{equation}
where $\mu$ is the cosine between a galaxy pair and the line of
sight, and DD, DR and RR are normalised and weighted data-data,
data-random and random-random galaxy pair counts respectively. The
correlation function is computed in bins of $\Delta r = 4 h^{-1}$Mpc
and $\Delta \mu =0.01$. 
Multipoles and wedges -
the two estimators that underpin the results in this paper - are constructed
from $\xi(r,\mu)$ following Section~\ref{sec:method_estimators}.

\subsection{Mock Catalogues} \label{sec:method_mocks}

We use 600 galaxy mock catalogues of \citet{2013MNRAS.428.1036M}
to estimate sample covariance matrices for all measurements in
this paper. These mocks are generated using a method similar to the
PTHalos mocks of \citet{2002MNRAS.329..629S} and recover the
amplitude of the clustering of halos to within 10 per cent. Full
details on the mocks can be found in \citet{2013MNRAS.428.1036M}. 
The mock catalogues correspond to a box at $z=0.55$
(and do not incorporate any evolution within the redshift of the
sample, which is expected to be small), include redshift-space
distortions, follow the observed sky completeness and reproduce the
radial number density of the observed galaxy sample.

\begin{figure*}
\vspace{0.4cm}
\includegraphics[width=3in]{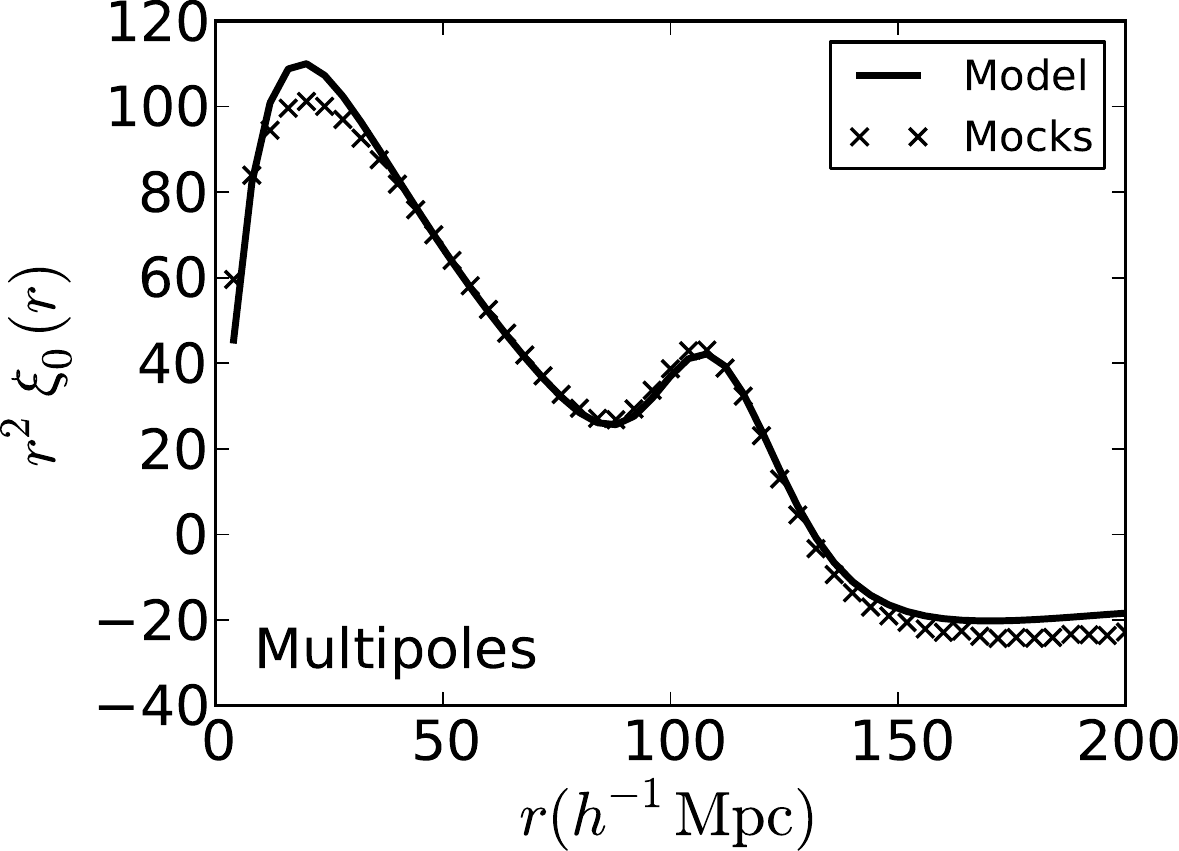}
\includegraphics[width=3in]{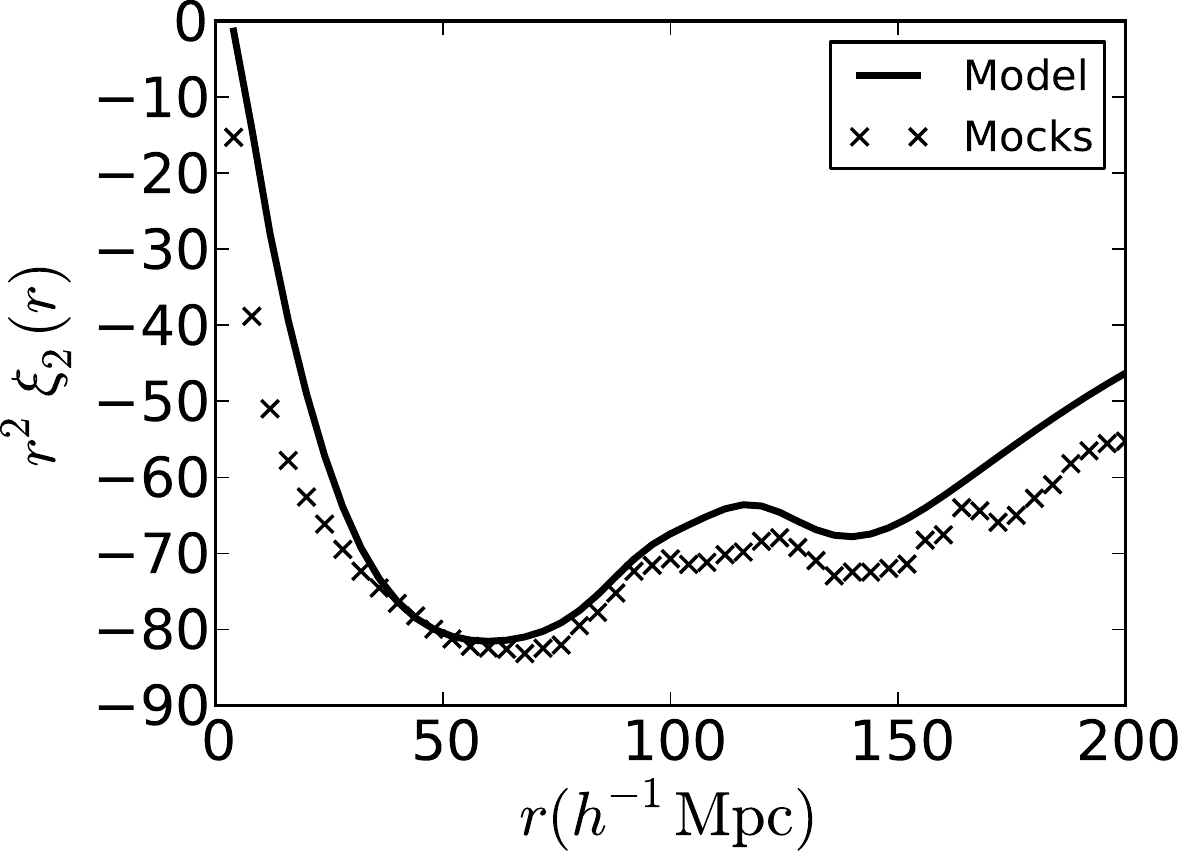}
\includegraphics[width=3in]{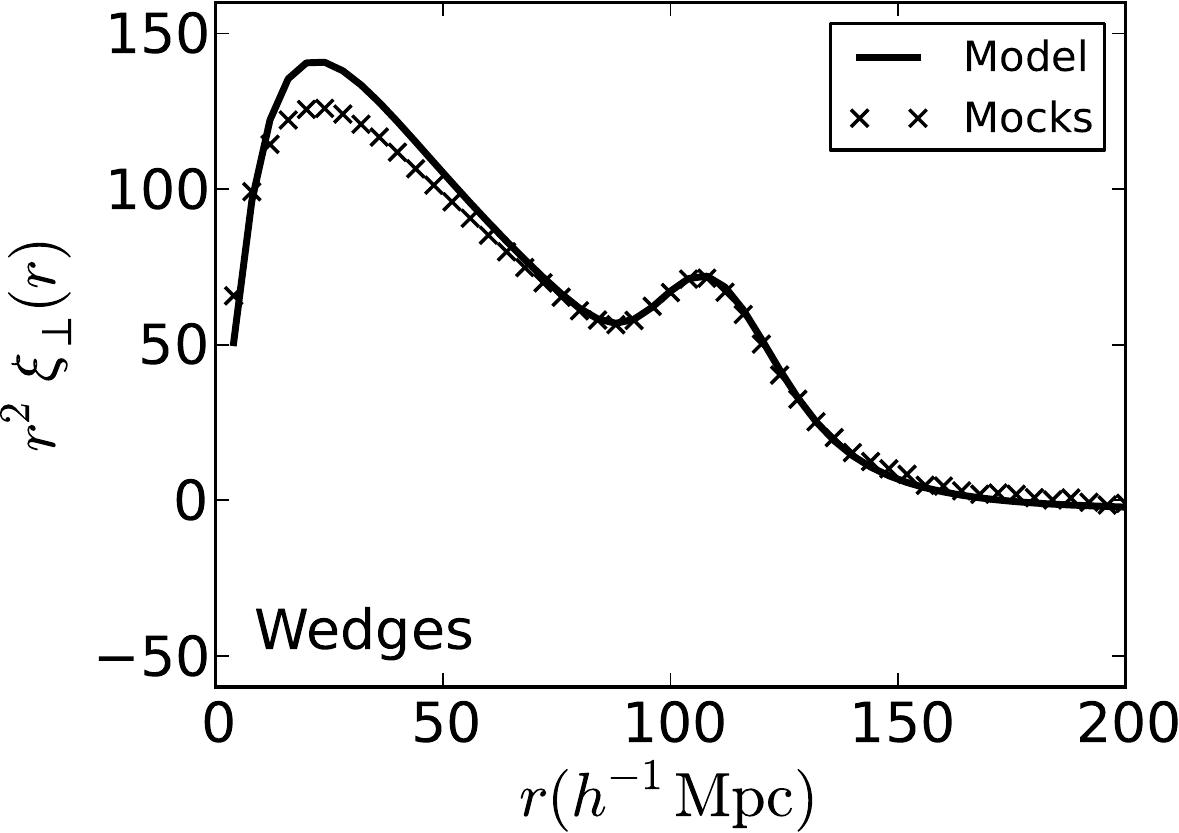}
\includegraphics[width=3in]{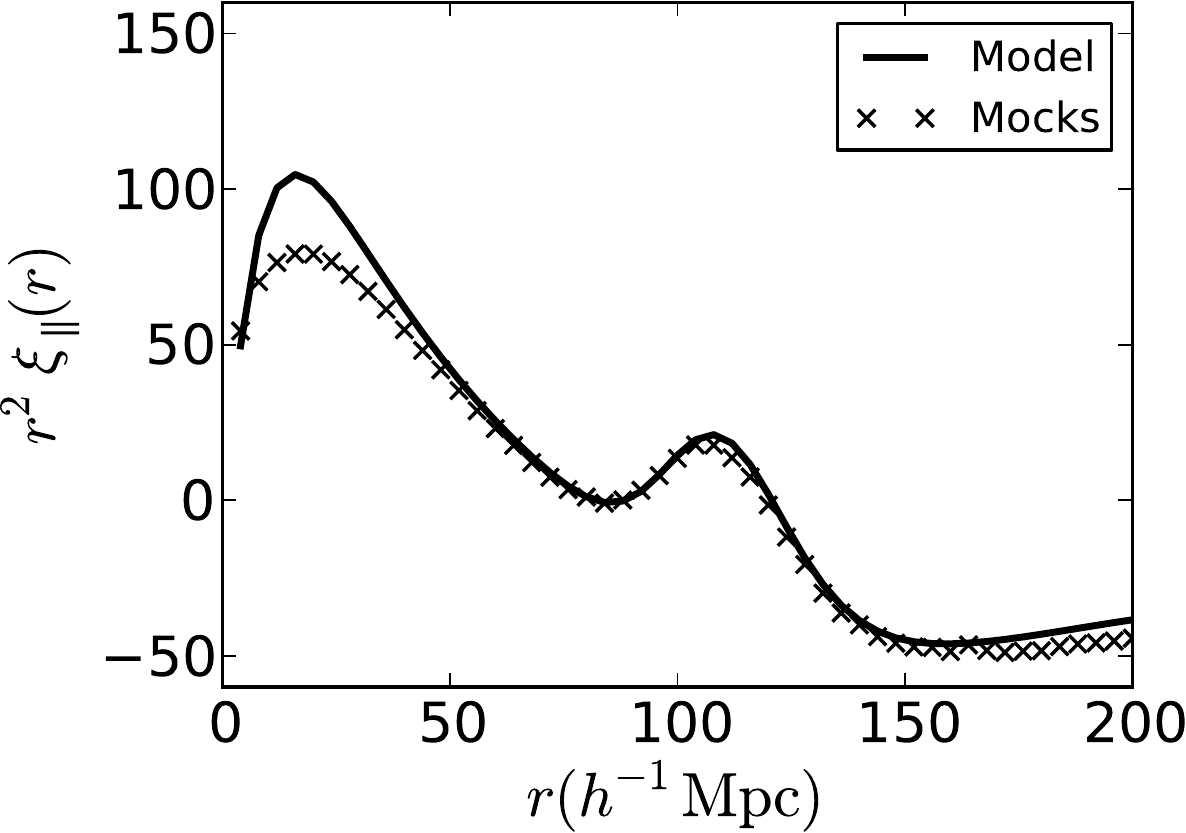}
\caption{Average of mocks (crosses) with our model of the correlation function (solid line)
overplotted. The upper panels show the monopole (left) and quadrupole (right) while the lower
panels plot the transverse (left) and radial (right) wedges.
No fit to the shape was done here, but the models were normalized 
to match the observed signals. 
}
\label{fig:mockmod}
\end{figure*}

Figure~\ref{fig:mockmod} shows the average monopole and quadrupole and
transverse and radial wedges of the correlation function over the 600
mocks (see \S~\ref{sec:method_estimators} for definitions). 

\subsection{Reconstruction} \label{sec:method_compute}

Following \citet{2012MNRAS.427.3435A}, we attempt to improve
the statistical sensitivity of the BAO measurement by reconstructing the linear density field, 
correcting for the effects
of non-linear structure growth around the BAO scale \citep{2007ApJ...664..675E}.
The reconstruction technique has been successfully
implemented on an anisotropic BAO analysis by \citet{2012arXiv1206.6732X}
using SDSS-II Luminous Red Galaxies at $z=0.35$, achieving an
improvement of a factor of 1.4 on the error on $D_A$ and of 1.2 on
the error on $H$, relative to the pre-reconstruction case.
\citet{2012MNRAS.427.3435A} successfully applied reconstruction on
the same dataset used here when measuring $D_V$ from spherically-averaged
two-point statistics. They observed only a slight reduction in the
error of $D_V$, when compared to the pre-reconstruction case, but at
a level consistent with mock galaxy catalogues.

The algorithm used in this paper is described in detail in 
\citet{2012MNRAS.427.2132P},
to which we refer the reader for full
details. Briefly, reconstruction uses the
density field to construct a displacement field that attempts to
recover a galaxy spatial distribution that more closely reproduces
the expected result from linear growth. A summary of the implementation
of the algorithm on the CMASS DR9 dataset (as used here) is given
in Section 4.1 of \citet{2012MNRAS.427.3435A}.

\begin{figure*}
\vspace{0.4cm}
\includegraphics[width=3in]{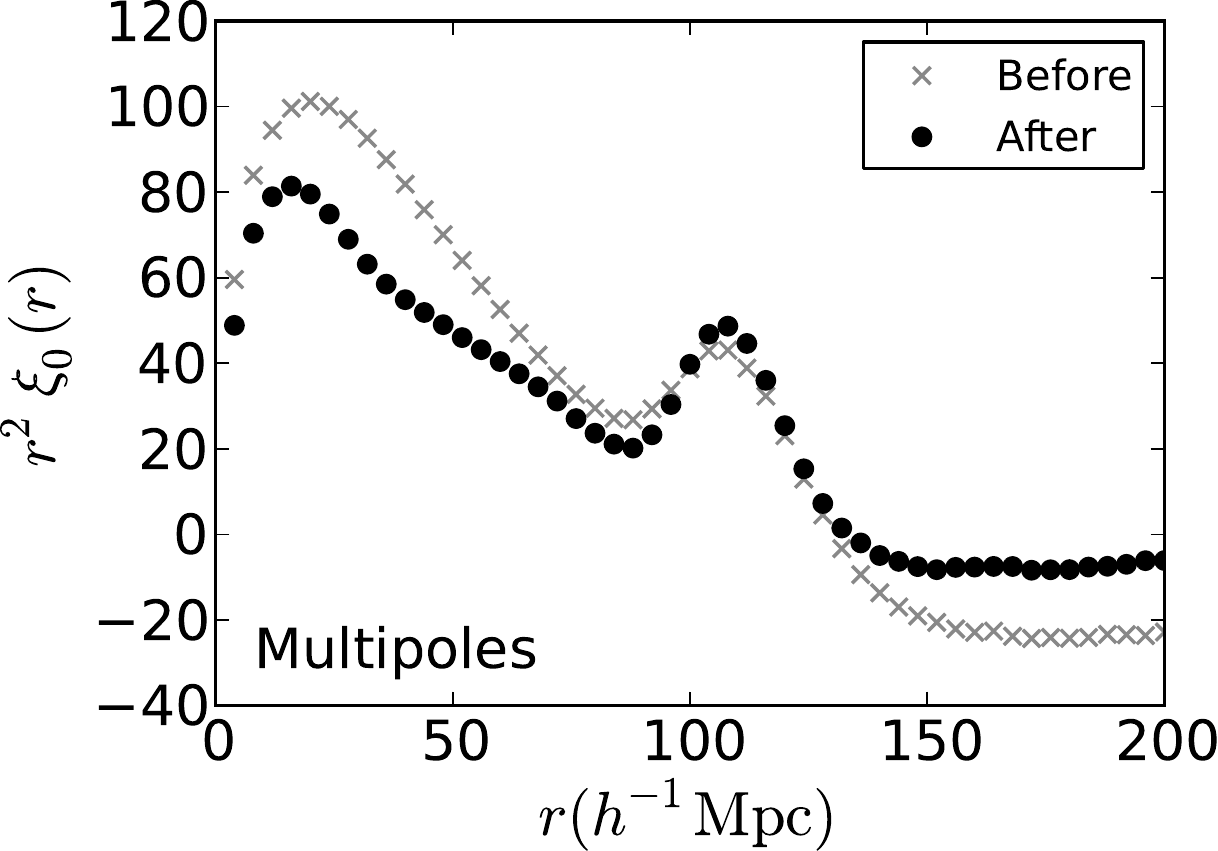}
\includegraphics[width=3in]{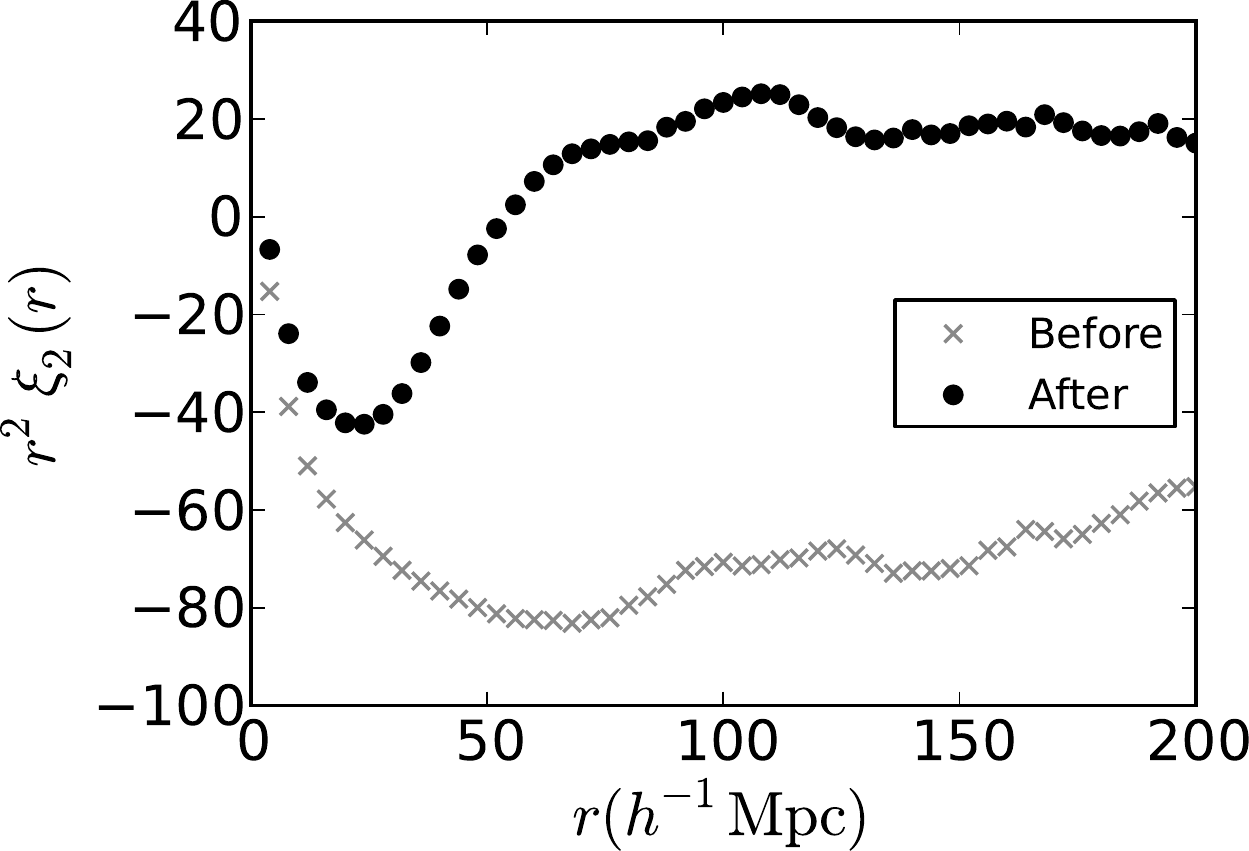}
\includegraphics[width=3in]{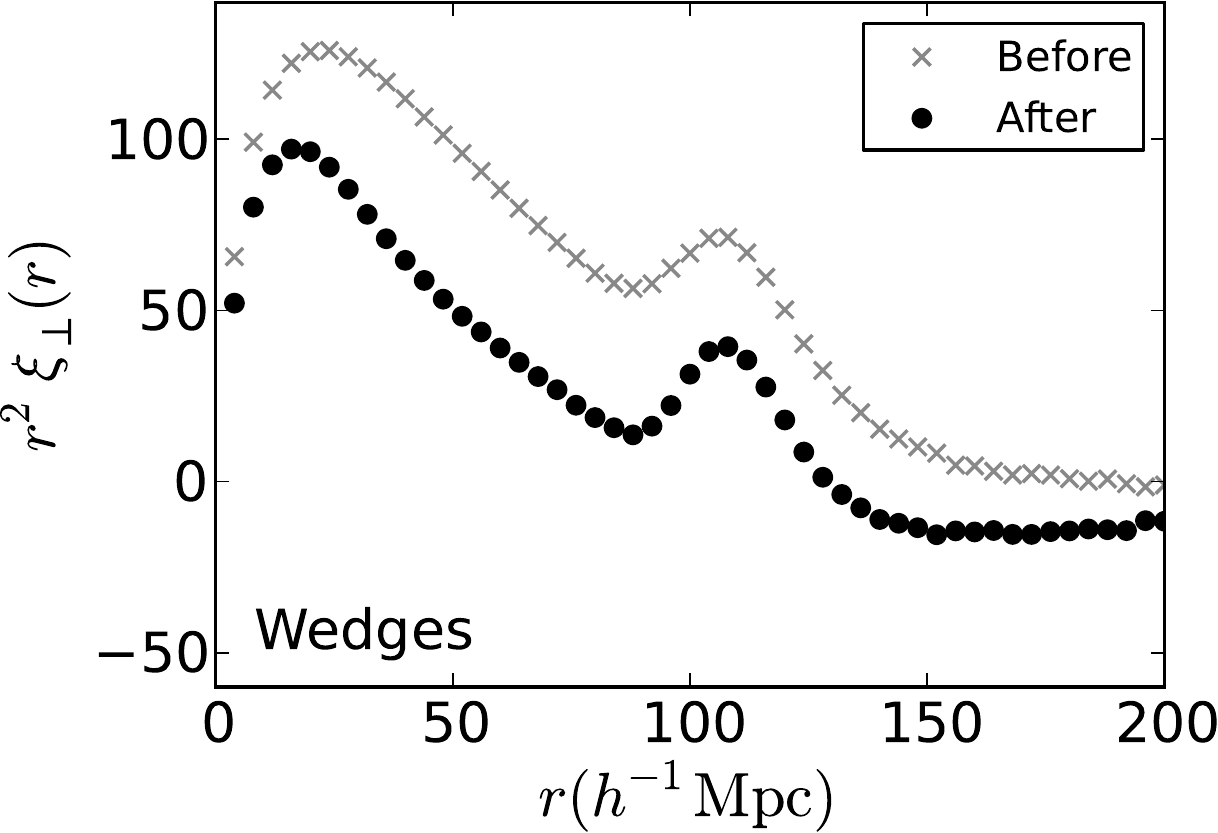}
\includegraphics[width=3in]{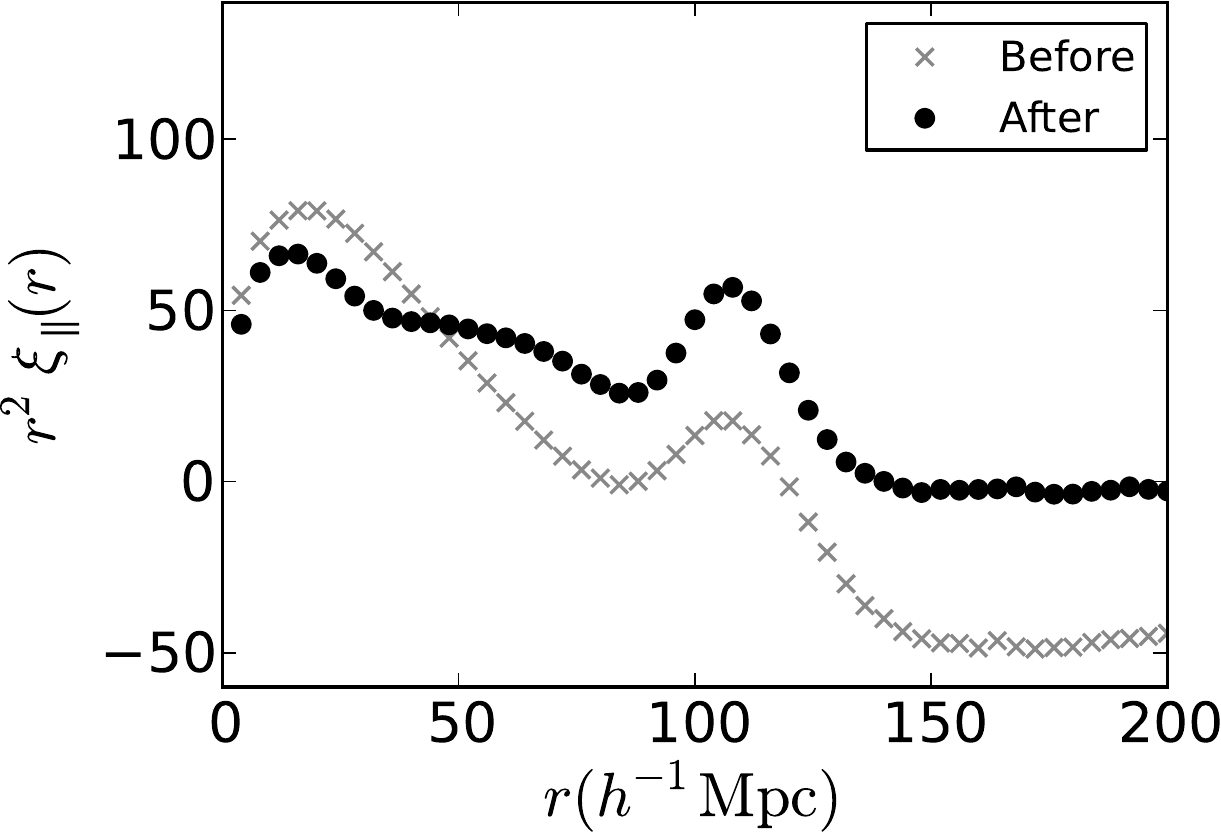}
\caption{Average of mocks before (grey) and after (black) reconstruction. We see
a sharpening of the acoustic feature in the monopole, and a drastic decrease in
amplitude of the quadrupole on large scales, which is consistent with the fact
that reconstruction removes large-scale redshift-space distortions. The
correlation function of both angular wedges show a clear sharpening of the
acoustic feature, a reduction of amplitude on large-scales in the transverse wedge and a corresponding
increase in the amplitude in the radial wedge.}
\label{fig:mockrec}
\end{figure*}

Figure~\ref{fig:mockrec} shows the average of the multipoles and
wedges of the correlation function before and after reconstruction.
Reconstruction sharpens the acoustic feature in the monopole, while 
decreasing the amplitude of the quadrupole, particularly at large scales where it goes close to
zero. These changes are manifested in the wedges as a sharpening of the BAO feature in both wedges 
as well as a decrease in the difference in amplitude between the transverse and radial wedge.
This is expected since reconstruction removes much of the
large-scale redshift-space distortions. Assuming the correct cosmology, an ideal
reconstruction algorithm would perfectly restore isotropy and eliminate the
quadrupole \footnote{Reconstruction only corrects for the dynamical quadrupole
induced by peculiar velocities. The incorrect cosmology would induce a
quadrupole through the Alcock-Paczynski test, even in the absence of this
dynamical quadrupole (see \citet{2008PhRvD..77l3540P},
\citet{2012MNRAS.419.3223K} and \citet{2012arXiv1206.6732X} for a detailed
discussion and illustrative examples.)} 
In the wedges, this would be manifest by
the transverse and radial wedge being the same. We depart from this ideal because of an 
imperfect treatment of of nonlinear evolution and small-scale effects, the survey geometry and imperfections 
in the implementation of the reconstruction algorithm itself. However, these
imperfections affect the broad band shape of the correlation function but do not
bias the location of the BAO feature, as we explicitly demonstrate below.

\section{Method} \label{sec:method}
\subsection{Parametrization}
\label{sec:method_param}

The choice of an incorrect cosmology distorts the BAO feature in the galaxy
correlation function, stretching it in both the
transverse and radial directions. The shift in the transverse direction
constrains the angular diameter distance relative to the sound horizon, 
$D_{A}(z)/r_{s}$, while the radial direction constrains the relative Hubble
parameter $cz/(H(z) r_s)$. As is standard in the BAO literature, when 
fitting for these, we parameterize with respect to a fiducial model
(indicated by a superscript ${\rm fid}$) :
\begin{equation}
\alpha_{\perp} = \frac{D_{A} (z) \rsf}{D^{\rm fid}_{A} r_{s}} \,,
\end{equation}
and 
\begin{equation}
\alpha_{||} = \frac{H^{\rm fid}(z) \rsf}{H(z) r_{s}} \,.
\end{equation}

An alternative parametrization is to decompose these shifts into isotropic and
anisotropic components. We define an isotropic shift $\alpha$
\begin{equation}
\alpha = \alpha_{\perp}^{2/3} \alpha_{||}^{1/3} \,, 
\end{equation}
and the anisotropic shift $\epsilon$ by
\begin{equation}
1+ \epsilon = \left( \frac{\alpha_{||}}{\alpha_{\perp}} \right)^{1/3} \,.
\end{equation}
For the fiducial cosmological model, we have 
$\alpha = \alpha_\perp = \alpha_\parallel = 1$ and $\epsilon = 0$.
For completeness, we note 
\begin{align}
\alpha_{\perp} &= \frac{\alpha}{1+\epsilon} \\
\alpha_{||} &= \alpha (1+\epsilon)^2  \,\,.
\end{align}

The majority of previous BAO results have restricted their
analysis to the isotropically averaged correlation function and have therefore presented their 
results in terms of $\alpha$.  In this work, the fitting of the multipoles uses
the $\alpha, \epsilon$ parametrization, while the clustering wedges use
$\alpha_{||}, \alpha_{\perp}$. While these are formally equivalent, the
choices of data fitting ranges and priors imply that different parametrizations
probe somewhat different volumes in model space, an issue we discuss in later
sections. Although we use different parametrizations, we transform 
to the $\alpha_{\perp},\alpha_{||}$ parametrization when presenting
results for ease of comparison. 

\subsection{Clustering Estimators : Multipoles and Wedges}
\label{sec:method_estimators}

Measuring both $D_{A}$ and $H$ requires an estimator of the full 2D correlation
function $\xi(s, \mu)$ where $s$ is the separation between two points and
$\mu$ the cosine of angle to the line of sight.
However, working with the full 2D correlation function is impractical, given
that we estimate our covariance matrix directly from the sample covariance of
the mock catalogues. We therefore compress the 2D correlation function into a
small number (2 in this paper) of angular moments and use these for our analysis.

The first set of these moments are the Legendre moments (hereafter referred to
as multipoles) :
\begin{equation}
\xi_{\ell}(r) = \frac{2 \ell +1}{2} \int_{-1}^{1} d\mu \, \xi(r, \mu)
L_{\ell}(\mu) \,\,,
\end{equation}
where $L_{\ell}(\mu)$ is the $\ell^{\rm th}$ Legendre polynomial.
We focus on the two lowest non-zero multipoles, the monopole ($\ell = 0$) and
the quadrupole ($\ell = 2$). Within linear theory and the plane-parallel
approximation, only the $\ell = 0, 2$ and $4$ multipoles are non-zero. However,
on these scales, the hexadecapole is both small and noisy; we neglect it
in our analysis. Furthermore, after reconstruction, the effect of redshift space
distortions is significantly reduced, further decreasing the influence of the 
hexadecapole. 

We also consider an alternate set of moments, referred to as clustering wedges
\citep{2012MNRAS.419.3223K}:
\begin{equation}
\xi_{\Delta\mu}(r)=
	\frac{1}{\Delta\mu}\int_{\mu_{\rm min}}^{\mu_{\rm min}+\Delta\mu}
	d\mu\,\xi(r,\mu)\,,
\end{equation}
For purposes of this study 
we choose $\Delta\mu=0.5$ such 
that we have a basis comprising of a 
``radial" component 
$\xi_{||}(s)\equiv\xi(\mu>0.5,s)$ 
and a ``transverse" component 
$\xi_{\perp}(s)\equiv\xi(\mu<0.5,s)$. 
As the clustering wedges are an alternative 
projected basis of $\xi(\mu,\vec{s})$, 
we do not expect tighter constraints but rather 
find these useful for testing for systematics, 
as well as other technical advantages. 
A full in-depth description 
of the method, and comparison to clustering 
multipoles is described in \citet{kazin13}.

\subsection{A Model for the Correlation Function}

Robustly estimating $D_A$ and $H$ from the correlation function requires a model
for the 2D correlation function. We start with the 2D power spectrum template :
\begin{equation}
P_t(k,\mu) = (1+\beta\mu^2)^2 F(k,\mu,\Sigma_s)P_{\rm dw}(k,\mu)
\label{eqn:tdp}
\end{equation}
where
\begin{equation}
F(k,\mu,\Sigma_s) = \frac{1}{(1+k^2\mu^2\Sigma_s^2)^2}
\label{eqn:fog}
\end{equation}
is a streaming model for the Finger-of-God (FoG) effect \citep{1994MNRAS.267.1020P}
and the $(1+\beta\mu^2)^2$ term is the Kaiser model for large-scale
redshift-space distortions \citep{1987MNRAS.227....1K}. Here $\Sigma_s$ is the
streaming scale which we set to $1\hMpc$ based on test fits to the
average mock correlation function. Note that there are currently two
similar Lorentzian models for FoG in the literature. The difference
arises from the following: 1) assuming that small-scale redshift space
distortions can be modeled by convolving the density field with an
exponential gives two powers of the Lorentzian in Fourier space as in
Equation (\ref{eqn:fog}), 2) assuming that the pairwise velocity field
is exponentially distributed results in only one power of the Lorentzian
as in \citet{1998ASSL..231..185H}. We let $\beta$ vary in our fits (note that
$\beta$ is degenerate with quadrupole bias). To limit the model from picking
unphysical values of $\beta$, we place a Gaussian prior centered on
$f/b\sim\Omega_m(z)^{0.55}/b = 0.25$ before reconstruction and 0 after
reconstruction with 0.2 standard deviation. The post-reconstruction
prior center of $\beta=0$ is chosen since we expect reconstruction to
remove large-scale redshift space distortions.

The de-wiggled power spectrum $P_{\rm dw}(k,\mu)$ is defined as
\begin{align}
P_{\rm dw}(k,\mu) =& [P_{\rm lin}(k) - P_{\rm nw}(k)] \nonumber \\
& \cdot \exp \bigg[
-\frac{k^2\mu^2\Sigma_\parallel^2 +k^2(1-\mu^2)\Sigma_\perp^2}{2} \bigg]
+P_{\rm nw}(k) \nonumber \\
\label{eq:template}
\end{align}
where $P_{\rm lin}(k)$ is the linear theory power spectrum and
$P_{\rm nw}(k)$ is a power spectrum without the acoustic oscillations
\citep{1998ApJ...496..605E}. $\Sigma_\parallel$ and $\Sigma_\perp$ are the radial and
transverse components of $\Sigma_{\rm nl}$, i.e.  $\Sigma_{\rm nl}^2 =
(\Sigma_\parallel^2 + \Sigma_\perp^2)/2$, where $\Sigma_{\rm nl}$
is the standard term used to damp the BAO to model the effects of
non-linear structure growth \citep{2007ApJ...664..660E}. Here, the damping is
anisotropic due to the Kaiser effect. We set $\Sigma_\perp=6\hMpc$
and $\Sigma_\parallel=11\hMpc$ before reconstruction and
$\Sigma_\perp=\Sigma_\parallel=3\hMpc$ after reconstruction as in
\citet{2012arXiv1206.6732X}.

Given this model of the 2D power spectrum, we decompose it into its 
Legendre moments, 
\begin{equation}
P_{\ell,t}(k) = \frac{2\ell+1}{2}\int^{1}_{-1} P_t(k, \mu)L_\ell(\mu) d\mu
\end{equation}
which can then be transformed to configuration space using
\begin{equation}
\xi_{\ell,t}(r) = i^\ell \int \frac{k^3 d\log(k)}{2\pi^2} P_{\ell,t}(k)
j_\ell(kr).
\end{equation}
Here, $j_\ell(kr)$ is the $\ell$-th spherical bessel function and
$L_\ell(\mu)$ is the $\ell$-th Legendre polynomial. We then synthesize the 
2D correlation function from these moments by :
\begin{equation}
\xi(r, \mu) = \sum_{\ell = 0}^{\ell_{\rm max}} \xi_{\ell}(r) L_{\ell}(\mu) \,\,.
\end{equation}
In this work, we truncate the above sum at $\ell_{\rm max} = 4$. 

In order to compare to data, we must map the observed $r_{\rm obs}, \mu_{\rm
obs}$ pairs (defined for a fiducial cosmology) to their true values $r, \mu$.
These transformations are most compactly written by working in transverse
($r_{\perp}$) and radial ($r_{\parallel}$) separations define by
\begin{align}
r^2 & = r_{\perp}^2  + r_{\parallel}^2 \\
\mu & = \frac{r_{\parallel}}{r}
\end{align}
We then simply have
\begin{align}
r_{\perp} & = \alpha_{\perp} r_{\perp, {\rm obs}} \\
r_{\parallel} & = \alpha_{\parallel} r_{\parallel, {\rm obs}} .
\end{align}
Expressions in terms of $r, \mu$ are in \citet{2012arXiv1206.6732X} and
\citet{kazin13}. One can then compute $\xi(r, \mu)_{\rm obs}$ and project on to
either the multipole or wedge basis.

Our final model for the correlation function includes nuisance parameters 
to absorb imperfections in the overall shape of the model due to mismatches
in cosmology or potential smooth systematic effects. In particular, we fit
\begin{align}
\xi_{0}(r) &= B_0^2 \xi_{0}(r) + A_0(r) \nonumber \\ 
\xi_{2}(r) &= \xi_{2}(r) + A_2(r)
\label{eqn:monoquadt}
\end{align}
and
\begin{align}
\xi_{\perp}(r) &= B_{\perp} \xi_{0, \perp}(r) + A_{\perp}(r) \nonumber \\ 
\xi_{\parallel}(r) &= B_{\parallel} \xi_{\parallel}(r) + A_{\parallel}(r)
\label{eqn:perppart}
\end{align}
where
\begin{equation}
A_{\ell}(r) = \frac{a_{\ell,1}}{r^2} + \frac{a_{\ell,2}}{r} + a_{\ell,3}; \,
\ell=0,2,\perp, \parallel \,.
\label{eqn:fida}
\end{equation}
Note that these correlation functions are all in observed coordinates; we just
suppress the ${\rm obs}$ subscripts for brevity.
The $A_\ell(r)$ marginalize
errors in broadband (shape) information (eg. scale-dependent bias and
redshift-space distortions)  through the $a_{\ell,
1}\ldots a_{\ell, 3}$ nuisance parameters \citep{2012MNRAS.427.2146X}. 
$B_0^2$ is a bias-like term that adjusts the amplitude of the model to fit the data. We
perform a rough normalization of the model to the data before fitting
so $B_0^2$ should be $\sim1$. To ensure $B_0^2$ is positive (a negative
value would be unphysical), we perform the fit in $\log(B_0^2)$ using a
Gaussian prior with standard deviation 0.4 centered at 0 as described
in \citet{2012MNRAS.427.2146X}. While the multipole analysis does not include an
analogous term for $\xi_2$, we allow $\beta$ to vary, effectively allowing the
amplitude of the quadrupole to change. In the case of the wedges analyses,
$\beta$ is kept fixed, but the amplitudes of both wedges are free to vary. 
No additional priors are imposed on these amplitudes.
The clustering wedges analyses fit 76 data points with 10 parameters, while the
multipole analyses fit 80 data points with 10 parameters.

We also place a 15\% tophat prior on $1+\epsilon$ to limit low S/N measurements 
from exploring large excursions in $\epsilon$. Such a prior should have no
impact for standard cosmological models. In order to demonstrate this, we sample
cosmologies with $\Omega_K$, $w_0$ and $w_a$ free from the WMAP7 posterior
distribution and compute $\epsilon$ for each case. The largest
(absolute) excursion is $\sim$ 8\% with 95\% of points
between$-0.058<\epsilon<0.045$, justifying the choice of our prior.

We assume a Gaussian likelihood for the correlation functions :
\begin{equation}
\chi^2 = (\vec{m} - \vec{d})^T C^{-1} (\vec{m}-\vec{d})
\end{equation}
where $\vec{m}$ is the model and $\vec{d}$ is the data. The inverse covariance
matrix is a scaled version of the inverse of the sample covariance matrix $C_s$
\citep{Muirhead, 2007A&A...464..399H}
\begin{equation}
C^{-1} = C^{-1}_s \frac{N_{\rm mocks} - N_{\rm bins} - 2}{N_{\rm mocks} - 1}
\end{equation}
with the factor correcting for the fact that the inverse of the sample
covariance matrix estimated from $N_{\rm mocks}$ is a biased estimate of the
inverse covariance matrix. 

The multipole and the clustering wedges analyses handle this likelihood surface
differently. The wedges analysis uses a Markov Chain Monte Carlo
algorithm to sample from the posterior distribution of $\alpha_{\perp}$ and $\alpha_{||}$,
marginalizing over all the remaining parameters. The multipole analysis maps out
the likelihood surface in $\alpha$ and $\epsilon$, analytically marginalizing 
over the linear parameters in the model, but using the maximum likelihood 
values for the non-linear parameters. In addition, to
suppress unphysical downturns in the $\chi^2$ distribution at small
$\alpha$ (corresponding to the BAO feature being moved to scales larger than the
range of the data being fit, see \citet{2012MNRAS.427.2146X} for more
details), we apply a Gaussian prior on $\log(\alpha)$ with a standard deviation
of 0.15. As we see below, in the limit of a well detected BAO feature, these
differences have a small (compared to our statistical errors) impact on the
derived distances. However, in the opposite limit of a poorly measured BAO
feature, these differences can be important. We explore this further in the next
section. Fortunately, the DR9 sample has a well defined BAO feature and we
obtain consistent results irrespective of the method.

\section{Mock Results} \label{sec:mock_results}

\subsection{Multipole Fits}
% \begin{figure*}
% \includegraphics[width=3in]{plots/alpha_snr}
% \includegraphics[width=3in]{plots/epsilon_snr}
% \caption{Histograms of $(D_A-\langle D_A \rangle)/\sigma_{D_A}$ (left)
% and $(H-\langle H \rangle)/\sigma_H$ (right) after reconstruction.
% \label{fig:snrhist}}
% \end{figure*}

\begin{figure}
\includegraphics[width=3in]{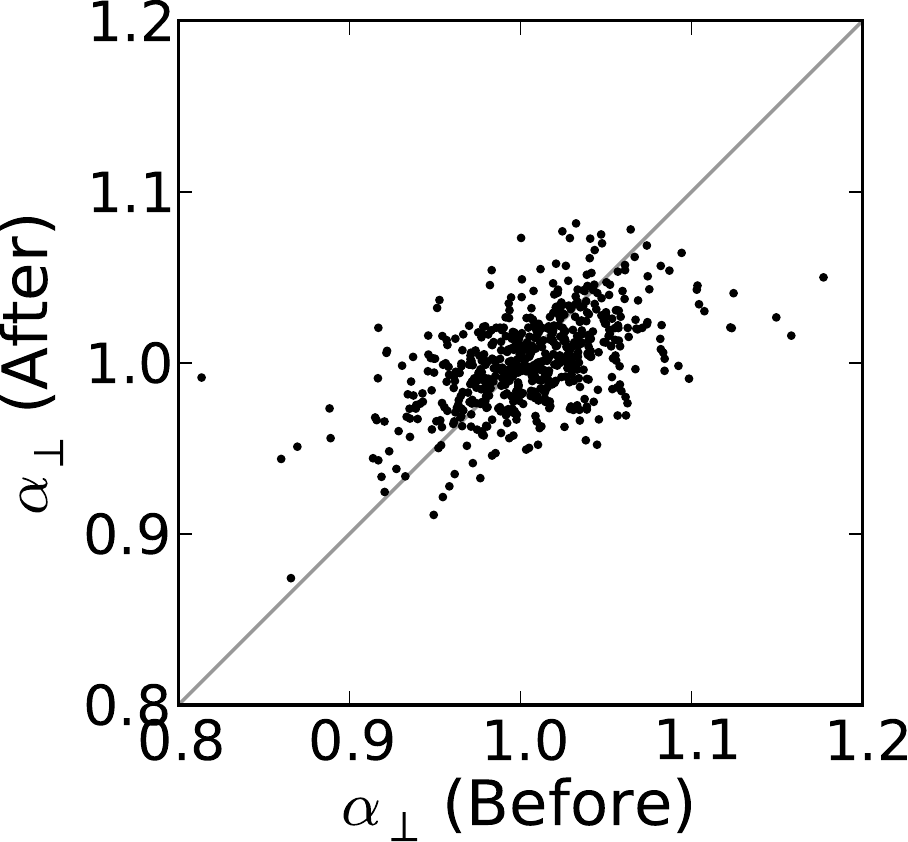}
\includegraphics[width=3in]{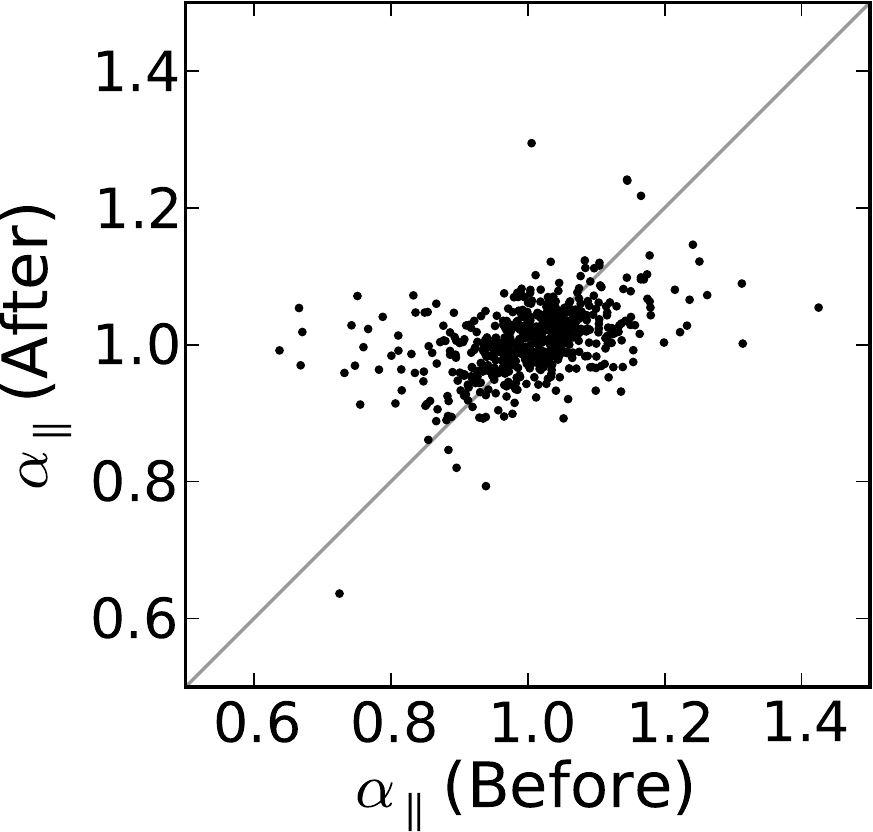}
\caption{A comparison of $\alpha_{\perp}$  and $\alpha_{\parallel}$
for the 600 mock catalogues before and after reconstruction. These
values have been derived from the multipole analysis. The points mostly lie on
the 1:1 line, but the number of outliers are reduced after reconstruction.}
\label{fig:reccomp}
\end{figure}

\begin{figure}
\includegraphics[width=3in]{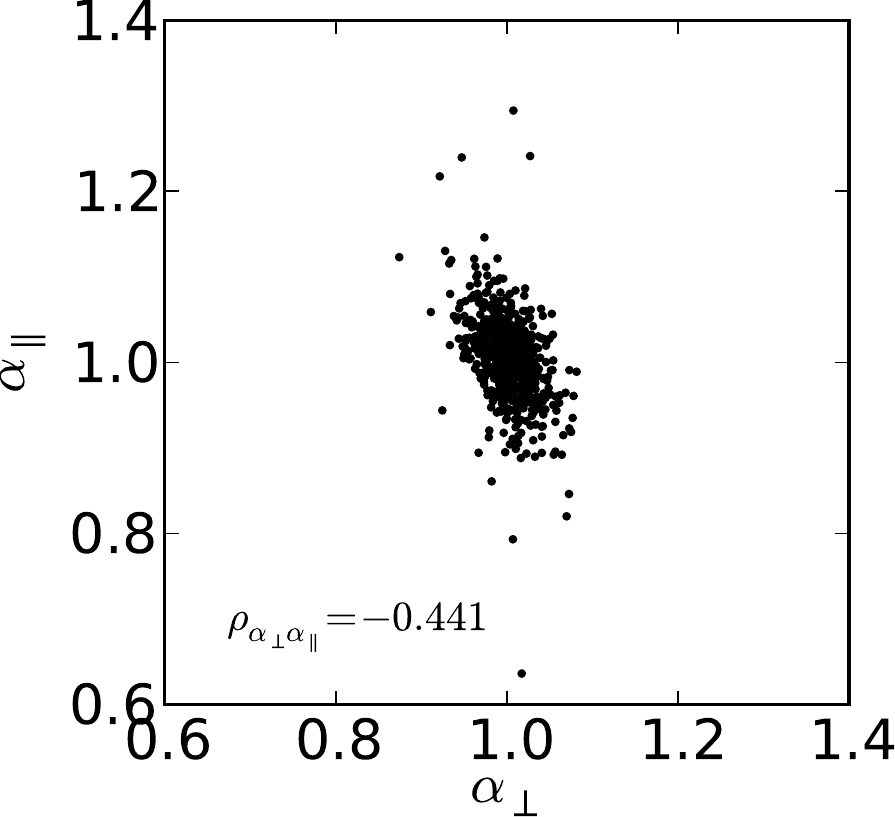}
\caption{The distribution of $\alpha_\perp$ versus $\alpha_\parallel$ from the
600 mock catalogues after reconstruction. As in Fig.~\ref{fig:reccomp}, these
values are derived from the multipole analysis. The estimates of the two
distances are anti-correlated, with a correlation coefficient of $\sim -0.44$.
Note that $H \sim 1/\alpha_{||}$.}
\label{fig:dah}
\end{figure}

\begin{figure}
\includegraphics[width=3in]{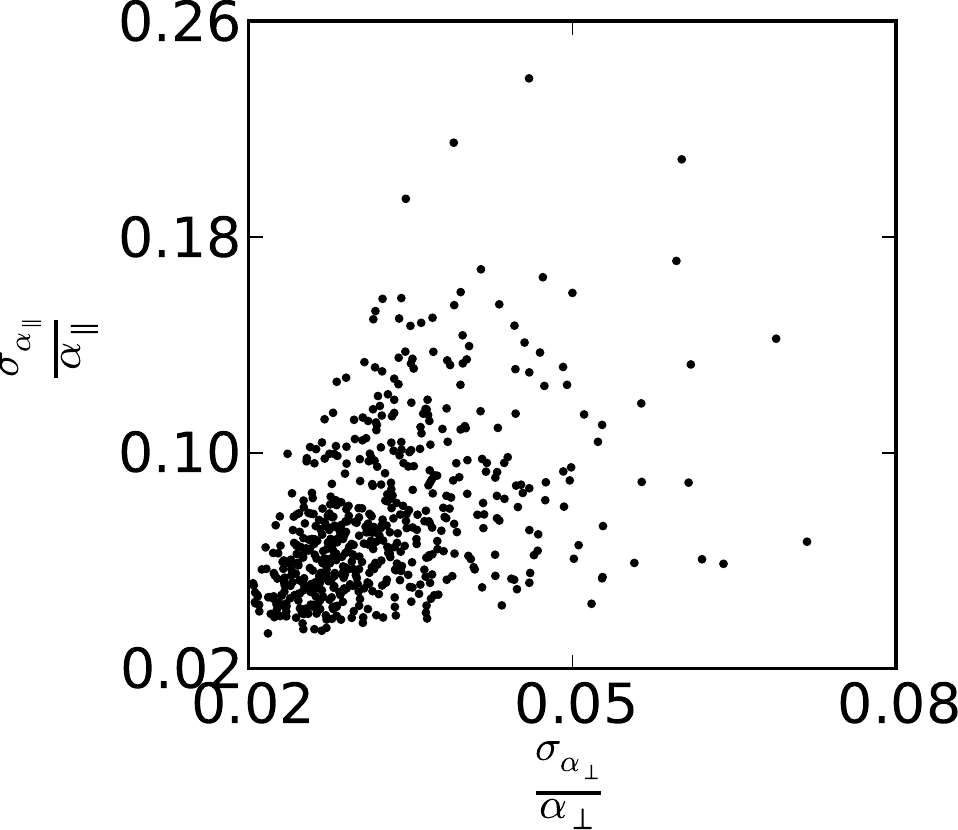}
\caption{The errors in estimated distances, $\sigma_{\alpha_\perp}/\alpha_\perp$
versus $\sigma_{\alpha_\parallel}/\alpha_\parallel$, for the mock catalogues.
The line of sight distance is more weakly constrained than the transverse
distance.}
\label{fig:sase}
\end{figure}

\begin{table*}
\caption{Fitting results from the multipole analysis of the mock catalogues
for various parameter choices.
The model is given in column 1. The median $\alpha_\perp$ is given 
in column 2 with the 16th/84th percentiles from the mocks given in 
column 3 (these are denoted as the quantiles in the text, hence 
the label Qtls in the table). The median $\alpha_\parallel$ is given in 
column 6 with corresponding quantiles in column 7. The median difference 
in $\alpha_\perp$ on a mock-by-mock basis between the model listed in 
column 1 and the fiducial model is given in column 4 with corresponding 
quantiles in column 5. The analogues for $\alpha_\parallel$ are given in 
columns 8 and 9. The mean $\chi^2$/dof is given in column 10.}
\label{tab:alphas}

\begin{tabular}{lccrcccrcc}

\hline
Model&
$\widetilde{\alpha_\perp}$&
Qtls&
$\widetilde{\Delta\alpha_\perp}$&
Qtls&
$\widetilde{\alpha_\parallel}$&
Qtls&
$\widetilde{\Delta\alpha_\parallel}$&
Qtls&
$\langle\chi^2\rangle/dof$\\

\hline
\multicolumn{10}{c}{Redshift Space without Reconstruction}\\
\hline

Fiducial $[f]$ &
$1.008$&
$^{+0.034}_{-0.037}$&
--&
--&
$1.006$&
$^{+0.072}_{-0.074}$&
--&
--&
60.06/70\\
\\[-1.5ex]
$(\Sigma_\perp,\Sigma_\parallel) \rightarrow (8,8) h^{-1}\rm{Mpc}$. &
$1.011$&
$^{+0.039}_{-0.038}$&
$0.005$&
$^{+0.006}_{-0.006}$&
$1.004$&
$^{+0.073}_{-0.088}$&
$-0.007$&
$^{+0.012}_{-0.013}$&
60.33/70\\
\\[-1.5ex]
$\Sigma_s \rightarrow 0 h^{-1}\rm{Mpc}$. &
$1.007$&
$^{+0.035}_{-0.037}$&
$0.000$&
$^{+0.000}_{-0.000}$&
$1.006$&
$^{+0.071}_{-0.075}$&
$-0.001$&
$^{+0.001}_{-0.001}$&
60.04/70\\
\\[-1.5ex]
$A_2(r)=poly2$. &
$1.007$&
$^{+0.035}_{-0.037}$&
$-0.000$&
$^{+0.002}_{-0.002}$&
$1.008$&
$^{+0.071}_{-0.075}$&
$0.001$&
$^{+0.006}_{-0.007}$&
60.92/71\\
\\[-1.5ex]
$A_2(r)=poly4$. &
$1.007$&
$^{+0.035}_{-0.038}$&
$0.000$&
$^{+0.003}_{-0.003}$&
$1.010$&
$^{+0.070}_{-0.083}$&
$-0.000$&
$^{+0.007}_{-0.007}$&
59.20/69\\
\\[-1.5ex]
$30<r<200\hMpc$ range. &
$1.012$&
$^{+0.040}_{-0.038}$&
$0.003$&
$^{+0.009}_{-0.007}$&
$0.987$&
$^{+0.075}_{-0.090}$&
$-0.017$&
$^{+0.014}_{-0.022}$&
68.87/80\\
\\[-1.5ex]
$70<r<200\hMpc$ range. &
$1.007$&
$^{+0.033}_{-0.039}$&
$-0.001$&
$^{+0.006}_{-0.007}$&
$1.010$&
$^{+0.071}_{-0.075}$&
$0.001$&
$^{+0.010}_{-0.012}$&
52.28/60\\
\\[-1.5ex]
$50<r<150\hMpc$ range. &
$1.007$&
$^{+0.035}_{-0.037}$&
$-0.001$&
$^{+0.008}_{-0.009}$&
$1.010$&
$^{+0.073}_{-0.090}$&
$0.000$&
$^{+0.017}_{-0.020}$&
39.34/44\\
\hline
\multicolumn{10}{c}{Redshift Space with Reconstruction}\\
\hline
Fiducial $[f]$ &
$1.001$&
$^{+0.025}_{-0.026}$&
--&
--&
$1.006$&
$^{+0.041}_{-0.045}$&
--&
--&
61.06/70\\
\\[-1.5ex]
$(\Sigma_\perp,\Sigma_\parallel) \rightarrow (2,4) h^{-1}\rm{Mpc}$. &
$1.001$&
$^{+0.024}_{-0.027}$&
$-0.001$&
$^{+0.001}_{-0.001}$&
$1.007$&
$^{+0.040}_{-0.045}$&
$0.001$&
$^{+0.002}_{-0.002}$&
61.13/70\\
\\[-1.5ex]
$\Sigma_s \rightarrow 0 h^{-1}\rm{Mpc}$. &
$1.001$&
$^{+0.025}_{-0.026}$&
$0.000$&
$^{+0.000}_{-0.000}$&
$1.006$&
$^{+0.041}_{-0.044}$&
$-0.000$&
$^{+0.001}_{-0.001}$&
60.99/70\\
\\[-1.5ex]
$A_2(r)=poly2$. &
$1.000$&
$^{+0.024}_{-0.026}$&
$-0.001$&
$^{+0.001}_{-0.002}$&
$1.006$&
$^{+0.043}_{-0.046}$&
$-0.000$&
$^{+0.002}_{-0.001}$&
63.40/71\\
\\[-1.5ex]
$A_2(r)=poly4$. &
$1.003$&
$^{+0.024}_{-0.026}$&
$0.002$&
$^{+0.002}_{-0.003}$&
$1.003$&
$^{+0.042}_{-0.046}$&
$-0.003$&
$^{+0.004}_{-0.005}$&
59.78/69\\
\\[-1.5ex]
$30<r<200\hMpc$ range. &
$1.004$&
$^{+0.025}_{-0.026}$&
$0.003$&
$^{+0.004}_{-0.004}$&
$1.008$&
$^{+0.040}_{-0.044}$&
$0.000$&
$^{+0.006}_{-0.005}$&
71.25/80\\
\\[-1.5ex]
$70<r<200\hMpc$ range. &
$1.002$&
$^{+0.023}_{-0.028}$&
$-0.001$&
$^{+0.005}_{-0.004}$&
$1.008$&
$^{+0.039}_{-0.044}$&
$0.002$&
$^{+0.008}_{-0.007}$&
52.48/60\\
\\[-1.5ex]
$50<r<150\hMpc$ range. &
$1.003$&
$^{+0.023}_{-0.026}$&
$0.000$&
$^{+0.006}_{-0.006}$&
$1.005$&
$^{+0.044}_{-0.047}$&
$-0.002$&
$^{+0.009}_{-0.011}$&
39.95/44\\
\hline
\end{tabular}

\end{table*}

We start by summarizing the results of analyzing the multipoles measured from
the mock catalogues; a corresponding discussion of the clustering wedges is in 
\citet{kazin13}. A summary of multipole results is in Table~\ref{tab:alphas}
and in Figs.~\ref{fig:reccomp} to \ref{fig:sase}. 

Figure~\ref{fig:reccomp} and the first line in Table~\ref{tab:alphas} show that
we recover the correct distances ($\alpha_{\perp} = \alpha_{\parallel} = 1$)
both before and after reconstruction. Reconstruction does reduce both the
scatter in the measurements and the number of outliers, reflecting the 
sharpening of the acoustic signal in the correlation function.

Even though we measure both $\alpha_{\perp}$ and $\alpha_{\parallel}$,
Figure~\ref{fig:dah} shows that these are correlated with a correlation
coefficient of -0.441 (-0.494 before reconstruction). 
Note that the sign of this correlation reverses when we consider $D_{A}$ and
$H$, since $H \sim 1/\alpha_{\parallel}$.

In Figure \ref{fig:sase}, we show the post-reconstruction
$\sigma_{\alpha_\perp}/\alpha_\perp$ versus
$\sigma_{\alpha_\parallel}/\alpha_\parallel$ values from each mock. 
The errors on $\alpha_\perp$ and $\alpha_\parallel$ are correlated as expected;
the errors are related to the strength of the BAO signal in any given
realization. Similar results are seen before reconstruction.

We also test the robustness of our fits by varying the
fiducial model parameters; the 
results of these are in Table \ref{tab:alphas}. 
We test cases in which $\Sigma_\perp$
and $\Sigma_\parallel$ are varied, $\Sigma_s$ is varied, the form of
$A_2(r)$ is varied and the range of data used in the fit is varied. 
In general, the recovered values of $\alpha_\perp$ and $\alpha_\parallel$ are consistent
with the fiducial model. The largest discrepancy arises in the
pre-reconstruction measurement of $\alpha_\parallel$ where we have extended the
fitting range down to $30\hMpc$. However, we know that our model at
small scales is not particularly well matched to the mocks, as we saw
in Figure \ref{fig:mockmod}, and hence the larger difference obtained
by fitting down to smaller scales is not surprising. The  
smaller discrepancies in both $\alpha_{\perp}$ and $\alpha_\parallel$ when the
other parameters are varied do not appear to be distinguishable in any
individual mock as indicated by the quantiles on $\Delta \alpha_\perp$ and
$\Delta \alpha_\parallel$. \citet{2012arXiv1206.6732X} discuss similar differences and
attribute them to disagreement between the model and data at small
scales. In addition, the mock catalogues used here are derived from a
perturbation theory based approach, so they may not be fully faithful
on small scales.

\subsection{Multipoles vs. Clustering Wedges}
\begin{figure*}
\includegraphics[width=0.85\linewidth]{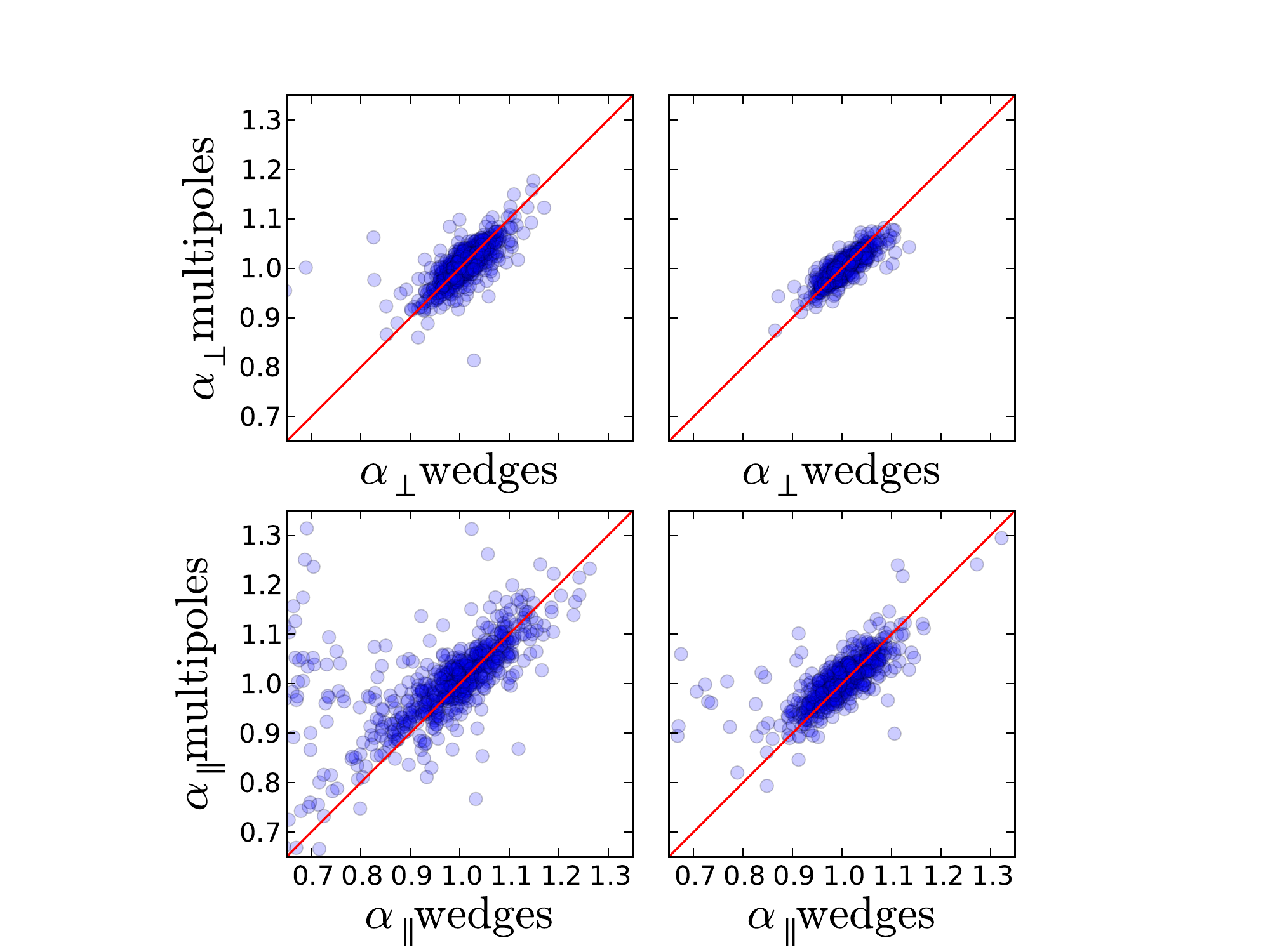}
\includegraphics[width=0.85\linewidth]{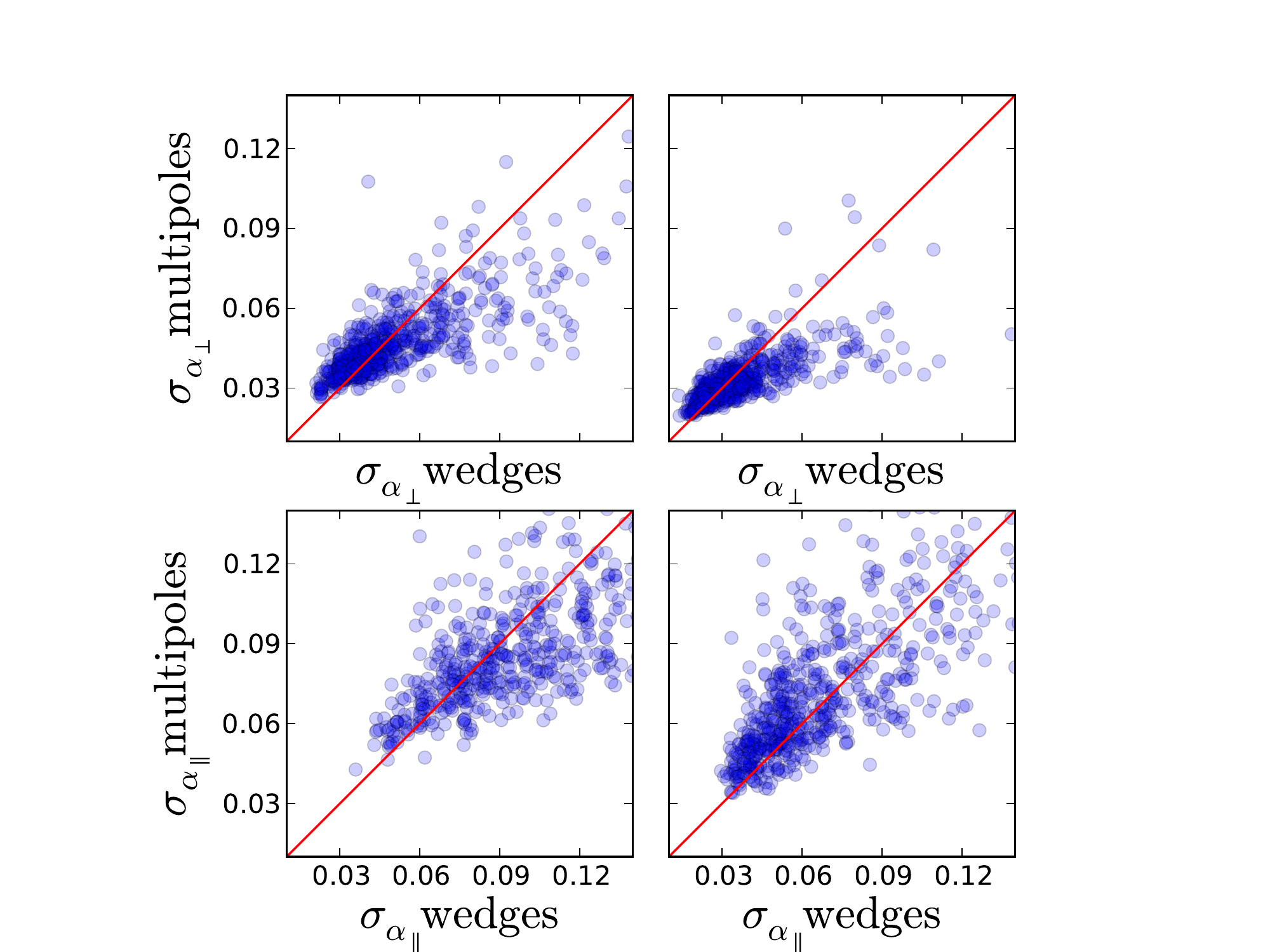}
\caption{Comparison between the measurements (top rows) and estimated errors (bottom rows) for
$\alpha_{\perp}$ and $\alpha_{||}$ obtained from the multipoles analysis and the corresponding 
results using the wedges technique, for all 600 PTHalos mocks. Left panels 
show the comparison before using reconstruction, and right panels show the 
comparison after reconstruction.}
\label{fig:comparison}
\end{figure*}

\begin{figure*}
\includegraphics[width=0.85\linewidth]{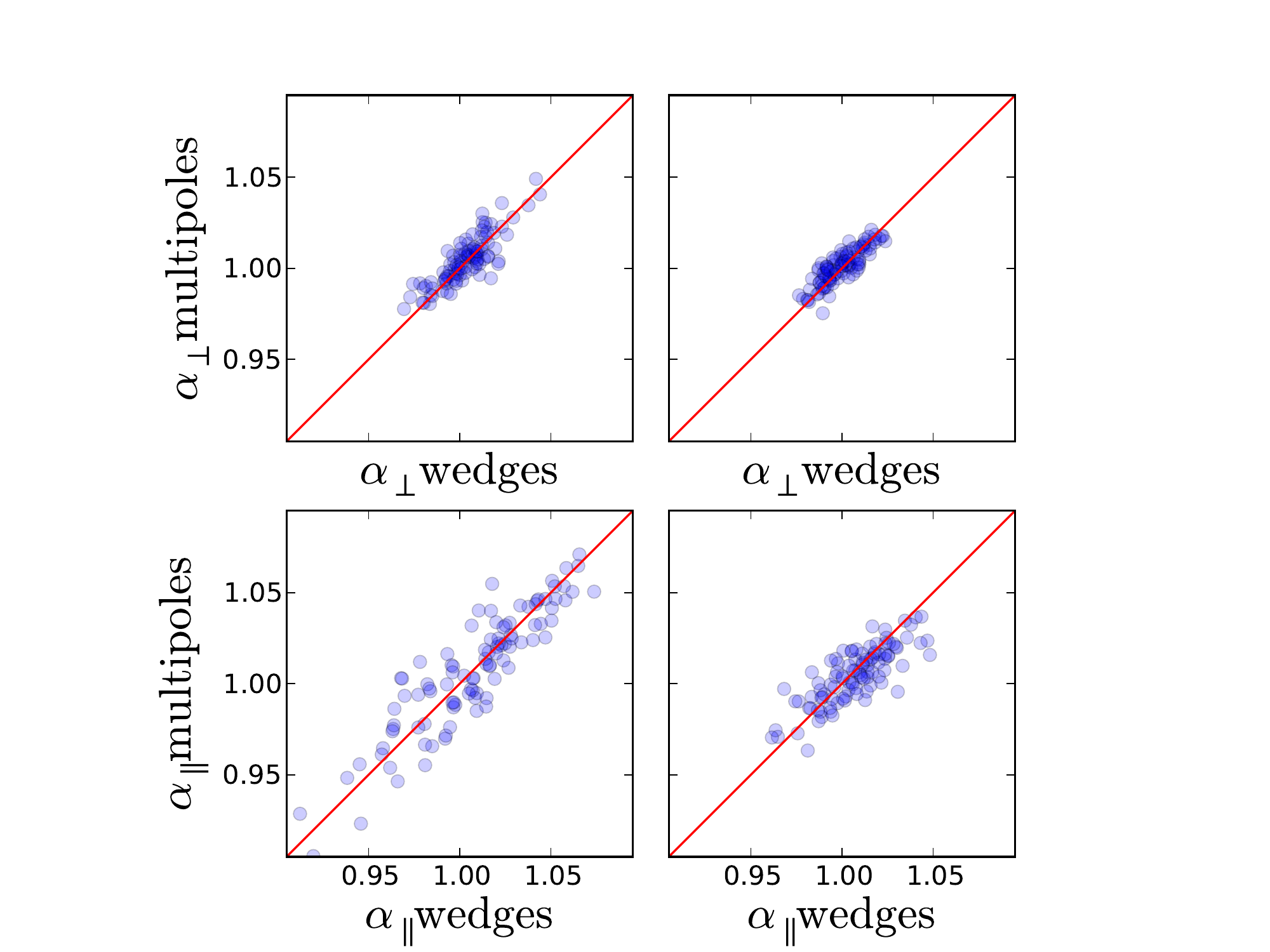}
\includegraphics[width=0.85\linewidth]{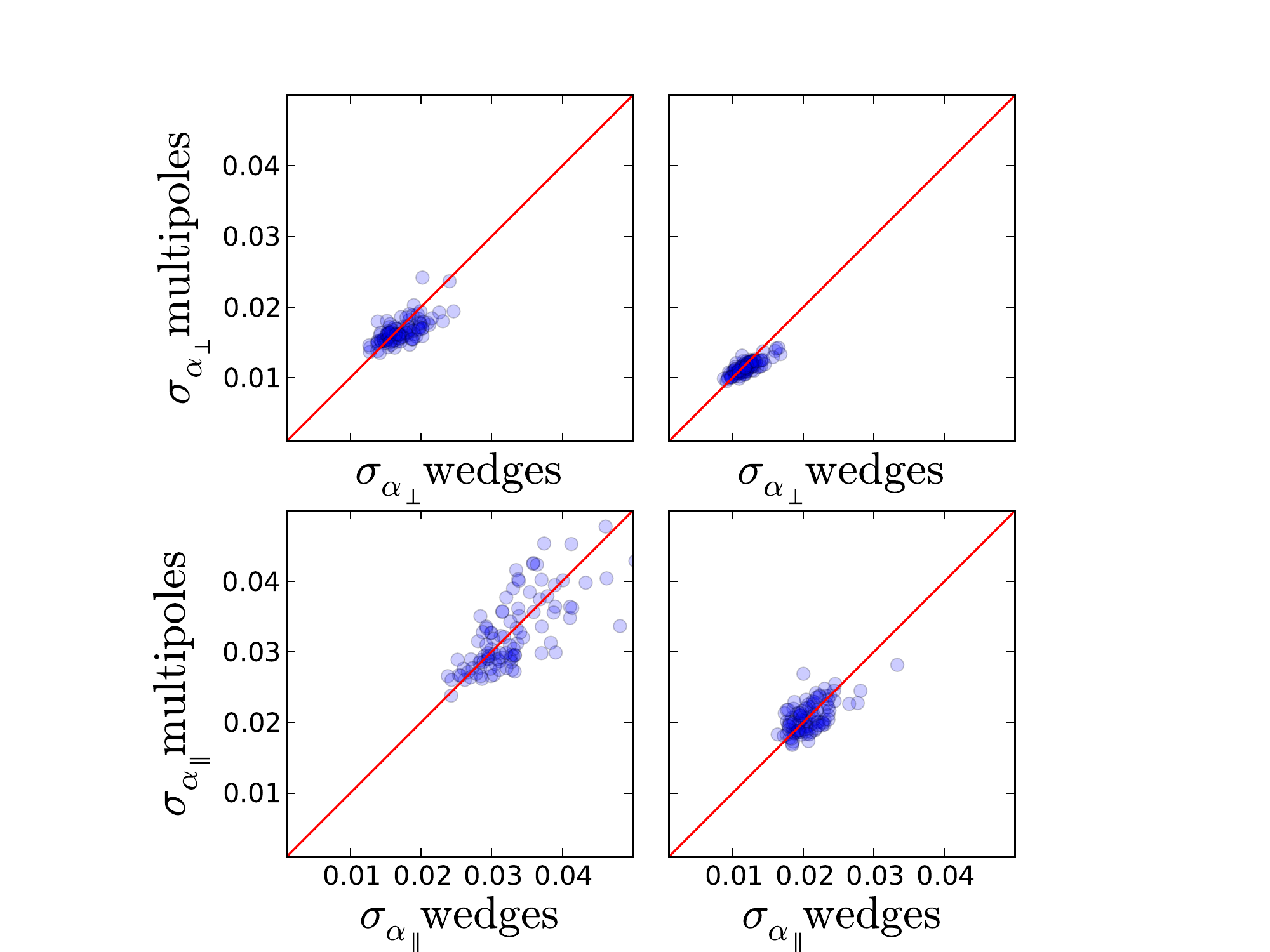}
\caption{Same as Figure~\ref{fig:comparison} but using the 100 groups of 
mocks, each of which is the average of six mocks, to increase the
signal-to-noise ratio of the BAO feature.
Note that in this case the agreement between the analysis using multipoles and wedges is much closer 
than in the non-stacked case.}
\label{fig:comparison_stacked}
\end{figure*}

We now turn to comparisons of the results obtained in the previous section with
the clustering wedges analysis in \citet{kazin13}. In the limit where multipoles
with $\ell \geq 4$ are negligible on large scales (as is our case), the
monopole/quadrupole and clustering wedges are just a basis rotation and one would expect similar results
from both. However, the marginalization of the broad band information and the various priors will 
impact the two differently. Furthermore, we adopt different
techniques (and codes) in both, so this comparison tests the robustness of these
approaches.

Figure~\ref{fig:comparison} and Tables~\ref{tab:mock_wedges_multipoles_summary}
and \ref{tab:mock_wedges_multipoles_comparison} summarize the results for
both the measured distance scales and the estimated errors. Both methods yield
identical results on average, but we note considerable scatter about this mean relation.
Examining the individual mocks in detail, we find that a majority of these
outliers correspond to realizations with a weak BAO detection. We quantify this
by comparing fits with and without a BAO feature in them. Before reconstruction,
23\% of the mocks have a $< 3\sigma$ detection of the BAO feature in them; after
reconstruction, this number drops to 4.6\%. This improvement is also manifest in
the right column of Figure~\ref{fig:comparison}.

We further test this idea by recasting the mocks into 100 sets,
each of which is the average of the correlation functions of 6 of our
DR9 mocks.  With an improvement in signal-to-noise ratio of a factor
of $\sqrt{6}$, the acoustic peak is expected to be well detected. We present
these results in Figure~\ref{fig:comparison_stacked}. 
There are none of the
catastrophic failures of Figure~\ref{fig:comparison} and very good agreement in
both the estimated distances and errors for these individual ``stacked''
realizations. This suggests that the information content in these two approaches
is indeed very similar.

\begin{table*}

\caption{Average results from the 600 mocks (top four rows) and the 
100 stacked mocks (bottow four rows). The table shows the median values of 
$\alpha_{\perp}$, $\alpha_{\parallel}$, $\sigma_{\alpha_{\perp}}$, and
$\sigma_{\alpha_{\parallel}}$, together with their 68\% 
confidence level region as shown by the 16$^{\mathrm{th}}$ and the 84$^{\mathrm{th}}$ 
percentiles in the mock ensemble. Results are shown for both multipoles and wedges, 
as well as pre- and post-reconstruction. As previously, tildes
represent median quantities.}
\label{tab:mock_wedges_multipoles_summary}
\begin{tabular}{lcccc}
\hline
                  & $\widetilde{\alpha_{\perp}}$ &
                  $\widetilde{\alpha_{\parallel}}$ &
                  $\widetilde{\sigma_{\alpha_{\perp}}}$ &
                  $\widetilde{\sigma_{\alpha_{\parallel}}}$ \\
\hline
                  & \multicolumn{4}{c}{original mocks} \\
\hline
wedges            & $1.010 ^{+0.040}_{-0.040} $ & $0.992 ^{+0.083}_{-0.124} $ & $0.044 ^{+0.032}_{-0.012} $ & $0.102 ^{+0.062}_{-0.033} $ \\
multipoles        & $1.008 ^{+0.035}_{-0.037} $ & $1.007 ^{+0.070}_{-0.076} $ & $0.044 ^{+0.016}_{-0.008} $ & $0.088 ^{+0.041}_{-0.020} $ \\
\hline
recon. wedges         & $1.000 ^{+0.034}_{-0.027} $ & $0.999 ^{+0.053}_{-0.052} $ & $0.032 ^{+0.018}_{-0.009} $ & $0.061 ^{+0.047}_{-0.018} $ \\
recon. multipoles     & $1.001 ^{+0.025}_{-0.026} $ & $1.006 ^{+0.041}_{-0.045} $ & $0.031 ^{+0.009}_{-0.005} $ & $0.067 ^{+0.037}_{-0.017} $ \\
\hline
                  & \multicolumn{4}{c}{stacked mocks} \\
\hline
wedges            & $1.003 ^{+0.012}_{-0.012} $ & $1.014 ^{+0.029}_{-0.038} $ & $0.017 ^{+0.003}_{-0.002} $ & $0.032 ^{+0.006}_{-0.004} $ \\
multipoles        & $1.004 ^{+0.013}_{-0.012} $ & $1.010 ^{+0.033}_{-0.034} $ & $0.016 ^{+0.002}_{-0.001} $ & $0.031 ^{+0.008}_{-0.004} $ \\
\hline
recon. wedges         & $1.000 ^{+0.012}_{-0.010} $ & $1.008 ^{+0.017}_{-0.020} $ & $0.012 ^{+0.001}_{-0.001} $ & $0.020 ^{+0.003}_{-0.002} $ \\
recon. multipoles     & $1.001 ^{+0.010}_{-0.009} $ & $1.006 ^{+0.014}_{-0.016} $ & $0.011 ^{+0.001}_{-0.001} $ & $0.020 ^{+0.002}_{-0.002} $ \\
\hline
\end{tabular}
\end{table*}

\begin{table*}

\caption{Average results from the 600 mocks (top two rows) and the 100 stacked mocks 
(bottow two rows). The table shows the median values of 
$\Delta\alpha_{\perp}$, $\Delta\alpha_{\parallel}$,
$\Delta\sigma_{\alpha_{\perp}}$, and $\Delta\sigma_{\alpha_{\parallel}}$, (where
$\Delta$ denotes the difference of the results using wedges minus the ones using multipoles) together with their 68\% 
confidence level region as shown by the 16$^{\mathrm{th}}$ and the 84$^{\mathrm{th}}$ percentiles 
in the mock ensemble. Results are shown for both multipoles and wedges, as well
as pre- and post-reconstruction. As previously, tildes represent median
quantities.}
\label{tab:mock_wedges_multipoles_comparison}
\begin{tabular}{lcccc}
\hline
                   &  $\widetilde{\Delta\alpha_{\perp}}$ &
                   $\widetilde{\Delta\alpha_{\parallel}}$ &
                   $\widetilde{\Delta\sigma_{\alpha_{\perp}}}$ &
                   $\widetilde{\Delta\sigma_{\alpha_{\parallel}}}$ \\
\hline
                  & \multicolumn{4}{c}{original mocks} \\
\hline
pre-recon.  & $+0.004 ^{+0.020}_{-0.023} $ & $-0.015 ^{+0.046}_{-0.053} $ & $-0.000 ^{+0.019}_{-0.008} $ & $+0.009 ^{+0.036}_{-0.019} $ \\
post-recon. & $+0.001 ^{+0.015}_{-0.014} $ & $-0.005 ^{+0.027}_{-0.027} $ & $+0.000 ^{+0.011}_{-0.005} $ & $-0.004 ^{+0.021}_{-0.016} $ \\
\hline
                  & \multicolumn{4}{c}{stacked mocks} \\
\hline
pre-recon.  & $-0.001 ^{+0.008}_{-0.007} $ & $+0.001 ^{+0.014}_{-0.015} $ & $+0.001 ^{+0.002}_{-0.002} $ & $+0.000 ^{+0.003}_{-0.004} $ \\
post-recon. & $-0.001 ^{+0.005}_{-0.006} $ & $+0.003 ^{+0.007}_{-0.012} $ & $+0.000 ^{+0.001}_{-0.001} $ & $+0.000 ^{+0.002}_{-0.002} $ \\
\hline
\end{tabular}
\end{table*}

\subsection{Isotropic vs anisotropic BAO measurements}
\label{sec:iso_vs_anisotropic}
\begin{figure}
\includegraphics[trim=0cm 0cm 2cm 8cm, width=0.45\textwidth]{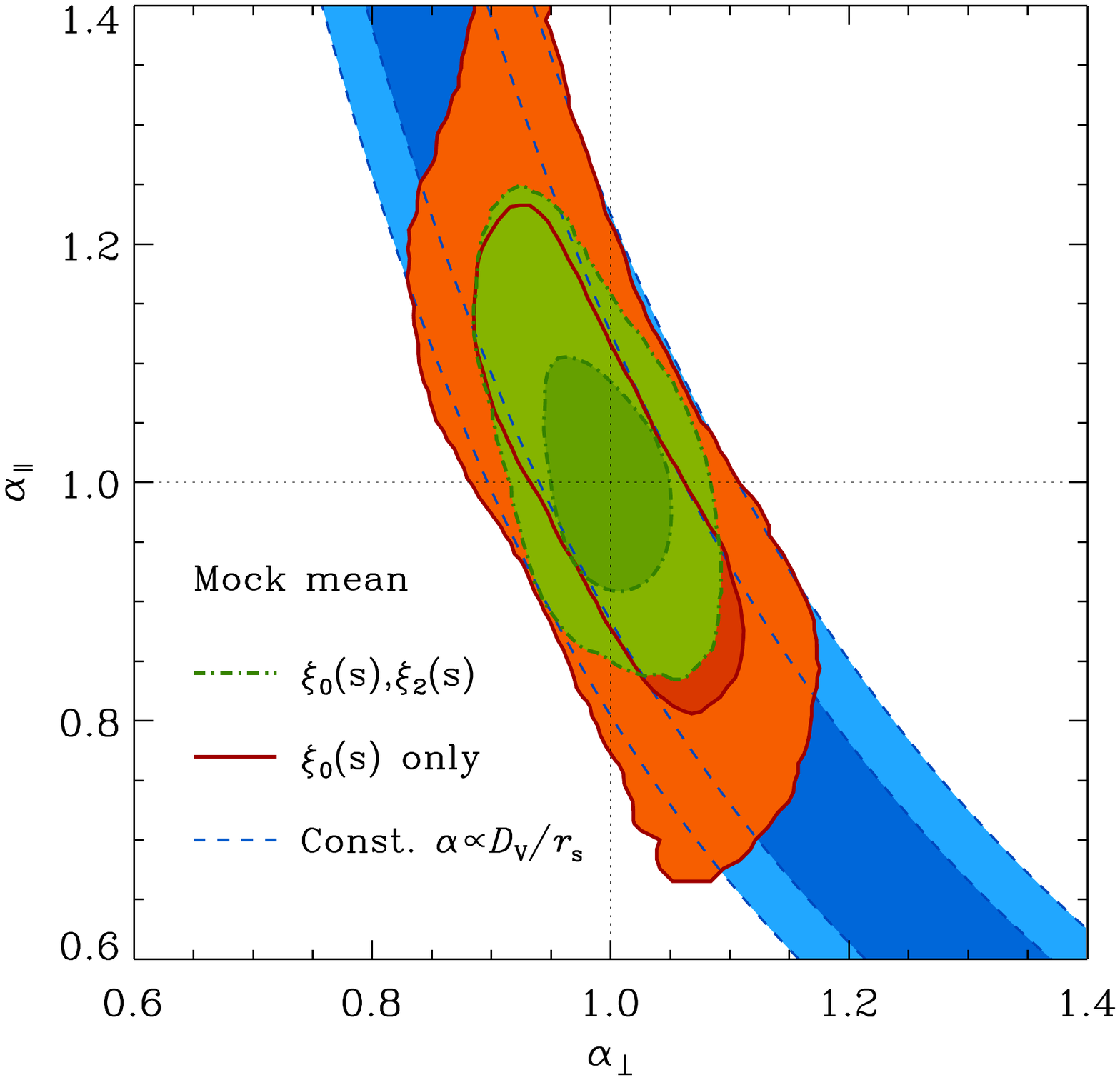}
\caption{
Comparison of the 65\% and 95\% constraints in the $\alpha_{\parallel}$--$\alpha_{\perp}$ plane 
obtained from the mean monopole of our mock catalogues (solid lines, orange),
and from its combination with the mean quadrupole (dot-dashed lines, green). The
constraints from $\xi_0(s)$ follow a degeneracy which is well described
by lines of constant $\alpha\propto D_{\rm V}/r_{\rm s}$, shown by the dashed
lines (blue). The extra information in the anisotropic BAO measurement helps to
break this degeneracy.
}
\label{fig:dah_mono} 
\end{figure}

We now compare the results obtained from anisotropic BAO measurements with those
derived from their isotropically averaged counterparts. 
As described in Section~\ref{sec:method}, spherically-averaged clustering
measurements are only sensitive to the isotropic shift $\alpha$, while anisotropic measurements provide extra constraints on the
distortion parameter $\epsilon$.
Figure~\ref{fig:dah_mono} compares the constraints on $\alpha_{\perp}$ and
$\alpha_{\parallel}$ obtained by analyzing $\xi_{0}$ and $\xi_2$ (dot-dashed
lines) with those obtained by analyzing $\xi_0$ alone (solid lines). To avoid
noise from particular realizations, we use the average of the mock catalogues
after reconstruction here. Analyzing the clustering wedges give essentially
identical results.

As expected, the constraints derived from $\xi_0(s)$ exhibit a strong degeneracy well described by lines of constant
$\alpha \propto D_{\rm V}/r_{\rm s}$, shown by the dashed lines; including
$\xi_2$ breaks this degeneracy. The degeneracy is not perfect
because large values of $\epsilon$ strongly distort the BAO feature in 
$\xi(s,\mu)$, causing a strong damping of the acoustic peak in the resulting
$\xi_0(s)$. As the peak can be almost completely erased, these values give poor fits to the data when compared to $\epsilon=0$.
We note that this requires going beyond the linear approximations used in
\citet{2008PhRvD..77l3540P} and \citet{2012arXiv1206.6732X}. However, these
constraints are weak and can be ignored in all practical applications.

\section{DR9 Results} \label{sec:data_results}
\subsection{Basic Results}

\begin{figure*}
\includegraphics[width=0.9\textwidth]{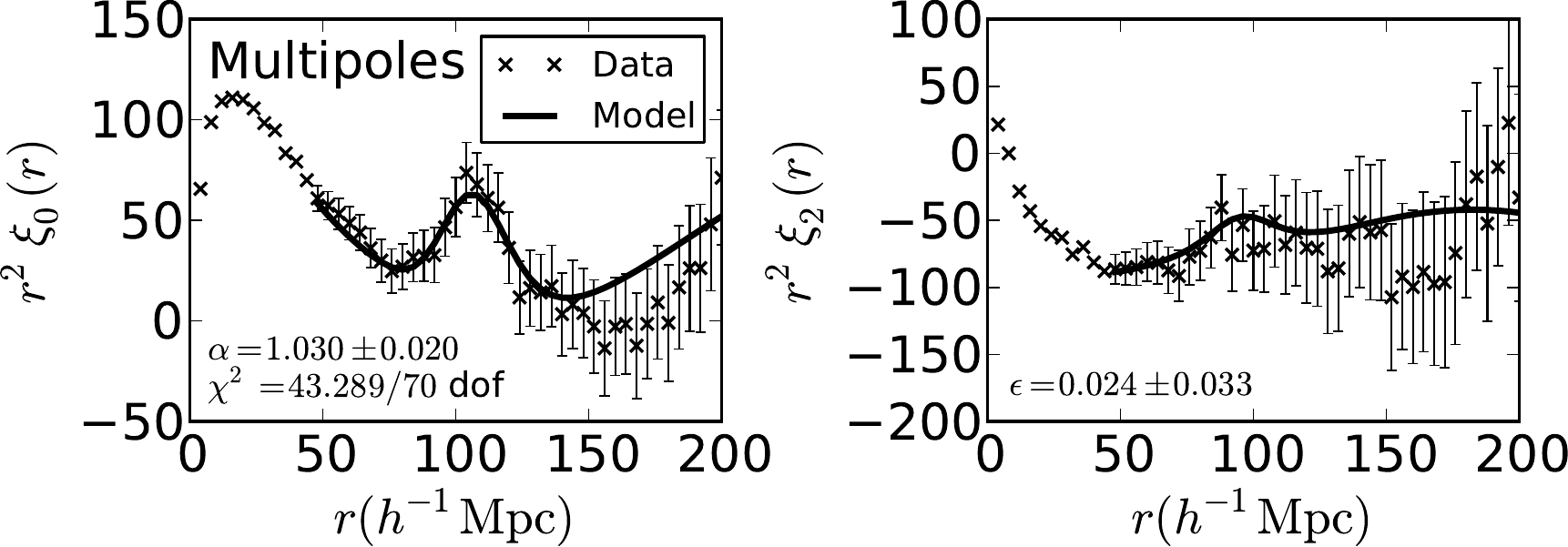}
\includegraphics[width=0.9\textwidth]{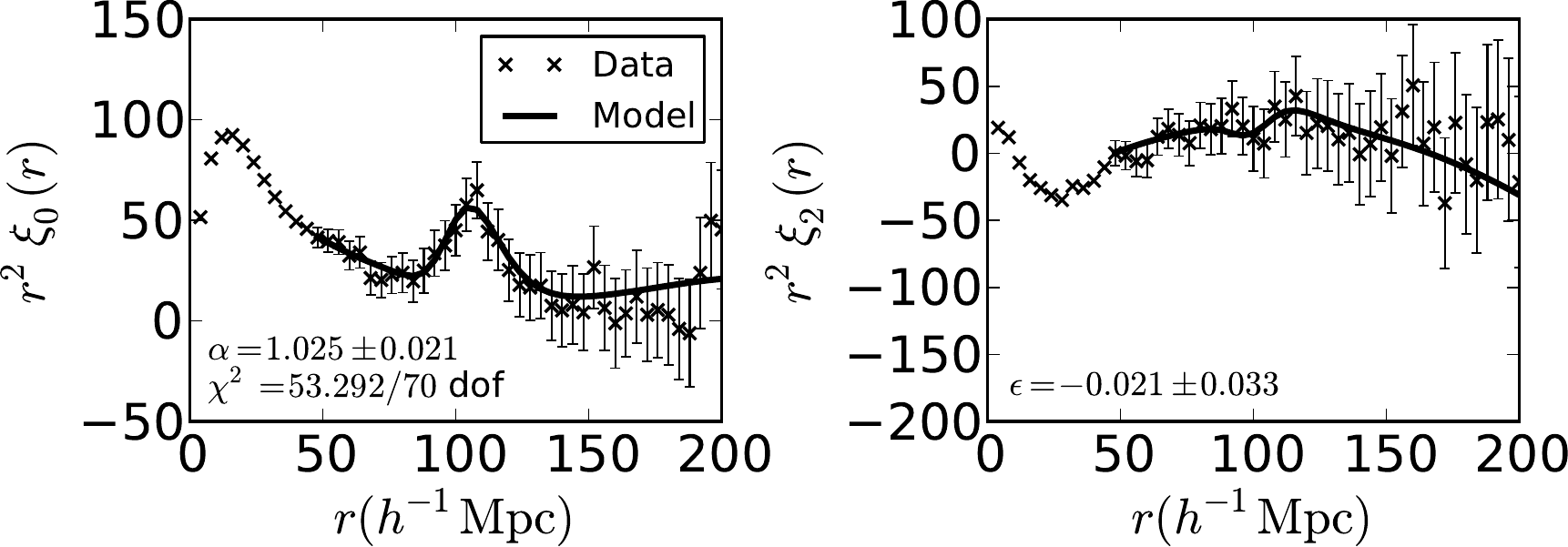}
\caption{DR9 data (multipoles) before (top) and after (bottom) reconstruction
with best-fit model (\S~\ref{sec:method}) overplotted. Note that the errors are
correlated between bins. The distance parameters ($\alpha, \epsilon)$) of the
best fit and the corresponding $\chi^2$ values are listed in the plots.}
\label{fig:dr9res_multipoles}
\end{figure*}

\begin{figure*}
\includegraphics[width=0.9\textwidth]{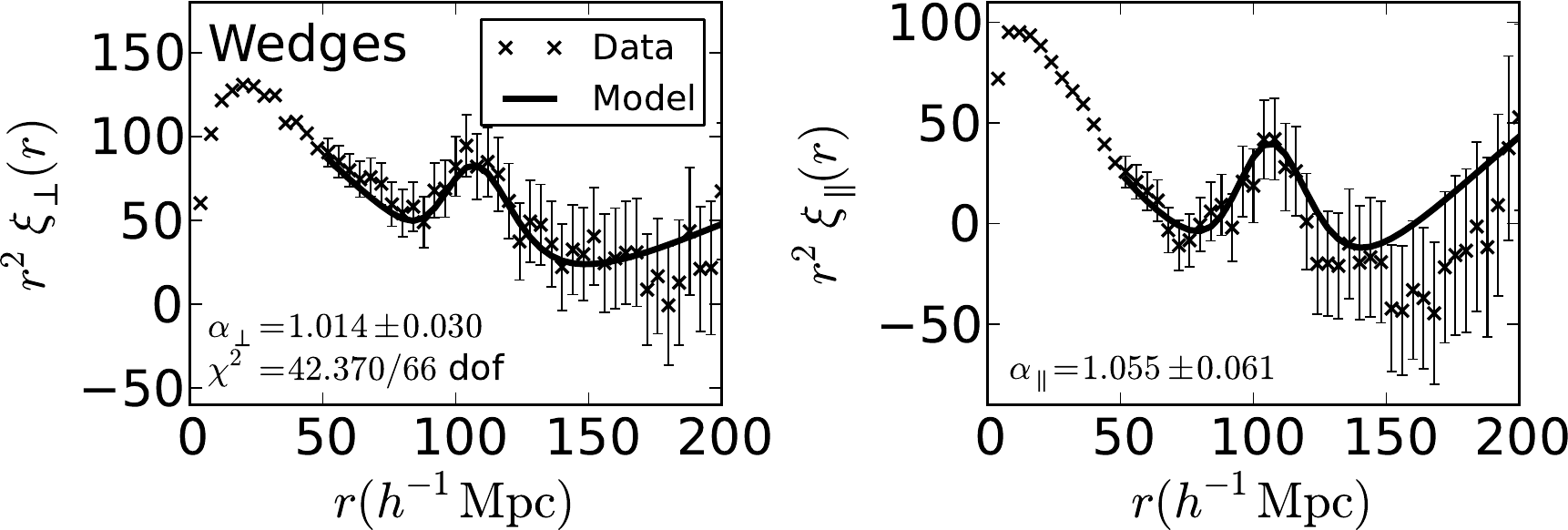}
\includegraphics[width=0.9\textwidth]{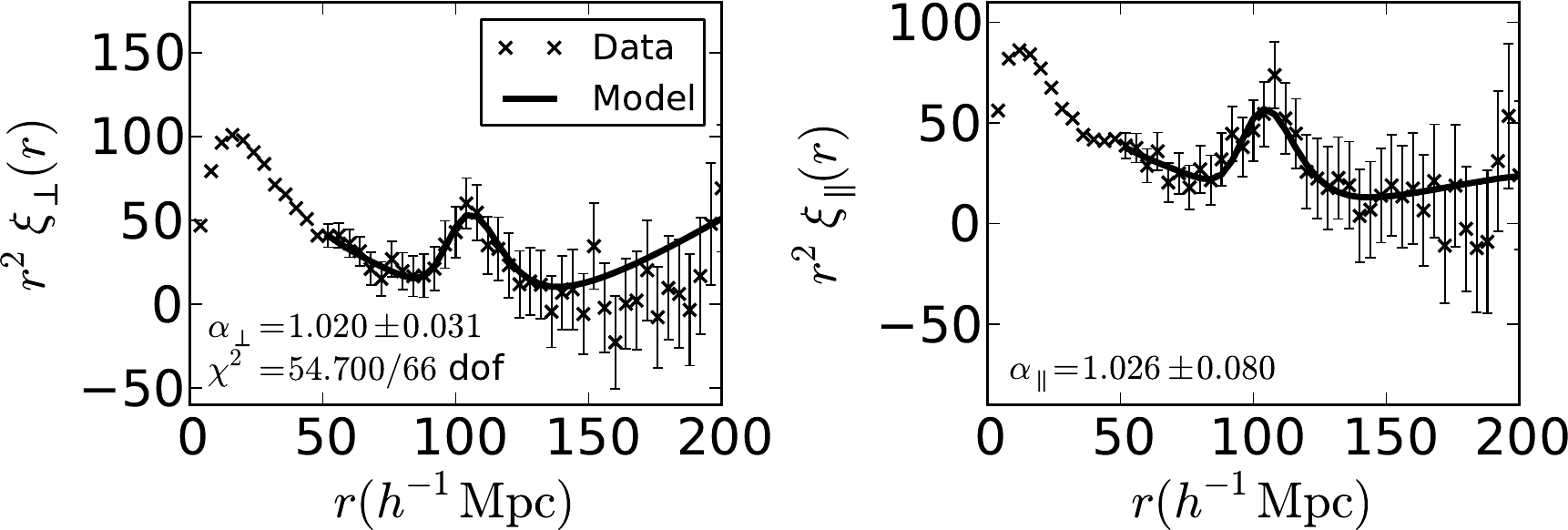}
\caption{As in Figure~\ref{fig:dr9res_multipoles} but for the clustering
wedges.} 
\label{fig:dr9res_wedges}
\end{figure*}

\begin{figure}
\includegraphics[width=3in]{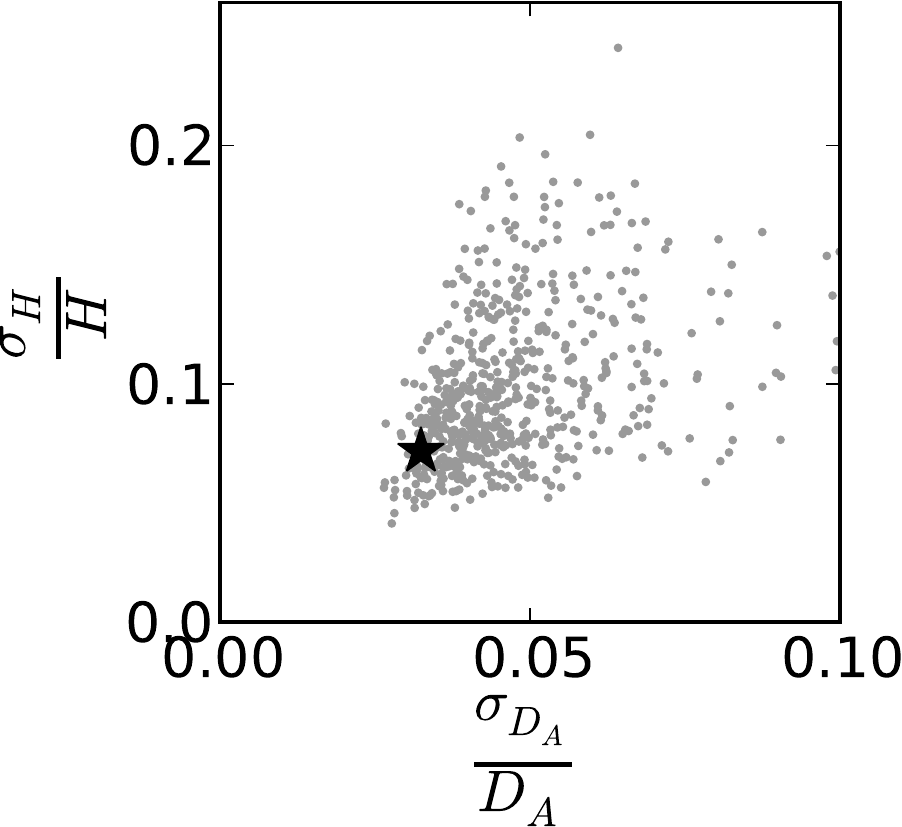}
\includegraphics[width=3in]{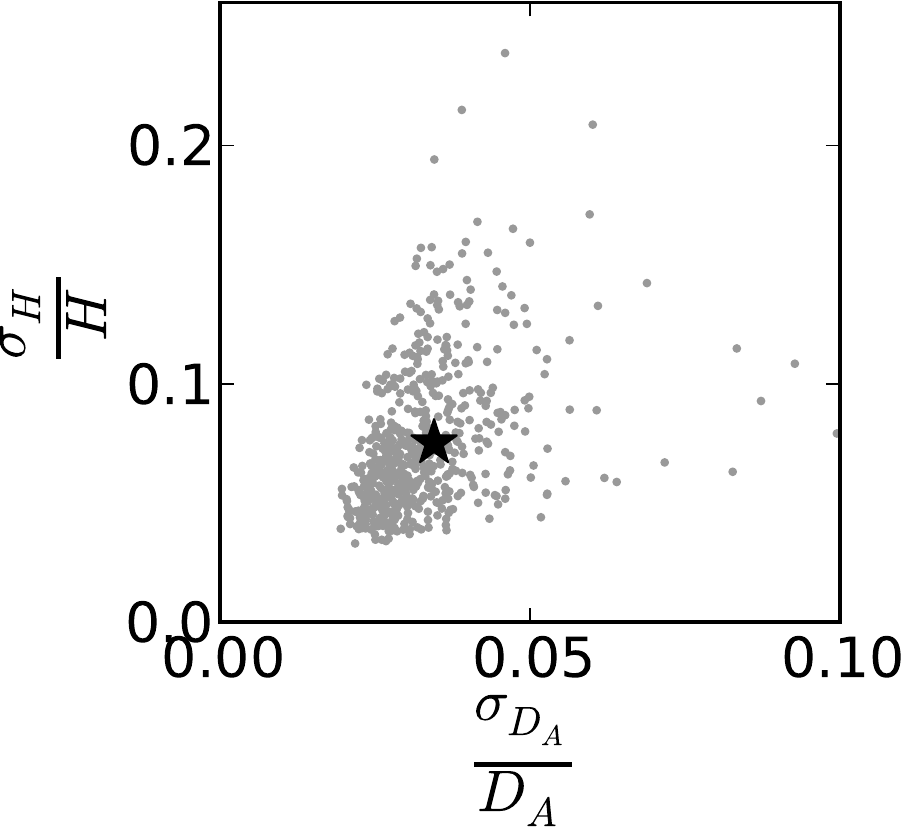}
\caption{The DR9 $\sigma_{D_A}/D_A$ and $\sigma_H/H$ values before
(left) and after (right) reconstruction overplotted on the mock values. The DR9
values are consistent with the distribution expected from the mock catalogues,
with the pre-reconstruction case on the better-constrained end and the
post-reconstruction case more average.}
\label{fig:dr9_sase}
\end{figure}

\begin{figure}
%\includegraphics[trim=0cm
%2cm 1cm 2cm, width=3in, clip=true]{plots/y}
\includegraphics[width=3in]{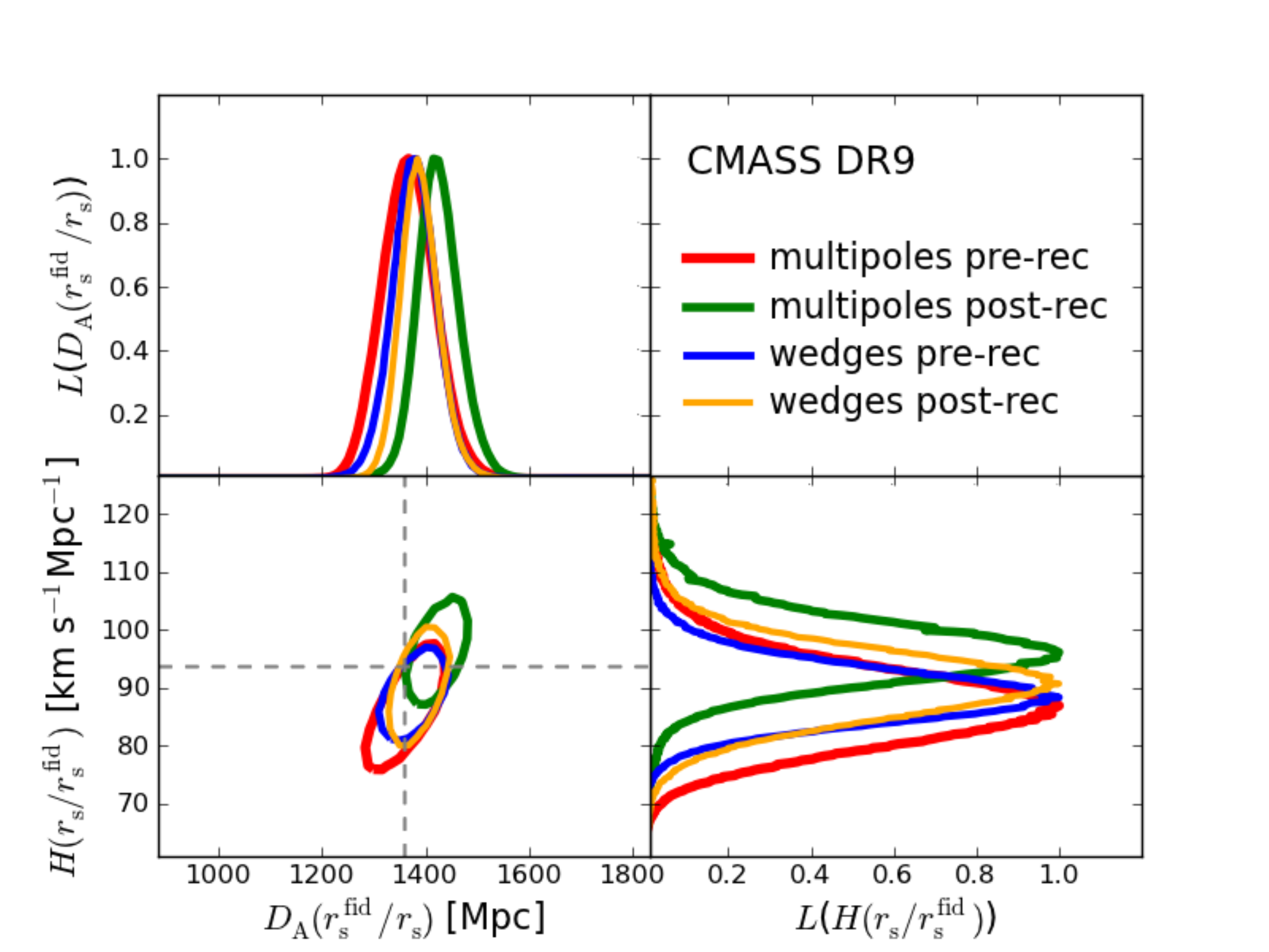}
%2cm 1cm 2cm, width=3in, clip=true]{plots/hda_all_combos_dewiggled_mcmcmethod}
\caption{Pre- and post-reconstruction 2D 68\% contours and 1D probability
distributions of $D_A$ and $H$ measured from the DR9 data, for both the
multipoles and wedges. For consistency, both the multipoles and wedges have been
analyzed with the MCMC code in \citet{kazin13}. The lines mark the fiducial
cosmology used in the analysis.
}
\label{fig:pdr9}
\end{figure}

\begin{table*}
\caption{DR9 fitting results for various models. The model is given in column 1.
The measured $D_A(z)(r_{\rm s}^{\rm fid}/r_{\rm s})$ values are given in column 2 and the measured $H(z)(r_{\rm s}/r_{\rm s}^{\rm fid})$ values are given in column 3. The $\chi^2$/dof is given in column 4. For DR9 CMASS, $z=0.57$ and in our fiducial cosmology $r_{\rm s}^{\rm fid}=153.19$Mpc.}
\label{tab:dr9_alphas}

\begin{tabular}{@{}lccc}

\hline
Model&
$D_A(z)(r_{\rm s}^{\rm fid}/r_{\rm s})$&
$H(z)(r_{\rm s}/r_{\rm s}^{\rm fid})$&
$\chi^2/dof$\\
&(Mpc)&(km/s/Mpc)&\\

\hline
\multicolumn{4}{c}{Redshift Space without Reconstruction}\\
\hline

Fiducial $[f]$ &
$1367 \pm   44$&
$86.6 \pm 6.2$&
43.29/70\\
\\[-1.5ex]
$(\Sigma_\perp,\Sigma_\parallel) \rightarrow (8,8) h^{-1}\rm{Mpc}$. &
$1371 \pm   50$&
$87.7 \pm 5.8$&
44.54/70\\
\\[-1.5ex]
$\Sigma_s \rightarrow 0 h^{-1}\rm{Mpc}$. &
$1367 \pm   44$&
$86.7 \pm 6.2$&
43.26/70\\
\\[-1.5ex]
$A_2(r)=poly2$. &
$1366 \pm   44$&
$86.4 \pm 6.1$&
43.72/71\\
\\[-1.5ex]
$A_2(r)=poly4$. &
$1367 \pm   44$&
$86.6 \pm 6.3$&
43.29/69\\
\\[-1.5ex]
$30<r<200\hMpc$ range. &
$1357 \pm   44$&
$84.8 \pm 5.7$&
56.14/80\\
\\[-1.5ex]
$70<r<200\hMpc$ range. &
$1365 \pm   44$&
$86.5 \pm 6.4$&
41.68/60\\
\hline
\multicolumn{4}{c}{Redshift Space with Reconstruction}\\
\hline
Fiducial $[f]$ &
$1424 \pm   43$&
$95.4 \pm 7.5$&
53.29/70\\
\\[-1.5ex]
$(\Sigma_\perp,\Sigma_\parallel) \rightarrow (2,4) h^{-1}\rm{Mpc}$. &
$1419 \pm   42$&
$94.9 \pm 7.6$&
53.20/70\\
\\[-1.5ex]
$\Sigma_s \rightarrow 0 h^{-1}\rm{Mpc}$. &
$1424 \pm   43$&
$95.6 \pm 7.5$&
53.34/70\\
\\[-1.5ex]
$A_2(r)=poly2$. &
$1422 \pm   43$&
$95.6 \pm 7.8$&
55.47/71\\
\\[-1.5ex]
$A_2(r)=poly4$. &
$1421 \pm   43$&
$94.9 \pm 7.6$&
52.76/69\\
\\[-1.5ex]
$30<r<200\hMpc$ range. &
$1433 \pm   46$&
$94.9 \pm 8.2$&
63.94/80\\
\\[-1.5ex]
$70<r<200\hMpc$ range. &
$1418 \pm   40$&
$95.4 \pm 7.1$&
42.92/60\\
\\[-1.5ex]
$50<r<150\hMpc$ range. &
$1405 \pm   39$&
$94.3 \pm 6.4$&
26.80/44\\
\hline
\end{tabular}

\end{table*}

We now apply the methods validated in the previous section to the DR9 data. We
assume the same fiducial cosmology as for the mock catalogs and use the same
models and covariance matrices in our fits. As in the previous section, we begin
by focusing on the multipole analysis and then compare with the companion
analyses.

The DR9 data and the best-fit model (\S~\ref{sec:method}) are shown in Figures
\ref{fig:dr9res_multipoles} for the multipoles and \ref{fig:dr9res_wedges} for
the wedges. Also shown are the best fit distance parameters, $\alpha, \epsilon$
for the multipoles and $\alpha_{\perp}, \alpha_{\parallel}$ for the wedges, as well as
the $\chi^2$-values of the fits. We remind the reader that although in most of
the discussion we present $\alpha_{\perp, \parallel}$ results (to aid
comparisons), the multipole analysis is done in $\alpha, \epsilon$ space. In all
cases, the models are good fits to the data.
As in \citet{2012MNRAS.427.3435A}, we do not see a significant improvement in the constraints after reconstruction.
These measurements imply $D_A(z=0.57)=1367\pm44\,{\rm Mpc}$ and
$H(z=0.57)=86.6\pm6.2\,{\rm km/s/Mpc}$ before reconstruction assuming a sound
horizon equal to the fiducial value $r_s=153.19$ Mpc. After reconstruction we
have $D_A(z=0.57)=1424\pm43\,{\rm Mpc}$ and $H(z=0.57)=95.4\pm7.5\,{\rm
km/s/Mpc}$ : a 3.0\% measurement of $D_A$ and a 7.9\% measurement of
$H$ at $z=0.57$.
The two values are correlated with a correlation coefficient $\rho_{D_A H} = 0.65$ 
before reconstruction and $\rho_{D_A H}=0.63$ after reconstruction. 
The difference from the expected value of $\rho_{D_A H} \sim 0.4$ (from the
mocks) is due to sample variance.
We also test the robustness of these results to
variations in the choices made in the fitting procedure. The results are
summarized in Table~\ref{tab:dr9_alphas}. Our results are insensitive to these
choices, similar to the mock catalogues.

Figure \ref{fig:dr9_sase} compares the DR9 $\sigma_{D_A}/D_A$ and
$\sigma_H/H$ values from the multipole analysis with the distribution estimated
from the mock catalogues. Before reconstruction, the DR9 data clearly lie 
towards the better constrained end of the
mocks; after reconstruction, they appear more average. 
Indeed, our mock results indicate that $\sigma_{D_A}/D_A$ and $\sigma_H/H$ are
actually larger after reconstruction $\sim10\%$ of the time. 
We conclude that these measurements are consistent with our expectations.

Figure~\ref{fig:pdr9} shows the 2D contours and marginalized 1D distributions 
in $\alpha_\perp$ and $\alpha_\parallel$ as measured by the wedges and
multipoles. The likelihoods agree well before reconstruction but shift slightly
after reconstruction. These differences are again consistent with the scatter
seen in the mock catalogues: Figures~\ref{fig:reccomp} comparing the multipole 
measurements before and after reconstruction and Figure~\ref{fig:comparison}
comparing the multipoles and the wedges.

% 
% Figure \ref{fig:dr9pr} shows the likelihoood contours before
% reconstruction for the 5 anisotropic analyses of DR9 CMASS clustering. The
% constraints on $D_A$ and $H$ from the current work are obtained through
% BAO-only $\chi^2$-analyses of the monopole and quadrupole correlation
% functions. The \citet{kazin13} contraints are based on analyses of
% the BAO from wedges using an MCMC approach.
% \citet{2012MNRAS.426.2719R}, \citet{chuang13} and \citet{sanchez13} attempt to
% use the full shape of the correlation function to constrain $D_A$ and $H$ using
% MCMC. \citet{2012MNRAS.426.2719R} and \citet{chuang13} use monopole and
% quadrupole data whereas \citet{sanchez13} uses wedges data. One can see that in general,
% the full-shape analyses give tighter constraints on $D_A$ and $H$ as
% one might expect. Overall, the contours from these 5 different methods
% agree remarkably well indicating the robustness of our results.
% Figure~\ref{fig:dr92dlikepostrecon} shows that this agreement continues
% after reconstruction, although we restrict here to the analyses which
% only fit for the position of the BAO feature, marginalizing out the shape 
% information.

\subsection{Consensus} \label{sec:consensus}
The results on the mock catalogues demonstrate that both the multipoles and
clustering wedges yield consistent results on average for $D_A$ and $H$.
Furthermore, the mock catalogues do not favour one analysis technique over the
other, either in terms of overall precision of the result or the robustness to
outliers. In order to reach a consensus value appropriate for cosmological fits,
we choose to average the log-likelihood surfaces obtained from both the
clustering wedges and multipole measurements after reconstruction. As
Figure~\ref{fig:pdr9} emphasizes, the core of these surfaces is
very similar and this averaging will yield results consistent with either of the
two individual approaches. Our consensus values are $H(0.57) = 92.9\pm7.8\,{\rm
km/s/Mpc}$ and $D_A(0.57) =1408\pm45{\rm Mpc}$ with a correlation coefficient of
$0.55$.
This correlation implies that using either value individually will
yield sub-optimal constraints; using them together requires correctly accounting for
the correlation between them.

Along with our statistical errors, we must also estimate any contribution from
systematic errors.  Systematic shifts in the acoustic scale are generally very
small because the large scale of the acoustic peak ensures that non-linear
gravitational effects are weak.  Our analysis method uses the marginalization
over a quadratic polynomial to remove systematic tilts from the measured
correlation functions.  The mock catalogs provide a careful check of the ability
of the fitting method to recover the input cosmology.  Table~\ref{tab:alphas}
shows this recovery to be better than 1\%: after reconstruction, we find for the
fiducial case a 0.1\% shift in $\alpha_\perp$ and a 0.6\% shift in
$\alpha_\parallel$ using the multipole method.  Other choices of fitting
parameters vary the results by $O(0.2\%)$.  The shifts in the wedges results are
similar.  \citet{kazin13} investigate the choice of fitting template
(Eq.~\ref{eq:template}) and find sub-percent dependence.  Hence, we conclude that
the systematic errors from the fitting methodology are small, of order 0.5\%.

Beyond this, astrophysical systematic shifts of the acoustic scale
are expected to be small.  \citet{2011ApJ...734...94M} showed that
a wide range of halo occupation distribution galaxy bias models
produced shifts of the acoustic scale of order 0.5\% or less.
Moreover, they found that the shifts vanished to within 0.1\% after
reconstruction was applied.  It is likely that reconstruction in
the DR9 survey geometry is less effective than it was in the
\citet{2011ApJ...734...94M} periodic box geometry, but we still
expect the shifts from galaxy bias to be below 0.5\%.  The only
astrophysical bias effect known to single out the acoustic scale
is the early universe streaming velocities identified by
\citet{2010PhRvD..82h3520T}. This effect can in principle
be detected with enough precision to negligibly affect the final
errors on the distance measurements \citep{2011JCAP...07..018Y}.
However, we have not yet assessed this size of the signal in BOSS data, although
it is not expected to be large given the vast difference in mass
scale between CDM mini-halos and those containing giant elliptical
galaxies.

We therefore estimate any systematic errors to be below 1\%, which is
negligible compared to our statistical errors.  Future work will
undoubtedly be able to further limit the systematic errors from both
fitting methodology and galaxy bias.

\subsection{Comparison with other Works} \label{sec:compare}
\begin{figure}
\includegraphics[width=3in]{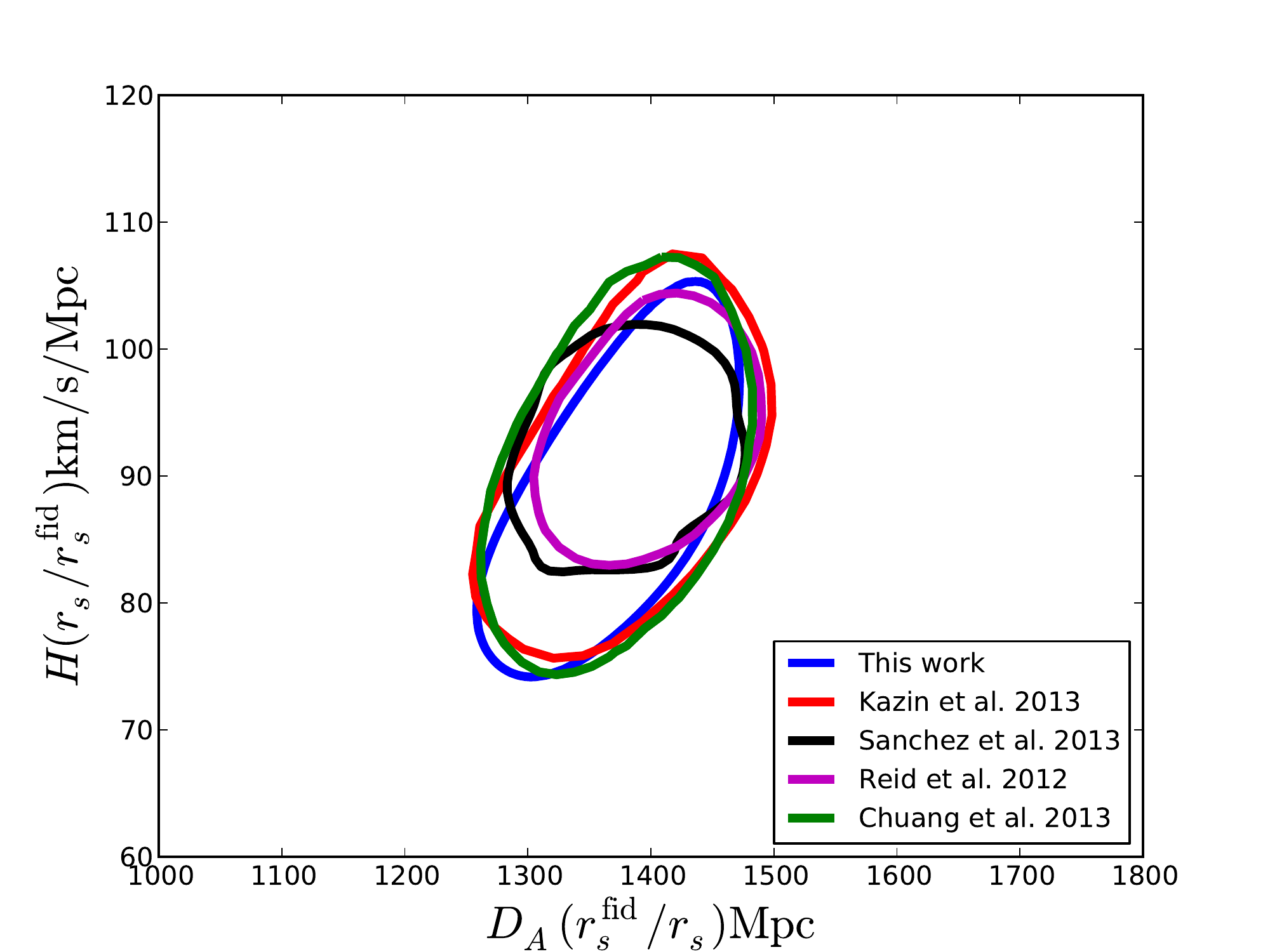}
\caption{ Pre-reconstruction joint likelihood
distributions (68\% confidence intervals) 
for $D_A(r_s^{\rm fid}/r_s)$ and  $H(r_s/r_s^{\rm fid}$) for different analyses of the CMASS DR9 data :
multipole-based analyses (blue, this work, purple, \citet{2012MNRAS.426.2719R}
and green, \citet{chuang13}) and the clustering wedges analysis (red,
\citet{kazin13} and black, \citet{sanchez13}).
This work and the companion work on wedges in \citet{kazin13} restrict to
fitting the BAO position only, while the remaining works fit the full shape of
the correlation function including the cosmological constraints from
redshift-space distortions. All of these agree well, with the full-shape methods
being generally more constraining than the BAO only methods.  }
\label{fig:dr9pr}
\end{figure}

% \begin{figure}
% \includegraphics[width=3in]{plots/2dlikelihood_comparison2}
% \caption{The same as Figure~\ref{fig:dr9pr} except now post
% reconstruction. We restrict here to the BAO only methods, since the broadband
% shape of the correlation function post-reconstruction is not well characterized.
% We show both the 68\% and 95\% contours  in this plot.}
% \label{fig:dr92dlikepostrecon}
% \end{figure}

\begin{table}
\caption{
Summary of the measurements of $D_{\rm A}(z)(\rsf/r_{\rm s})$,
$H(z)(r_{\rm s}/\rsf)$, and their cross-correlation, $\rho_{D_A H}$, from the
CMASS DR9 data.
The upper and middle sections of the table list the values   
obtained in this work from the pre- and post-reconstruction 
analyses of multipoles and clustering wedges, respectively. Our consensus values, defined in Section~\ref{sec:consensus}, are also given.
For comparison, the lower section of the table lists the results obtained in our companion papers, 
Kazin et al.(2013), S\'anchez et al.(2013) and Chuang et al.(2013).
All values correspond to the mean redshift of the sample, $z=0.57$.
\label{tab:dr9keyres}}
\begin{tabular}{lccc}
\hline
& $D_{\rm A}(z)(\rsf/r_{\rm s})$ & $H(z)(r_{\rm s}/\rsf)$ &
$\rho_{D_A H}$  \\[0.5mm] 
\hline
\multicolumn{4}{c}{Before Reconstruction} \\[0.5mm]
\hline
$(\xi_0(s),\xi_2(s))$                   & $ 1367\pm44 $    & $ 86.6 \pm 6.2 $ & 0.65 \\[0.5mm]
$(\xi_{\perp}(s),\xi_{\parallel}(s))$   & $1379\pm42$      & $88.3\pm5.1$     & 0.52 \\[0.5mm]
\hline
\multicolumn{4}{c}{After Reconstruction}  \\[0.5mm]
\hline
$(\xi_0(s),\xi_2(s))$                   & $1424\pm43$      & $95.4\pm7.5 $    & 0.63 \\[0.5mm]
$(\xi_{\perp}(s),\xi_{\parallel}(s))$   & $1386\pm36$      & $90.6\pm6.7$     & 0.50 \\[0.5mm]
Consensus                               & $1408\pm45$      & $92.9\pm7.8$     & 0.55   \\[0.5mm]
\hline
\multicolumn{4}{c}{Companion analyses}    \\[0.5mm]
\hline
Kazin et al.                            & $1386\pm39$      & $90.3\pm6.1$     & 0.48 \\[0.5mm]
S\'anchez et al.                        & $1379\pm39$      & $91.0\pm4.1$     & 0.30 \\[0.5mm]
Chuang et al.                           & $1371\pm41$      & $88.9\pm6.1$     & 0.49 \\[0.5mm]
Reid et al. 							& $1395\pm39$ 	   & $92.7\pm4.5$     & 0.24 \\[0.5mm]
\hline
\end{tabular}
\end{table}

Figure~\ref{fig:dr9pr} shows a comparison of the
two-dimensional 68\% confidence limits from our constraints on $D_{\rm A}(z)(r_{\rm s}^{\rm fid}/r_{\rm s})$ and
$H(z)(r_{\rm s}/r_{\rm s}^{\rm fid})$ and those of our companion papers: 
\citet{kazin13}, \citet{chuang13} and \citet{sanchez13} as well as the previous
work by \citet{2012MNRAS.426.2719R}.  
The corresponding one-dimensional marginalized constraints on these quantities
are listed in Table~\ref{tab:dr9keyres} showing excellent consistency. 

These analyses are based on different statistics and modelling details.
\citet{kazin13} explore the geometric constraints inferred from the BAO signal in both clustering wedges and
multipoles, by means of the de-wiggled template analysed here and an alternative
form based on renormalized perturbation theory \citep{2006PhRvD..73f3519C}.
\citet{chuang13} and \citet{sanchez13} exploit the information encoded in the full shape of these measurements
to derive cosmological constraints. While \citet{kazin13} and \citet{chuang13} follow the same approach applied
here and treat $D_{\rm A}$ and $H$ as free parameters (i.e. without adopting a specific relation between
their values), \citet{sanchez13} treats these quantities as derived parameters, with their values computed
in the context of the cosmological models being tested. The consistency of the derived constraints on 
$D_{\rm A}(z=0.57)\left(r_{\rm s}^{\rm fid}/r_{\rm s}\right)$ and
$H(z=0.57)\left(r_{\rm s}/r_{\rm s}^{\rm fid}\right)$ demonstrates the robustness of our results with 
respect to these differences in the implemented methodologies.

\citet{2012MNRAS.426.2719R} used the full shape of the  monopole-quadrupole pair of the SDSS-DR9 CMASS sample
to extract information from the Alcock-Paczynski test and the growth of structures. 
Based on these measurements they constrained the parameter combinations 
$D_{\rm V}(z)\left(r_{\rm s}^{\rm fid}/r_{\rm s}\right) = 2072\pm38\,{\rm Mpc}$ and 
$F(z)\equiv(1+z)D_{\rm A}(z)H(z)/c=0.675_{-0.038}^{0.042}$ at $z=0.57$.
From our consensus anisotropic BAO measurements we infer the constraints 
$D_{\rm V}(z=0.57)\left(r_{\rm s}^{\rm fid}/r_{\rm s}\right) = 2076\pm58\,{\rm Mpc}$ and 
$F(z=0.57)=0.692\pm0.087$, in excellent agreement with the results of \citet{2012MNRAS.426.2719R}.

\begin{figure}
\includegraphics[trim=0cm 0cm 2cm 8cm, width=0.45\textwidth]{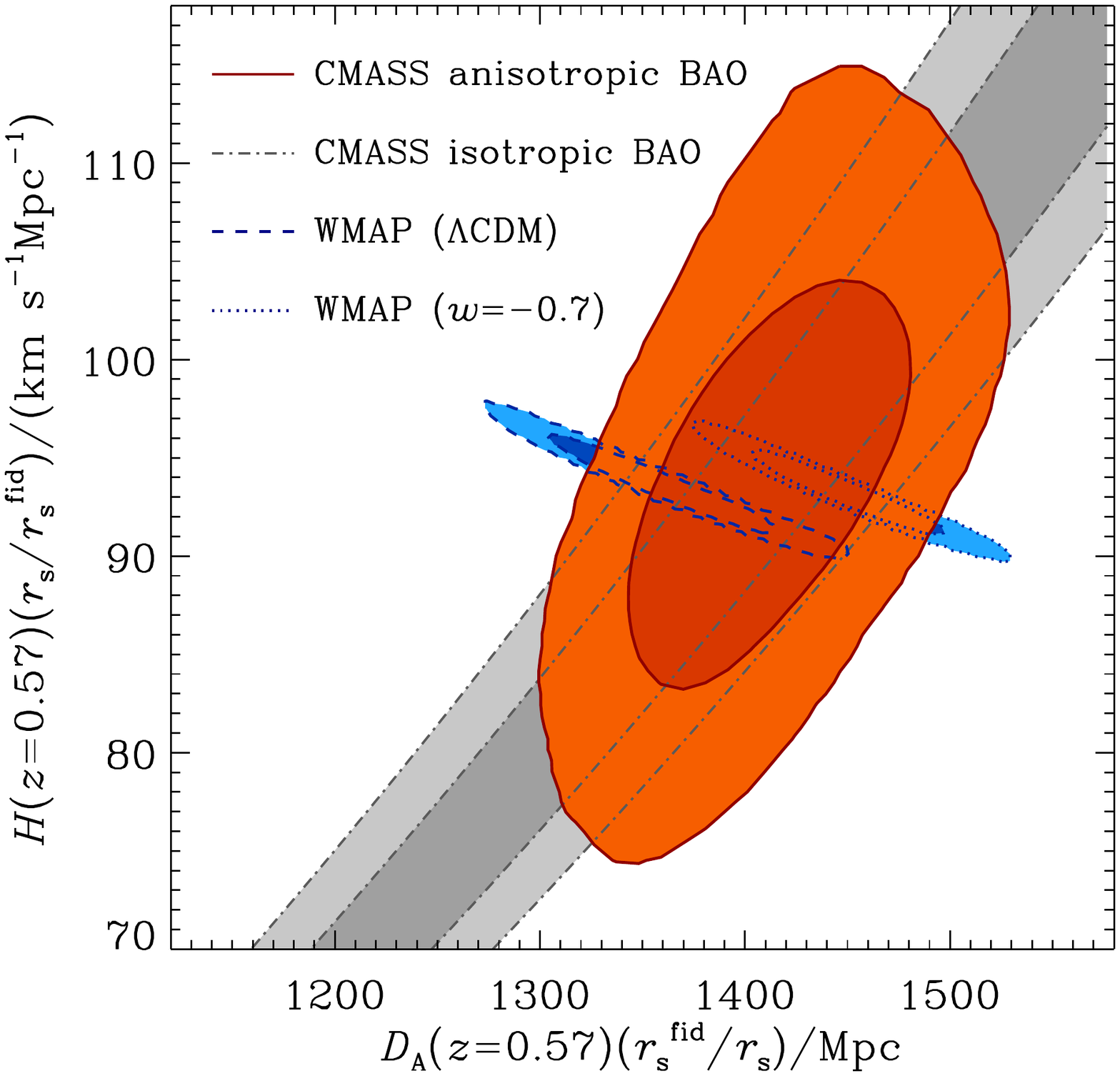}
\caption{ Comparison of the 65\% and 95\% constraints in the $D_{\rm
A}(z=0.57)\left(r_{\rm s}^{\rm fid}/r_{\rm s}\right)$--$H(z=0.57)\left(r_{\rm s}/r_{\rm s}^{\rm fid}\right)$ plane from
the CMASS consensus anisotropic BAO constraints described in
Section~\ref{sec:consensus} (solid lines) and those of the isotropic BAO
measurements of \citet{2012MNRAS.427.3435A} (dot-dashed lines). The WMAP prediction for
these parameters assuming a flat $\Lambda$CDM model (dashed lines), shows good
agreement with the CMASS constraints. Note that the CMASS constraints do {\it
not} assume $w=-1$ or flatness. The WMAP prediction obtained assuming a dark
energy equation of state $w = -0.7$ is also shown (dotted lines).
}
\label{fig:comp} 
\end{figure}

\citet{2012MNRAS.427.3435A} studied the isotropic BAO signal using the same galaxy sample studied here. 
As discussed in Section~\ref{sec:method}, spherically-averaged BAO measurements
constrain the ratio $D_{\rm V}(z)/r_{\rm s}$.
By combining the results obtained from the post-reconstruction CMASS correlation function and power spectrum,
\citet{2012MNRAS.427.3435A} obtained a consensus constraint of
$D_{\rm V}(z=0.57)\left(r_{\rm s}^{\rm fid}/r_{\rm s}\right)=2094\pm33\, {\rm Mpc}$.
This result corresponds to the constraints shown by the dot-dashed lines in
Figure~\ref{fig:comp}, which are in good agreement with the ones derived here.
The comparison of these results illustrates the extra information provided by anisotropic BAO measurements, 
which breaks the degeneracy between $D_A$ and $H$ obtained from isotropic BAO analyses.

Assuming a flat $\Lambda$CDM cosmology, the information provided by CMB observations is
enough to obtain a precise prediction of the values of 
$D_{\rm A}(z=0.57)\left(r_{\rm s}^{\rm fid}/r_{\rm s}\right)$ and 
$H(z=0.57)\left(r_{\rm s}/r_{\rm s}^{\rm fid}\right)$. The dashed lines in
Figure~\ref{fig:comp} correspond to the predictions for these quantities derived
under the assumption of a $\Lambda$CDM model from the WMAP observations of
\citet{2012arXiv1212.5225B} (computed as described in
Section~\ref{sec:cosmology}).
The anisotropic BAO constraints inferred from the CMASS sample are in good agreement with the
$\Lambda$CDM WMAP predictions. This is a clear indication of the consistency between these datasets and
their agreement with the standard $\Lambda$CDM model.

The CMB predictions are strongly dependent on the assumptions about dark energy or curvature. 
For any choice of $\Omega_k$ and $w(z)$, WMAP selects a different small region in the 
$D_{\rm A}(z=0.57)\left(r_{\rm s}^{\rm fid}/r_{\rm s}\right)$--$H(z=0.57)\left(r_{\rm s}/r_{\rm s}^{\rm fid}\right)$
plane. This is illustrated by the dotted contours in Figure~\ref{fig:comp},
which correspond to the WMAP prediction obtained assuming a flat universe with dark energy equation of state parameter $w=-0.7$. 
If the assumptions about curvature and dark energy are relaxed, i.e., these parameters are allowed
to vary freely, the region allowed by the CMB becomes significantly larger. Then, the combination 
of the CMB predictions with the BAO measurements can be used to constrain 
these cosmological parameters. In the next section we will explore the cosmological implications
of the combination of these datasets.

\section{Cosmological Implications} \label{sec:cosmology}

In this section we explore the constraints on the cosmological parameters
in different cosmological models from an analysis of galaxy BAO and CMB data, 
highlighting the improvements obtained from the anisotropic multipole analysis 
of the BOSS DR9 CMASS galaxy sample introduced in this paper. 

In Table~\ref{tab:cosmo} we summarize our main cosmological constraints
for different cosmological models: 
$\Lambda$CDM in which the Universe is flat and dark energy is represented by a cosmological constant, 
$o$CDM in which the spatial curvature ($\Omega_k$) is a free parameter, 
$w$CDM in which we allow the equation of state of dark energy ($w$) to vary, and 
$ow$CDM in which we let both parameters vary. 
Different columns represent different combinations of CMB and BAO datasets. 
The CMB data comes from the final data release of WMAP \citep[WMAP9; ][]{2012arXiv1212.5226H}. 
We combine CMB data with BAO constraints from DR7 (SDSS-II LRGs) and DR9 (BOSS CMASS) galaxies. 
The isotropic BAO constraints include SDSS-II LRGs from \citet{2012MNRAS.427.2132P}, and CMASS 
galaxies from \citet{2012MNRAS.427.3435A}, with anisotropic BAO data from SDSS-II LRGs in 
\citet{2012arXiv1206.6732X} and from CMASS galaxies (this work). 

As seen in Figure~\ref{fig:comp}, the cosmological information contained in the
anisotropic clustering data breaks the degeneracy present in the isotropic case
between the angular diameter distance and the Hubble parameter (orange contours
versus gray band respectively). Moreover, these distance measurements allow us
to constrain cosmological parameters such as the spatial curvature $\Omega_k$ or
the dark energy equation of state $w$. The blue contours in Figure~\ref{fig:comp} 
show CMB constraints assuming a $\Lambda$CDM model where we change the equation of state of dark energy 
to $w=-0.7$ from $w=-1$ (which is the case for a cosmological constant). We can 
see that the locus of the allowed parameter space is clearly different in each of these 
cases given the size of these error ellipses. We note that the distance constraints 
from the anisotropic BAO analysis are less tight and hence they benefit from the complementarity 
of other cosmological probes, such as the CMB. This complementarity allows for 
precision measurements of cosmological parameters. The allowed parameter space can 
could be further reduced by combining information from anisotropic 
clustering from surveys covering different redshift ranges: such as the low-redshift 
BAO measurements of the 6dF Galaxy Survey \citep{2011MNRAS.416.3017B} to the 
high-redshift Lyman $\alpha$ forest clustering results 
\citep{2012arXiv1211.2616B,2013arXiv1301.3459S,2013arXiv1301.3456K}.

\begin{figure}
\includegraphics[width=0.9\linewidth]{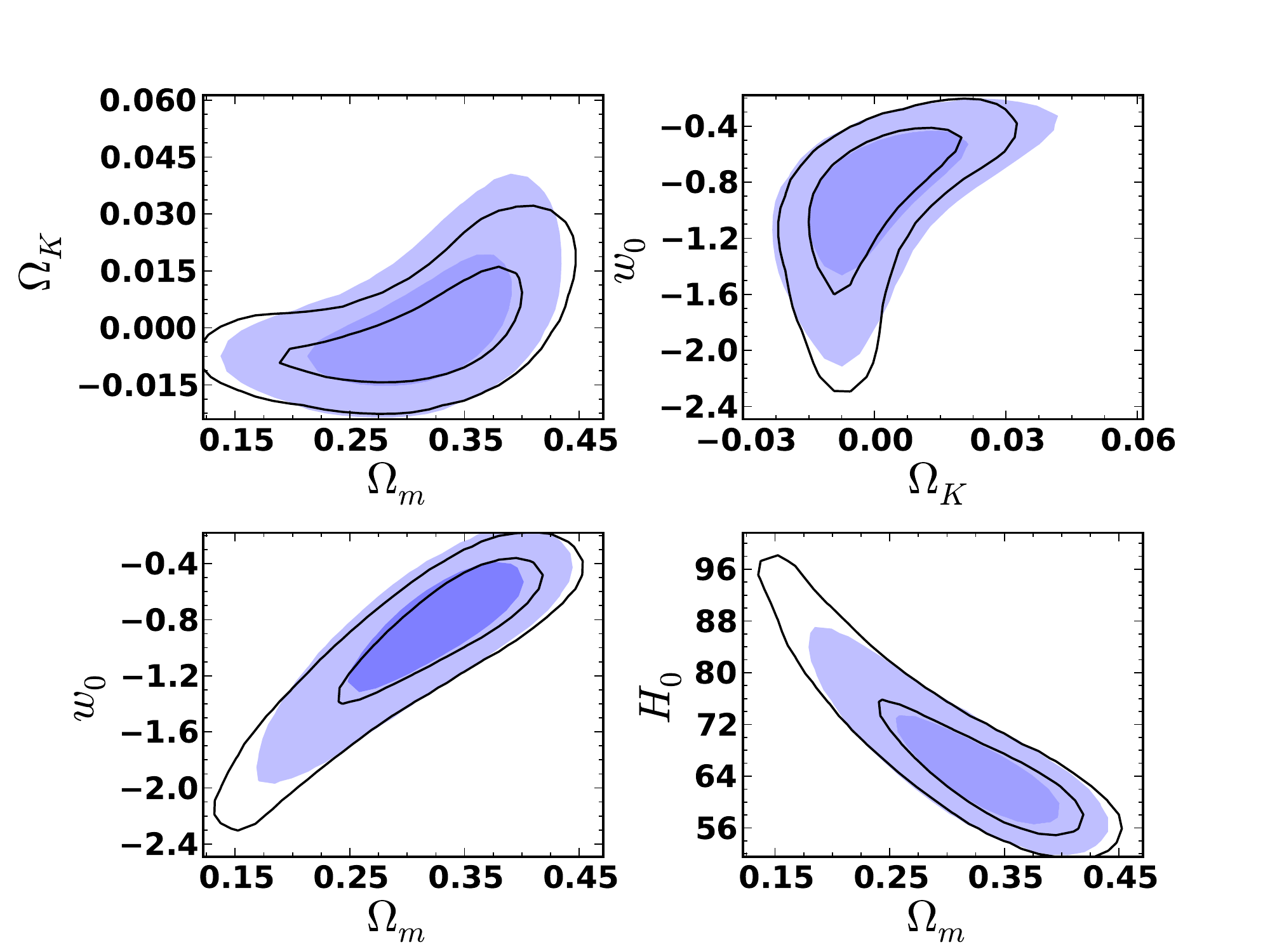}
\caption{Constraints on the cosmological parameters of the $ow$CDM model when
combining WMAP9 data with the anisotropic BAO data from CMASS presented in this
work (filled blue contours). For comparison, the constrained parameter space from
the combination of WMAP9 data with the isotropic BAO analysis in
\citet{2012MNRAS.427.3435A} is shown as black contours.
}
\label{fig:cosmo2Dani}
\end{figure}

We find an improvement in the cosmological constraints in the $ow$CDM cosmological 
model from the anisotropic BAO analysis versus the spherically-averaged
isotropic BAO analysis. These differences are apparent in
Figure~\ref{fig:cosmo2Dani} where the BAO information is combined with CMB data. Plotted here are the likelihood contours of
two cosmological parameters (from the set $\Omega_k$, $w$, $\Omega_M$,
and $H_0$) while marginalizing over the remaining cosmological parameters
in the $ow$CDM model. We can see that the allowed parameter space enclosed
by the 68\% and 95\% confidence level contours is smaller in the anisotropic case,
indicating that the anisotropic analysis provides a clear advantage in breaking 
the degeneracy between curvature and the equation of state of dark energy.

In Table~\ref{tab:wedgesmulti} we compare cosmological constraints
from the multipoles technique (this work) with the wedges technique 
discussed in \S4. We find that the wedges analysis \citep{kazin13} shows 
a marginally larger error bar in the cosmological parameters as compared 
to the multipoles technique. The table also compares the cosmological 
constraints using the consensus likelihood: the average of the 
log-likelihood from both multipoles and wedges.  We find that both techniques 
show consistent results.  When combined with CMB data, none of these results 
deviate significantly from a flat Universe with $\Omega_k=0.0$ or a cosmological constant with $w=-1$.

%\begin{figure}
%\includegraphics[width=0.9\linewidth]{plots/owcdm_wmapdr9iso_2D}
%\caption{Constraints on the cosmological parameters of the $ow$CDM model when combining WMAP9 data with 
%the isotropic BAO data from CMASS presented in \citet{2012MNRAS.427.3435A}. 
%}j
%\label{fig:cosmo2Diso}
%\end{figure}

\begin{table*}
\caption{Cosmological constraints from anisotropic BAO in CMASS DR9 data. Different rows show constraints on 
different cosmological models. Columns indicate different combinations of CMB and BAO 
datasets, where 'isotropic' indicates the isotropic BAO analysis of \citet{2012MNRAS.427.3435A}, 
and 'anisotropic' corresponds to the anisotropic BAO analysis presented here. The DR7 data include 
the analysis of the SDSS-II LRGs for the isotropic BAO in \citet{2012MNRAS.427.2132P} and 
the anisotropic results in \citet{2012arXiv1206.6732X}.  The Hubble constant $H_0$ is in units of km s$^{-1}$ Mpc$^{-1}$.
}
\label{tab:cosmo}
\begin{tabular}{llccccc}
\hline
Cosmological         & Parameter &   WMAP9   &    WMAP9          &   WMAP9          &   WMAP9        &   WMAP9          \\
model                &           &    +DR9           &   +DR9           &   +DR7+DR9     &   +DR7+DR9       \\
                     &           &    (isotropic)    &   (anisotropic)  &   (isotropic)  &   (anisotropic)  \\
\hline
$\Lambda$CDM & & & & & & \\
     & $\Omega_M$  &   0.300$\pm$0.016 & 0.295$\pm$0.017 & 0.296$\pm$0.012 & 0.290$\pm$0.012 & 0.280$\pm$0.026 \\
     & $H_0$       &   68.3 $\pm$ 1.3 & 68.8 $\pm$1.4 & 68.7 $\pm$1.0 & 69.1$\pm$1.0 & 70.0 $\pm$2.2 \\
\hline
$o$CDM & & & & & & \\
     & $\Omega_M$ &  0.304$\pm$0.016 & 0.298$\pm$0.016 & 0.293$\pm$0.012 & 0.290$\pm$0.012 & 0.507$\pm$0.236 \\
     & $H_0$      &  67.1 $\pm$1.5 & 67.8 $\pm$1.7 & 68.2 $\pm$1.1 & 68.7$\pm$1.2 & 56.2 $\pm$12.4  \\
     & $\Omega_k$ &  -0.006 $\pm$0.005 & -0.005$\pm$0.005 & -0.004$\pm$0.005 & -0.003$\pm$0.005 & -0.056$\pm$0.060 \\
\hline
$w$CDM & & & & & & \\
     & $\Omega_M$ &  0.333$\pm$0.041 & 0.313$\pm$0.042 & 0.297$\pm$0.027 & 0.297$\pm$0.022 & 0.302$\pm$0.096 \\
     & $H_0$      &  64.5 $\pm$5.0 & 66.6 $\pm$5.6 & 68.5 $\pm$4.0 & 68.1$\pm$3.2 & 69.9 $\pm$11.5 \\
     & $w$        & -0.84 $\pm$0.21 & -0.90 $\pm$0.22 & -0.99 $\pm$0.19 & -0.95$\pm$0.15 & -0.99 $\pm$0.35 \\
\hline
$ow$CDM & & & & & & \\
      & $\Omega_M$ &  0.310$\pm$0.070 & 0.314$\pm$0.058 & 0.269$\pm$0.045 & 0.284$\pm$0.039 & 0.596$\pm$0.254 \\
      & $H_0$      &  67.7 $\pm$9.6 & 66.7 $\pm$7.5 & 72.1 $\pm$6.7 & 69.7$\pm$5.3 & 51.9 $\pm$12.6 \\
      & $\Omega_k$ &  +0.000 $\pm$0.011 & +0.002$\pm$0.013 & -0.005$\pm$0.007 & -0.002$\pm$0.008 & -0.072$\pm$0.066 \\
      & $w$        &  -0.99 $\pm$0.44 & -0.92 $\pm$0.37 & -1.19 $\pm$0.34 & -1.05$\pm$0.30 & -1.02 $\pm$0.53 \\
\hline
\end{tabular}
\end{table*}

\begin{table*}
\caption{Comparison of the cosmological constraints from the analysis of wedges,
multipoles, and the consensus likelihood from both techniques, using anisotropic
BAO CMASS DR9 data. Different rows show constraints on different cosmological
models. The Hubble constant $H_0$ is in units of km s$^{-1}$ Mpc$^{-1}$.
}
\label{tab:wedgesmulti}
\begin{tabular}{llccc}
\hline
Cosmological & Parameter   &   WMAP9+DR9    &   WMAP9+DR9      &   WMAP9+DR9  \\
model        &             &   (consensus)  &   (multipoles)   &   (wedges)   \\
\hline
$\Lambda$CDM & & & & \\
     & $\Omega_M$  &  0.295$\pm$0.017 & 0.298$\pm$0.016 & 0.291$\pm$0.017 \\
     & $H_0$       &  68.8 $\pm$1.4 & 68.5 $\pm$1.3 & 69.1 $\pm$1.4 \\
\hline
$o$CDM & & & & \\
     & $\Omega_M$ &  0.298$\pm$0.016 & 0.301$\pm$0.016 & 0.296$\pm$0.019 \\
     & $H_0$      & 67.8 $\pm$1.7 & 67.5 $\pm$1.6 & 68.0 $\pm$2.0  \\
     & $\Omega_k$ & -0.005$\pm$0.005 & -0.005$\pm$0.005 & -0.005$\pm$0.005 \\
\hline
$w$CDM & & & & \\
     & $\Omega_M$ &  0.313$\pm$0.042 & 0.326$\pm$0.033 & 0.307$\pm$0.043  \\
     & $H_0$      &  66.6 $\pm$5.6 & 64.9 $\pm$4.0 & 67.3 $\pm$5.8 \\
     & $w$ &  -0.90 $\pm$0.22 & -0.84 $\pm$0.17 & -0.93 $\pm$0.23  \\
\hline
$ow$CDM & & & & \\
      & $\Omega_M$ &  0.314$\pm$0.058 & 0.327$\pm$0.050  & 0.297$\pm$0.059  \\
      & $H_0$      &  66.7 $\pm$7.5   & 65.0 $\pm$6.2    & 69.0 $\pm$8.1 \\
      & $\Omega_k$ & +0.002$\pm$0.013 & +0.002$\pm$0.011 & +0.000$\pm$0.011 \\
      & $w$        & -0.92 $\pm$0.37  & -0.85 $\pm$0.31  & -1.03 $\pm$0.39  \\ 
\hline
\end{tabular}
\end{table*}

\section{Conclusions} \label{sec:conclusions}
In this paper, we have presented a detailed analysis of the anisotropic
measurement of the baryon acoustic peak in the SDSS-III BOSS DR9
sample of $0.43<z<0.7$ galaxies.  The baryon acoustic oscillations
provide a robust standard ruler by which to measure the cosmological
distance scale.  One of the important opportunities of the BAO
method is its ability to measure the angular diameter distance and
Hubble parameter separately at higher redshift.  The BOSS DR9 sample
is large enough to provide a detection of the acoustic peak
both along and across the line of sight.

Our analysis has relied on two separate methods by which to measure
the acoustic peak in the anisotropic correlation function.  The first
uses the monopole and quadrupole of the anisotropic clustering, following
the methods of \citet{2012arXiv1206.6732X}.  The second separates the correlation
function into two bins of the angle between the separation vector of 
the pair and the line-of-sight, following \citet{2012MNRAS.419.3223K}.  The latter
analysis is further described in \citet{kazin13}.  In both cases, we
fit a model of the correlation function to the data, using
reconstruction to sharpen the acoustic peak and mock catalogs to
define the covariance matrix of the observables.  The fit is able
to vary the position of the acoustic peak in both the line-of-sight
and transverse directions, thereby measuring $H(z)$ and $D_A(z)$,
respectively.  The fitting includes a marginalization over broadband
functions in both directions, thereby isolating the acoustic peak
information from possible uncertainties in scale-dependent bias,
redshift distortions, the reconstruction method, and systematic
clustering errors.

From these fits, we define a consensus value of  
$H(0.57) (r_s/r_s^{\rm fid}) = 92.9 \pm 7.8$ km/s/Mpc (8.4\%) and 
$D_A(0.57) (r_s^{\rm fid}/r_s) = 1408 \pm 45$ Mpc (3.2\%).
These two measurements have a correlation coefficient of 0.55 that
should be taken into account when measuring parameters of cosmological
models.
We note that the sound horizon $r_s$ is constrained to $\sim$0.7\% rms 
from current CMB data in simple adiabatic cold dark matter models
\citep{2012arXiv1212.5225B,2012arXiv1212.5226H}; 
hence, the uncertainty in $r_s$ is subdominant for the usual fits.

Our results are highly consistent with the analysis of the spherically
averaged acoustic peak in \citet{2012MNRAS.427.3435A}, which yielded a
measurement of $D_V \propto D_A^{2/3}/H^{1/3}$.  We find
(Figure~\ref{fig:dah_mono}) that fitting the anisotropic model to only the
monopole data returns constraints elongated in the $D_V$ direction, justifying
the treatment in \citet{2012MNRAS.427.3435A}.  We further find that when
using the anisotropic data, we get constraints on $D_V$ similar
to that of \citet{2012MNRAS.427.3435A}.

The cosmological parameter measurements we achieve from our measurement
of $H(0.57)$ and $D_A(0.57)$ are similar to those found from
$D_V(0.57)$ in \citet{2012MNRAS.427.3435A}.  Because of those similarities,
we have presented only a small sampling of cosmologies; further
analyses can be found in \citet{2012MNRAS.427.3435A}.  We find strong 
consistency with the standard flat $w=-1$ cosmological model. 
The fact that separating $D_V$ into separate $D_A$ and $H$ information
does not improve the cosmological parameter fits is, we believe,
largely due to the relatively low redshift of the data set: as
$z\rightarrow0$, $D_A$ and $H$ provide degenerate information in
all cosmological models.   

The analyses here and in \citet{kazin13} are focused only on the anisotropic
acoustic peak.  
In addition, \citet{2012MNRAS.426.2719R}, \citet{sanchez13}, and \citet{chuang13} 
have studied the full shape of the anisotropic large-scale clustering
in the BOSS DR9 CMASS sample.  Such studies require more assumptions
about the modeling of galaxy bias and redshift distortions, but offer
stronger constraints on $H(z)$ through the use of the Alcock-Paczynski
effect on the broadband clustering signal.  The conclusions reached
are consistent with those here from the acoustic peak alone.

The anisotropic measurement of the baryon acoustic oscillations has now
been performed at three distinct redshifts: $z=0.35$ with the SDSS-II
Luminous Red Galaxy sample, $z=2.3$ with the SDSS-III BOSS Ly$\alpha$ forest
sample, and this analysis at $z=0.57$.  All three have found strong
consistency with the standard cosmological model of a spatially
flat Universe with acceleration driven by a cosmological constant.
These results represent only the first third of the SDSS-III BOSS 
data set but mark an important milestone for BAO studies. 
These anisotropic methods and measurements define a clear
path for the ambitious surveys of the coming decade.

\section{Acknowledgments} 
Numerical computations for the PTHalos mocks were done on the Sciama
High Performance Compute (HPC) cluster which is supported by the ICG,
SEPNet and the University of Portsmouth. Analyses were partially supported
by were supported by facilities and staff
of the Yale University Faculty of Arts and Sciences High Performance
Computing Center.

We acknowledge the use of the Legacy Archive for Microwave Background
Data Analysis (LAMBDA), part of the High Energy Astrophysics Science
Archive Center (HEASARC). HEASARC/LAMBDA is a service of the
Astrophysics Science Division at the NASA Goddard Space Flight
Center.

Funding for SDSS-III has been provided by the Alfred P. Sloan Foundation, the
Participating Institutions, the National Science Foundation, and the U.S.
Department of Energy Office of Science. The SDSS-III web site is
\texttt{http://www.sdss3.org/}.

SDSS-III is managed by the Astrophysical Research Consortium for the
Participating Institutions of the SDSS-III Collaboration including the
University of Arizona, the Brazilian Participation Group, Brookhaven National
Laboratory, University of Cambridge, Carnegie Mellon University, University of
Florida, the French Participation Group, the German Participation Group, Harvard
University, the Instituto de Astrofisica de Canarias, the Michigan State/Notre
Dame/JINA Participation Group, Johns Hopkins University, Lawrence Berkeley
National Laboratory, Max Planck Institute for Astrophysics, Max Planck Institute
for Extraterrestrial Physics, New Mexico State University, New York University,
Ohio State University, Pennsylvania State University, University of Portsmouth,
Princeton University, the Spanish Participation Group, University of Tokyo,
University of Utah, Vanderbilt University, University of Virginia, University of
Washington, and Yale University.

\bibliographystyle{mn2elong}
\bibliography{dr9_aniso,dr9_aniso_preprints,dr9_aniso_companions}

\begin{thebibliography}{92}
\expandafter\ifx\csname natexlab\endcsname\relax\def\natexlab#1{#1}\fi

\bibitem[{{Ahn} {et~al}\mbox{.}(2012){Ahn}, {Alexandroff}, {Allende Prieto},
  {Anderson}, {Anderton}, {Andrews}, {Aubourg}, {Bailey}, {Balbinot}, {Barnes},
  \& et~al.}]{2012ApJS..203...21A}
{Ahn} C.~P. {et~al.}, 2012, \apjs, 203, 21

\bibitem[{{Aihara} {et~al}\mbox{.}(2011){Aihara}, {Allende Prieto}, {An},
  {Anderson}, {Aubourg}, {Balbinot}, {Beers}, {Berlind}, {Bickerton},
  {Bizyaev}, {Blanton}, {Bochanski}, {Bolton}, {Bovy}, {Brandt}, {Brinkmann},
  {Brown}, {Brownstein}, {Busca}, {Campbell}, {Carr}, {Chen}, {Chiappini},
  {Comparat}, {Connolly}, {Cortes}, {Croft}, {Cuesta}, {da Costa}, {Davenport},
  {Dawson}, {Dhital}, {Ealet}, {Ebelke}, {Edmondson}, {Eisenstein},
  {Escoffier}, {Esposito}, {Evans}, {Fan}, {Femen{\'{\i}}a Castell{\'a}},
  {Font-Ribera}, {Frinchaboy}, {Ge}, {Gillespie}, {Gilmore}, {Gonz{\'a}lez
  Hern{\'a}ndez}, {Gott}, {Gould}, {Grebel}, {Gunn}, {Hamilton}, {Harding},
  {Harris}, {Hawley}, {Hearty}, {Ho}, {Hogg}, {Holtzman}, {Honscheid}, {Inada},
  {Ivans}, {Jiang}, {Johnson}, {Jordan}, {Jordan}, {Kazin}, {Kirkby}, {Klaene},
  {Knapp}, {Kneib}, {Kochanek}, {Koesterke}, {Kollmeier}, {Kron}, {Lampeitl},
  {Lang}, {Le Goff}, {Lee}, {Lin}, {Long}, {Loomis}, {Lucatello}, {Lundgren},
  {Lupton}, {Ma}, {MacDonald}, {Mahadevan}, {Maia}, {Makler}, {Malanushenko},
  {Malanushenko}, {Mandelbaum}, {Maraston}, {Margala}, {Masters}, {McBride},
  {McGehee}, {McGreer}, {M{\'e}nard}, {Miralda-Escud{\'e}}, {Morrison},
  {Mullally}, {Muna}, {Munn}, {Murayama}, {Myers}, {Naugle}, {Neto}, {Nguyen},
  {Nichol}, {O'Connell}, {Ogando}, {Olmstead}, {Oravetz}, {Padmanabhan},
  {Palanque-Delabrouille}, {Pan}, {Pandey}, {P{\^a}ris}, {Percival},
  {Petitjean}, {Pfaffenberger}, {Pforr}, {Phleps}, {Pichon}, {Pieri}, {Prada},
  {Price-Whelan}, {Raddick}, {Ramos}, {Reyl{\'e}}, {Rich}, {Richards}, {Rix},
  {Robin}, {Rocha-Pinto}, {Rockosi}, {Roe}, {Rollinde}, {Ross}, {Ross},
  {Rossetto}, {S{\'a}nchez}, {Sayres}, {Schlegel}, {Schlesinger}, {Schmidt},
  {Schneider}, {Sheldon}, {Shu}, {Simmerer}, {Simmons}, {Sivarani}, {Snedden},
  {Sobeck}, {Steinmetz}, {Strauss}, {Szalay}, {Tanaka}, {Thakar}, {Thomas},
  {Tinker}, {Tofflemire}, {Tojeiro}, {Tremonti}, {Vandenberg}, {Vargas
  Maga{\~n}a}, {Verde}, {Vogt}, {Wake}, {Wang}, {Weaver}, {Weinberg}, {White},
  {White}, {Yanny}, {Yasuda}, {Yeche}, \& {Zehavi}}]{2011ApJS..193...29A}
{Aihara} H. {et~al.}, 2011, \apjs, 193, 29

\bibitem[{{Alcock} \& {Paczynski}(1979)}]{1979Natur.281..358A}
{Alcock} C., {Paczynski} B., 1979, \nat, 281, 358

\bibitem[{{Anderson} {et~al}\mbox{.}(2012){Anderson}, {Aubourg}, {Bailey},
  {Bizyaev}, {Blanton}, {Bolton}, {Brinkmann}, {Brownstein}, {Burden},
  {Cuesta}, {da Costa}, {Dawson}, {de Putter}, {Eisenstein}, {Gunn}, {Guo},
  {Hamilton}, {Harding}, {Ho}, {Honscheid}, {Kazin}, {Kirkby}, {Kneib},
  {Labatie}, {Loomis}, {Lupton}, {Malanushenko}, {Malanushenko}, {Mandelbaum},
  {Manera}, {Maraston}, {McBride}, {Mehta}, {Mena}, {Montesano}, {Muna},
  {Nichol}, {Nuza}, {Olmstead}, {Oravetz}, {Padmanabhan},
  {Palanque-Delabrouille}, {Pan}, {Parejko}, {P{\^a}ris}, {Percival},
  {Petitjean}, {Prada}, {Reid}, {Roe}, {Ross}, {Ross}, {Samushia},
  {S{\'a}nchez}, {Schlegel}, {Schneider}, {Sc{\'o}ccola}, {Seo}, {Sheldon},
  {Simmons}, {Skibba}, {Strauss}, {Swanson}, {Thomas}, {Tinker}, {Tojeiro},
  {Maga{\~n}a}, {Verde}, {Wagner}, {Wake}, {Weaver}, {Weinberg}, {White}, {Xu},
  {Y{\`e}che}, {Zehavi}, \& {Zhao}}]{2012MNRAS.427.3435A}
{Anderson} L. {et~al.}, 2012, \mnras, 427, 3435

\bibitem[{{Angulo} {et~al}\mbox{.}(2008){Angulo}, {Baugh}, {Frenk}, \&
  {Lacey}}]{2008MNRAS.383..755A}
{Angulo} R.~E., {Baugh} C.~M., {Frenk} C.~S., {Lacey} C.~G., 2008, \mnras, 383,
  755

\bibitem[{{Bennett} {et~al}\mbox{.}(2012){Bennett}, {Larson}, {Weiland},
  {Jarosik}, {Hinshaw}, {Odegard}, {Smith}, {Hill}, {Gold}, {Halpern},
  {Komatsu}, {Nolta}, {Page}, {Spergel}, {Wollack}, {Dunkley}, {Kogut},
  {Limon}, {Meyer}, {Tucker}, \& {Wright}}]{2012arXiv1212.5225B}
{Bennett} C.~L. {et~al.}, 2012, ArXiv e-prints

\bibitem[{{Beutler} {et~al}\mbox{.}(2011){Beutler}, {Blake}, {Colless},
  {Jones}, {Staveley-Smith}, {Campbell}, {Parker}, {Saunders}, \&
  {Watson}}]{2011MNRAS.416.3017B}
{Beutler} F. {et~al.}, 2011, \mnras, 416, 3017

\bibitem[{{Blake} {et~al}\mbox{.}(2007){Blake}, {Collister}, {Bridle}, \&
  {Lahav}}]{2007MNRAS.374.1527B}
{Blake} C., {Collister} A., {Bridle} S., {Lahav} O., 2007, \mnras, 374, 1527

\bibitem[{{Blake} {et~al}\mbox{.}(2011{\natexlab{a}}){Blake}, {Davis}, {Poole},
  {Parkinson}, {Brough}, {Colless}, {Contreras}, {Couch}, {Croom},
  {Drinkwater}, {Forster}, {Gilbank}, {Gladders}, {Glazebrook}, {Jelliffe},
  {Jurek}, {Li}, {Madore}, {Martin}, {Pimbblet}, {Pracy}, {Sharp}, {Wisnioski},
  {Woods}, {Wyder}, \& {Yee}}]{2011MNRAS.415.2892B}
{Blake} C. {et~al.}, 2011{\natexlab{a}}, \mnras, 415, 2892

\bibitem[{{Blake} \& {Glazebrook}(2003)}]{2003ApJ...594..665B}
{Blake} C., {Glazebrook} K., 2003, \apj, 594, 665

\bibitem[{{Blake} {et~al}\mbox{.}(2011{\natexlab{b}}){Blake}, {Kazin},
  {Beutler}, {Davis}, {Parkinson}, {Brough}, {Colless}, {Contreras}, {Couch},
  {Croom}, {Croton}, {Drinkwater}, {Forster}, {Gilbank}, {Gladders},
  {Glazebrook}, {Jelliffe}, {Jurek}, {Li}, {Madore}, {Martin}, {Pimbblet},
  {Poole}, {Pracy}, {Sharp}, {Wisnioski}, {Woods}, {Wyder}, \&
  {Yee}}]{2011MNRAS.418.1707B}
{Blake} C. {et~al.}, 2011{\natexlab{b}}, \mnras, 418, 1707

\bibitem[{{Bolton} {et~al}\mbox{.}(2012){Bolton}, {Schlegel}, {Aubourg},
  {Bailey}, {Bhardwaj}, {Brownstein}, {Burles}, {Chen}, {Dawson}, {Eisenstein},
  {Gunn}, {Knapp}, {Loomis}, {Lupton}, {Maraston}, {Muna}, {Myers}, {Olmstead},
  {Padmanabhan}, {P{\^a}ris}, {Percival}, {Petitjean}, {Rockosi}, {Ross},
  {Schneider}, {Shu}, {Strauss}, {Thomas}, {Tremonti}, {Wake}, {Weaver}, \&
  {Wood-Vasey}}]{2012AJ....144..144B}
{Bolton} A.~S. {et~al.}, 2012, \aj, 144, 144

\bibitem[{{Bond} \& {Efstathiou}(1987)}]{1987MNRAS.226..655B}
{Bond} J.~R., {Efstathiou} G., 1987, \mnras, 226, 655

\bibitem[{{Busca} {et~al}\mbox{.}(2012){Busca}, {Delubac}, {Rich}, {Bailey},
  {Font-Ribera}, {Kirkby}, {Le Goff}, {Pieri}, {Slosar}, {Aubourg}, {Bautista},
  {Bizyaev}, {Blomqvist}, {Bolton}, {Bovy}, {Brewington}, {Borde}, {Brinkmann},
  {Carithers}, {Croft}, {Dawson}, {Ebelke}, {Eisenstein}, {Hamilton}, {Ho},
  {Hogg}, {Honscheid}, {Lee}, {Lundgren}, {Malanushenko}, {Malanushenko},
  {Margala}, {Maraston}, {Mehta}, {Miralda-Escud{\'e}}, {Myers}, {Nichol},
  {Noterdaeme}, {Olmstead}, {Oravetz}, {Palanque-Delabrouille}, {Pan},
  {P{\^a}ris}, {Percival}, {Petitjean}, {Roe}, {Rollinde}, {Ross}, {Rossi},
  {Schlegel}, {Schneider}, {Shelden}, {Sheldon}, {Simmons}, {Snedden},
  {Tinker}, {Viel}, {Weaver}, {Weinberg}, {White}, {Y{\`e}che}, \&
  {York}}]{2012arXiv1211.2616B}
{Busca} N.~G. {et~al.}, 2012, ArXiv e-prints

\bibitem[{{Chuang} \& {Wang}(2012)}]{2012MNRAS.426..226C}
{Chuang} C.-H., {Wang} Y., 2012, \mnras, 426, 226

\bibitem[{{Chuang} {et~al}\mbox{.}(2012){Chuang}, {Wang}, \&
  {Hemantha}}]{2012MNRAS.423.1474C}
{Chuang} C.-H., {Wang} Y., {Hemantha} M.~D.~P., 2012, \mnras, 423, 1474

\bibitem[{{Chuang} {et~al}\mbox{.}(2013){Chuang} {et~al.}}]{chuang13}
{Chuang} C.-H., {et~al.}, 2013, \mnras\, submitted

\bibitem[{{Cole} {et~al}\mbox{.}(2005){Cole}, {Percival}, {Peacock}, {Norberg},
  {Baugh}, {Frenk}, {Baldry}, {Bland-Hawthorn}, {Bridges}, {Cannon}, {Colless},
  {Collins}, {Couch}, {Cross}, {Dalton}, {Eke}, {De Propris}, {Driver},
  {Efstathiou}, {Ellis}, {Glazebrook}, {Jackson}, {Jenkins}, {Lahav}, {Lewis},
  {Lumsden}, {Maddox}, {Madgwick}, {Peterson}, {Sutherland}, \&
  {Taylor}}]{2005MNRAS.362..505C}
{Cole} S. {et~al.}, 2005, \mnras, 362, 505

\bibitem[{{Crocce} \& {Scoccimarro}(2006)}]{2006PhRvD..73f3519C}
{Crocce} M., {Scoccimarro} R., 2006, \prd, 73, 063519

\bibitem[{{Dawson} {et~al}\mbox{.}(2013){Dawson}, {Schlegel}, {Ahn},
  {Anderson}, {Aubourg}, {Bailey}, {Barkhouser}, {Bautista}, {Beifiori},
  {Berlind}, {Bhardwaj}, {Bizyaev}, {Blake}, {Blanton}, {Blomqvist}, {Bolton},
  {Borde}, {Bovy}, {Brandt}, {Brewington}, {Brinkmann}, {Brown}, {Brownstein},
  {Bundy}, {Busca}, {Carithers}, {Carnero}, {Carr}, {Chen}, {Comparat},
  {Connolly}, {Cope}, {Croft}, {Cuesta}, {da Costa}, {Davenport}, {Delubac},
  {de Putter}, {Dhital}, {Ealet}, {Ebelke}, {Eisenstein}, {Escoffier}, {Fan},
  {Filiz Ak}, {Finley}, {Font-Ribera}, {G{\'e}nova-Santos}, {Gunn}, {Guo},
  {Haggard}, {Hall}, {Hamilton}, {Harris}, {Harris}, {Ho}, {Hogg}, {Holder},
  {Honscheid}, {Huehnerhoff}, {Jordan}, {Jordan}, {Kauffmann}, {Kazin},
  {Kirkby}, {Klaene}, {Kneib}, {Le Goff}, {Lee}, {Long}, {Loomis}, {Lundgren},
  {Lupton}, {Maia}, {Makler}, {Malanushenko}, {Malanushenko}, {Mandelbaum},
  {Manera}, {Maraston}, {Margala}, {Masters}, {McBride}, {McDonald}, {McGreer},
  {McMahon}, {Mena}, {Miralda-Escud{\'e}}, {Montero-Dorta}, {Montesano},
  {Muna}, {Myers}, {Naugle}, {Nichol}, {Noterdaeme}, {Nuza}, {Olmstead},
  {Oravetz}, {Oravetz}, {Owen}, {Padmanabhan}, {Palanque-Delabrouille}, {Pan},
  {Parejko}, {P{\^a}ris}, {Percival}, {P{\'e}rez-Fournon},
  {P{\'e}rez-R{\`a}fols}, {Petitjean}, {Pfaffenberger}, {Pforr}, {Pieri},
  {Prada}, {Price-Whelan}, {Raddick}, {Rebolo}, {Rich}, {Richards}, {Rockosi},
  {Roe}, {Ross}, {Ross}, {Rossi}, {Rubi{\~n}o-Martin}, {Samushia},
  {S{\'a}nchez}, {Sayres}, {Schmidt}, {Schneider}, {Sc{\'o}ccola}, {Seo},
  {Shelden}, {Sheldon}, {Shen}, {Shu}, {Slosar}, {Smee}, {Snedden}, {Stauffer},
  {Steele}, {Strauss}, {Streblyanska}, {Suzuki}, {Swanson}, {Tal}, {Tanaka},
  {Thomas}, {Tinker}, {Tojeiro}, {Tremonti}, {Vargas Maga{\~n}a}, {Verde},
  {Viel}, {Wake}, {Watson}, {Weaver}, {Weinberg}, {Weiner}, {West}, {White},
  {Wood-Vasey}, {Yeche}, {Zehavi}, {Zhao}, \& {Zheng}}]{2013AJ....145...10D}
{Dawson} K.~S. {et~al.}, 2013, \aj, 145, 10

\bibitem[{{Doi} {et~al}\mbox{.}(2010){Doi}, {Tanaka}, {Fukugita}, {Gunn},
  {Yasuda}, {Ivezi{\'c}}, {Brinkmann}, {de Haars}, {Kleinman}, {Krzesinski}, \&
  {French Leger}}]{2010AJ....139.1628D}
{Doi} M. {et~al.}, 2010, \aj, 139, 1628

\bibitem[{{Eisenstein}(2002)}]{2002ASPC..280...35E}
{Eisenstein} D., 2002, in Astronomical Society of the Pacific Conference
  Series, Vol. 280, Next Generation Wide-Field Multi-Object Spectroscopy,
  {Brown} M.~J.~I., {Dey} A., eds., p.~35

\bibitem[{{Eisenstein} \& {Hu}(1998)}]{1998ApJ...496..605E}
{Eisenstein} D.~J., {Hu} W., 1998, \apj, 496, 605

\bibitem[{{Eisenstein} {et~al}\mbox{.}(2007{\natexlab{a}}){Eisenstein}, {Seo},
  {Sirko}, \& {Spergel}}]{2007ApJ...664..675E}
{Eisenstein} D.~J., {Seo} H.-J., {Sirko} E., {Spergel} D.~N.,
  2007{\natexlab{a}}, \apj, 664, 675

\bibitem[{{Eisenstein} {et~al}\mbox{.}(2007{\natexlab{b}}){Eisenstein}, {Seo},
  \& {White}}]{2007ApJ...664..660E}
{Eisenstein} D.~J., {Seo} H.-J., {White} M., 2007{\natexlab{b}}, \apj, 664, 660

\bibitem[{{Eisenstein} {et~al}\mbox{.}(2005){Eisenstein}, {Zehavi}, {Hogg},
  {Scoccimarro}, {Blanton}, {Nichol}, {Scranton}, {Seo}, {Tegmark}, {Zheng},
  {Anderson}, {Annis}, {Bahcall}, {Brinkmann}, {Burles}, {Castander},
  {Connolly}, {Csabai}, {Doi}, {Fukugita}, {Frieman}, {Glazebrook}, {Gunn},
  {Hendry}, {Hennessy}, {Ivezi{\'c}}, {Kent}, {Knapp}, {Lin}, {Loh}, {Lupton},
  {Margon}, {McKay}, {Meiksin}, {Munn}, {Pope}, {Richmond}, {Schlegel},
  {Schneider}, {Shimasaku}, {Stoughton}, {Strauss}, {SubbaRao}, {Szalay},
  {Szapudi}, {Tucker}, {Yanny}, \& {York}}]{2005ApJ...633..560E}
{Eisenstein} D.~J. {et~al.}, 2005, \apj, 633, 560

\bibitem[{{Feldman} {et~al}\mbox{.}(1994){Feldman}, {Kaiser}, \&
  {Peacock}}]{1994ApJ...426...23F}
{Feldman} H.~A., {Kaiser} N., {Peacock} J.~A., 1994, \apj, 426, 23

\bibitem[{{Fukugita} {et~al}\mbox{.}(1996){Fukugita}, {Ichikawa}, {Gunn},
  {Doi}, {Shimasaku}, \& {Schneider}}]{1996AJ....111.1748F}
{Fukugita} M., {Ichikawa} T., {Gunn} J.~E., {Doi} M., {Shimasaku} K.,
  {Schneider} D.~P., 1996, \aj, 111, 1748

\bibitem[{{Gazta{\~n}aga} {et~al}\mbox{.}(2009){Gazta{\~n}aga}, {Cabr{\'e}}, \&
  {Hui}}]{2009MNRAS.399.1663G}
{Gazta{\~n}aga} E., {Cabr{\'e}} A., {Hui} L., 2009, \mnras, 399, 1663

\bibitem[{{Gunn} {et~al}\mbox{.}(1998){Gunn}, {Carr}, {Rockosi}, {Sekiguchi},
  {Berry}, {Elms}, {de Haas}, {Ivezi{\'c}}, {Knapp}, {Lupton}, {Pauls},
  {Simcoe}, {Hirsch}, {Sanford}, {Wang}, {York}, {Harris}, {Annis}, {Bartozek},
  {Boroski}, {Bakken}, {Haldeman}, {Kent}, {Holm}, {Holmgren}, {Petravick},
  {Prosapio}, {Rechenmacher}, {Doi}, {Fukugita}, {Shimasaku}, {Okada}, {Hull},
  {Siegmund}, {Mannery}, {Blouke}, {Heidtman}, {Schneider}, {Lucinio}, \&
  {Brinkman}}]{1998AJ....116.3040G}
{Gunn} J.~E. {et~al.}, 1998, \aj, 116, 3040

\bibitem[{{Gunn} {et~al}\mbox{.}(2006){Gunn}, {Siegmund}, {Mannery}, {Owen},
  {Hull}, {Leger}, {Carey}, {Knapp}, {York}, {Boroski}, {Kent}, {Lupton},
  {Rockosi}, {Evans}, {Waddell}, {Anderson}, {Annis}, {Barentine}, {Bartoszek},
  {Bastian}, {Bracker}, {Brewington}, {Briegel}, {Brinkmann}, {Brown}, {Carr},
  {Czarapata}, {Drennan}, {Dombeck}, {Federwitz}, {Gillespie}, {Gonzales},
  {Hansen}, {Harvanek}, {Hayes}, {Jordan}, {Kinney}, {Klaene}, {Kleinman},
  {Kron}, {Kresinski}, {Lee}, {Limmongkol}, {Lindenmeyer}, {Long}, {Loomis},
  {McGehee}, {Mantsch}, {Neilsen}, {Neswold}, {Newman}, {Nitta}, {Peoples},
  {Pier}, {Prieto}, {Prosapio}, {Rivetta}, {Schneider}, {Snedden}, \&
  {Wang}}]{2006AJ....131.2332G}
{Gunn} J.~E. {et~al.}, 2006, \aj, 131, 2332

\bibitem[{{Hamilton}(1998)}]{1998ASSL..231..185H}
{Hamilton} A.~J.~S., 1998, in Astrophysics and Space Science Library, Vol. 231,
  The Evolving Universe, {Hamilton} D., ed., p. 185

\bibitem[{{Hartlap} {et~al}\mbox{.}(2007){Hartlap}, {Simon}, \&
  {Schneider}}]{2007A&A...464..399H}
{Hartlap} J., {Simon} P., {Schneider} P., 2007, \aap, 464, 399

\bibitem[{{Hinshaw} {et~al}\mbox{.}(2012){Hinshaw}, {Larson}, {Komatsu},
  {Spergel}, {Bennett}, {Dunkley}, {Nolta}, {Halpern}, {Hill}, {Odegard},
  {Page}, {Smith}, {Weiland}, {Gold}, {Jarosik}, {Kogut}, {Limon}, {Meyer},
  {Tucker}, {Wollack}, \& {Wright}}]{2012arXiv1212.5226H}
{Hinshaw} G. {et~al.}, 2012, ArXiv e-prints

\bibitem[{{Hogg}(1999)}]{1999astro.ph..5116H}
{Hogg} D.~W., 1999, ArXiv Astrophysics e-prints

\bibitem[{{Hu} \& {Haiman}(2003)}]{2003PhRvD..68f3004H}
{Hu} W., {Haiman} Z., 2003, \prd, 68, 063004

\bibitem[{{Hu} \& {Sugiyama}(1996)}]{1996ApJ...471..542H}
{Hu} W., {Sugiyama} N., 1996, \apj, 471, 542

\bibitem[{{Huff} {et~al}\mbox{.}(2007){Huff}, {Schulz}, {White}, {Schlegel}, \&
  {Warren}}]{2007APh....26..351H}
{Huff} E., {Schulz} A.~E., {White} M., {Schlegel} D.~J., {Warren} M.~S., 2007,
  Astroparticle Physics, 26, 351

\bibitem[{{H{\"u}tsi}(2006)}]{2006A&A...449..891H}
{H{\"u}tsi} G., 2006, \aap, 449, 891

\bibitem[{{H{\"u}tsi}(2010)}]{2010MNRAS.401.2477H}
{H{\"u}tsi} G., 2010, \mnras, 401, 2477

\bibitem[{{Jackson}(1972)}]{1972MNRAS.156P...1J}
{Jackson} J.~C., 1972, \mnras, 156, 1P

\bibitem[{{Kaiser}(1987)}]{1987MNRAS.227....1K}
{Kaiser} N., 1987, \mnras, 227, 1

\bibitem[{{Kazin} {et~al}\mbox{.}(2013){Kazin} {et~al.}}]{kazin13}
{Kazin} E., {et~al.}, 2013, \mnras\, submitted

\bibitem[{{Kazin} {et~al}\mbox{.}(2010){Kazin}, {Blanton}, {Scoccimarro},
  {McBride}, {Berlind}, {Bahcall}, {Brinkmann}, {Czarapata}, {Frieman}, {Kent},
  {Schneider}, \& {Szalay}}]{2010ApJ...710.1444K}
{Kazin} E.~A. {et~al.}, 2010, \apj, 710, 1444

\bibitem[{{Kazin} {et~al}\mbox{.}(2012){Kazin}, {S{\'a}nchez}, \&
  {Blanton}}]{2012MNRAS.419.3223K}
{Kazin} E.~A., {S{\'a}nchez} A.~G., {Blanton} M.~R., 2012, \mnras, 419, 3223

\bibitem[{{Kirkby} {et~al}\mbox{.}(2013){Kirkby}, {Margala}, {Slosar},
  {Bailey}, {Busca}, {Delubac}, {Rich}, {Blomqvist}, {Brownstein}, {Carithers},
  {Croft}, {Dawson}, {Font-Ribera}, {Miralda-Escud{\'e}}, {Myers}, {Nichol},
  {Palanque-Delabrouille}, {P{\^a}ris}, {Petitjean}, {Rossi}, {Schlegel},
  {Schneider}, {Viel}, {Weinberg}, \& {Y{\`e}che}}]{2013arXiv1301.3456K}
{Kirkby} D. {et~al.}, 2013, ArXiv e-prints

\bibitem[{{Landy} \& {Szalay}(1993)}]{1993ApJ...412...64L}
{Landy} S.~D., {Szalay} A.~S., 1993, \apj, 412, 64

\bibitem[{{Lewis} {et~al}\mbox{.}(2000){Lewis}, {Challinor}, \&
  {Lasenby}}]{2000ApJ...538..473L}
{Lewis} A., {Challinor} A., {Lasenby} A., 2000, \apj, 538, 473

\bibitem[{{Linder}(2003)}]{2003PhRvD..68h3504L}
{Linder} E.~V., 2003, \prd, 68, 083504

\bibitem[{{Lupton} {et~al}\mbox{.}(2001){Lupton}, {Gunn}, {Ivezi{\'c}},
  {Knapp}, {Kent}, \& {Yasuda}}]{2001ASPC..238..269L}
{Lupton} R., {Gunn} J.~E., {Ivezi{\'c}} Z., {Knapp} G.~R., {Kent} S., {Yasuda}
  N., 2001, in Astronomical Society of the Pacific Conference Series, Vol. 238,
  Astronomical Data Analysis Software and Systems X, {Harnden} Jr. F.~R.,
  {Primini} F.~A., {Payne} H.~E., eds., p. 269

\bibitem[{{Manera} {et~al}\mbox{.}(2013){Manera}, {Scoccimarro}, {Percival},
  {Samushia}, {McBride}, {Ross}, {Sheth}, {White}, {Reid}, {S{\'a}nchez}, {de
  Putter}, {Xu}, {Berlind}, {Brinkmann}, {Maraston}, {Nichol}, {Montesano},
  {Padmanabhan}, {Skibba}, {Tojeiro}, \& {Weaver}}]{2013MNRAS.428.1036M}
{Manera} M. {et~al.}, 2013, \mnras, 428, 1036

\bibitem[{{Mehta} {et~al}\mbox{.}(2012){Mehta}, {Cuesta}, {Xu}, {Eisenstein},
  \& {Padmanabhan}}]{2012MNRAS.427.2168M}
{Mehta} K.~T., {Cuesta} A.~J., {Xu} X., {Eisenstein} D.~J., {Padmanabhan} N.,
  2012, \mnras, 427, 2168

\bibitem[{{Mehta} {et~al}\mbox{.}(2011){Mehta}, {Seo}, {Eckel}, {Eisenstein},
  {Metchnik}, {Pinto}, \& {Xu}}]{2011ApJ...734...94M}
{Mehta} K.~T., {Seo} H.-J., {Eckel} J., {Eisenstein} D.~J., {Metchnik} M.,
  {Pinto} P., {Xu} X., 2011, \apj, 734, 94

\bibitem[{{Meiksin} {et~al}\mbox{.}(1999){Meiksin}, {White}, \&
  {Peacock}}]{1999MNRAS.304..851M}
{Meiksin} A., {White} M., {Peacock} J.~A., 1999, \mnras, 304, 851

\bibitem[{{Miller} {et~al}\mbox{.}(2001){Miller}, {Nichol}, \&
  {Batuski}}]{2001ApJ...555...68M}
{Miller} C.~J., {Nichol} R.~C., {Batuski} D.~J., 2001, \apj, 555, 68

\bibitem[{Muirhead(1982)}]{Muirhead}
Muirhead R.~J., 1982, Aspects of multivariate statistical theory, Wiley series
  in probability and mathematical statistics. Probability and mathematical
  statistics. John Wiley \& Sons, New York

\bibitem[{{Okumura} {et~al}\mbox{.}(2008){Okumura}, {Matsubara}, {Eisenstein},
  {Kayo}, {Hikage}, {Szalay}, \& {Schneider}}]{2008ApJ...676..889O}
{Okumura} T., {Matsubara} T., {Eisenstein} D.~J., {Kayo} I., {Hikage} C.,
  {Szalay} A.~S., {Schneider} D.~P., 2008, \apj, 676, 889

\bibitem[{{Padmanabhan} {et~al}\mbox{.}(2008){Padmanabhan}, {Schlegel},
  {Finkbeiner}, {Barentine}, {Blanton}, {Brewington}, {Gunn}, {Harvanek},
  {Hogg}, {Ivezi{\'c}}, {Johnston}, {Kent}, {Kleinman}, {Knapp}, {Krzesinski},
  {Long}, {Neilsen}, {Nitta}, {Loomis}, {Lupton}, {Roweis}, {Snedden},
  {Strauss}, \& {Tucker}}]{2008ApJ...674.1217P}
{Padmanabhan} N. {et~al.}, 2008, \apj, 674, 1217

\bibitem[{{Padmanabhan} {et~al}\mbox{.}(2007){Padmanabhan}, {Schlegel},
  {Seljak}, {Makarov}, {Bahcall}, {Blanton}, {Brinkmann}, {Eisenstein},
  {Finkbeiner}, {Gunn}, {Hogg}, {Ivezi{\'c}}, {Knapp}, {Loveday}, {Lupton},
  {Nichol}, {Schneider}, {Strauss}, {Tegmark}, \& {York}}]{2007MNRAS.378..852P}
{Padmanabhan} N. {et~al.}, 2007, \mnras, 378, 852

\bibitem[{{Padmanabhan} \& {White}(2008)}]{2008PhRvD..77l3540P}
{Padmanabhan} N., {White} M., 2008, \prd, 77, 123540

\bibitem[{{Padmanabhan} \& {White}(2009)}]{2009PhRvD..80f3508P}
{Padmanabhan} N., {White} M., 2009, \prd, 80, 063508

\bibitem[{{Padmanabhan} {et~al}\mbox{.}(2012){Padmanabhan}, {Xu}, {Eisenstein},
  {Scalzo}, {Cuesta}, {Mehta}, \& {Kazin}}]{2012MNRAS.427.2132P}
{Padmanabhan} N., {Xu} X., {Eisenstein} D.~J., {Scalzo} R., {Cuesta} A.~J.,
  {Mehta} K.~T., {Kazin} E., 2012, \mnras, 427, 2132

\bibitem[{{Peacock} \& {Dodds}(1994)}]{1994MNRAS.267.1020P}
{Peacock} J.~A., {Dodds} S.~J., 1994, \mnras, 267, 1020

\bibitem[{{Peebles} \& {Yu}(1970)}]{1970ApJ...162..815P}
{Peebles} P.~J.~E., {Yu} J.~T., 1970, \apj, 162, 815

\bibitem[{{Percival} {et~al}\mbox{.}(2007){Percival}, {Cole}, {Eisenstein},
  {Nichol}, {Peacock}, {Pope}, \& {Szalay}}]{2007MNRAS.381.1053P}
{Percival} W.~J., {Cole} S., {Eisenstein} D.~J., {Nichol} R.~C., {Peacock}
  J.~A., {Pope} A.~C., {Szalay} A.~S., 2007, \mnras, 381, 1053

\bibitem[{{Percival} {et~al}\mbox{.}(2010){Percival}, {Reid}, {Eisenstein},
  {Bahcall}, {Budavari}, {Frieman}, {Fukugita}, {Gunn}, {Ivezi{\'c}}, {Knapp},
  {Kron}, {Loveday}, {Lupton}, {McKay}, {Meiksin}, {Nichol}, {Pope},
  {Schlegel}, {Schneider}, {Spergel}, {Stoughton}, {Strauss}, {Szalay},
  {Tegmark}, {Vogeley}, {Weinberg}, {York}, \& {Zehavi}}]{2010MNRAS.401.2148P}
{Percival} W.~J. {et~al.}, 2010, \mnras, 401, 2148

\bibitem[{{Perlmutter} {et~al}\mbox{.}(1999){Perlmutter}, {Aldering},
  {Goldhaber}, {Knop}, {Nugent}, {Castro}, {Deustua}, {Fabbro}, {Goobar},
  {Groom}, {Hook}, {Kim}, {Kim}, {Lee}, {Nunes}, {Pain}, {Pennypacker},
  {Quimby}, {Lidman}, {Ellis}, {Irwin}, {McMahon}, {Ruiz-Lapuente}, {Walton},
  {Schaefer}, {Boyle}, {Filippenko}, {Matheson}, {Fruchter}, {Panagia},
  {Newberg}, {Couch}, \& {Supernova Cosmology Project}}]{1999ApJ...517..565P}
{Perlmutter} S. {et~al.}, 1999, \apj, 517, 565

\bibitem[{{Pier} {et~al}\mbox{.}(2003){Pier}, {Munn}, {Hindsley}, {Hennessy},
  {Kent}, {Lupton}, \& {Ivezi{\'c}}}]{2003AJ....125.1559P}
{Pier} J.~R., {Munn} J.~A., {Hindsley} R.~B., {Hennessy} G.~S., {Kent} S.~M.,
  {Lupton} R.~H., {Ivezi{\'c}} {\v Z}., 2003, \aj, 125, 1559

\bibitem[{{Reid} {et~al}\mbox{.}(2012){Reid}, {Samushia}, {White}, {Percival},
  {Manera}, {Padmanabhan}, {Ross}, {S{\'a}nchez}, {Bailey}, {Bizyaev},
  {Bolton}, {Brewington}, {Brinkmann}, {Brownstein}, {Cuesta}, {Eisenstein},
  {Gunn}, {Honscheid}, {Malanushenko}, {Malanushenko}, {Maraston}, {McBride},
  {Muna}, {Nichol}, {Oravetz}, {Pan}, {de Putter}, {Roe}, {Ross}, {Schlegel},
  {Schneider}, {Seo}, {Shelden}, {Sheldon}, {Simmons}, {Skibba}, {Snedden},
  {Swanson}, {Thomas}, {Tinker}, {Tojeiro}, {Verde}, {Wake}, {Weaver},
  {Weinberg}, {Zehavi}, \& {Zhao}}]{2012MNRAS.426.2719R}
{Reid} B.~A. {et~al.}, 2012, \mnras, 426, 2719

\bibitem[{{Riess} {et~al}\mbox{.}(1998){Riess}, {Filippenko}, {Challis},
  {Clocchiatti}, {Diercks}, {Garnavich}, {Gilliland}, {Hogan}, {Jha},
  {Kirshner}, {Leibundgut}, {Phillips}, {Reiss}, {Schmidt}, {Schommer},
  {Smith}, {Spyromilio}, {Stubbs}, {Suntzeff}, \&
  {Tonry}}]{1998AJ....116.1009R}
{Riess} A.~G. {et~al.}, 1998, \aj, 116, 1009

\bibitem[{{Ross} {et~al}\mbox{.}(2012){Ross}, {Percival}, {S{\'a}nchez},
  {Samushia}, {Ho}, {Kazin}, {Manera}, {Reid}, {White}, {Tojeiro}, {McBride},
  {Xu}, {Wake}, {Strauss}, {Montesano}, {Swanson}, {Bailey}, {Bolton}, {Dorta},
  {Eisenstein}, {Guo}, {Hamilton}, {Nichol}, {Padmanabhan}, {Prada},
  {Schlegel}, {Maga{\~n}a}, {Zehavi}, {Blanton}, {Bizyaev}, {Brewington},
  {Cuesta}, {Malanushenko}, {Malanushenko}, {Oravetz}, {Parejko}, {Pan},
  {Schneider}, {Shelden}, {Simmons}, {Snedden}, \&
  {Zhao}}]{2012MNRAS.424..564R}
{Ross} A.~J. {et~al.}, 2012, \mnras, 424, 564

\bibitem[{{Samushia} {et~al}\mbox{.}(2013){Samushia}, {Reid}, {White},
  {Percival}, {Cuesta}, {Lombriser}, {Manera}, {Nichol}, {Schneider},
  {Bizyaev}, {Brewington}, {Malanushenko}, {Malanushenko}, {Oravetz}, {Pan},
  {Simmons}, {Shelden}, {Snedden}, {Tinker}, {Weaver}, {York}, \&
  {Zhao}}]{2013MNRAS.429.1514S}
{Samushia} L. {et~al.}, 2013, \mnras, 429, 1514

\bibitem[{{Sanchez} {et~al}\mbox{.}(2013){Sanchez} {et~al.}}]{sanchez13}
{Sanchez} A., {et~al.}, 2013, \mnras\, submitted

\bibitem[{{Schlegel} {et~al}\mbox{.}(1998){Schlegel}, {Finkbeiner}, \&
  {Davis}}]{1998ApJ...500..525S}
{Schlegel} D.~J., {Finkbeiner} D.~P., {Davis} M., 1998, \apj, 500, 525

\bibitem[{{Scoccimarro} \& {Sheth}(2002)}]{2002MNRAS.329..629S}
{Scoccimarro} R., {Sheth} R.~K., 2002, \mnras, 329, 629

\bibitem[{{Seo} {et~al}\mbox{.}(2010){Seo}, {Eckel}, {Eisenstein}, {Mehta},
  {Metchnik}, {Padmanabhan}, {Pinto}, {Takahashi}, {White}, \&
  {Xu}}]{2010ApJ...720.1650S}
{Seo} H.-J. {et~al.}, 2010, \apj, 720, 1650

\bibitem[{{Seo} \& {Eisenstein}(2003)}]{2003ApJ...598..720S}
{Seo} H.-J., {Eisenstein} D.~J., 2003, \apj, 598, 720

\bibitem[{{Seo} \& {Eisenstein}(2007)}]{2007ApJ...665...14S}
{Seo} H.-J., {Eisenstein} D.~J., 2007, \apj, 665, 14

\bibitem[{{Seo} {et~al}\mbox{.}(2012){Seo}, {Ho}, {White}, {Cuesta}, {Ross},
  {Saito}, {Reid}, {Padmanabhan}, {Percival}, {de Putter}, {Schlegel},
  {Eisenstein}, {Xu}, {Schneider}, {Skibba}, {Verde}, {Nichol}, {Bizyaev},
  {Brewington}, {Brinkmann}, {Nicolaci da Costa}, {Gott}, {Malanushenko},
  {Malanushenko}, {Oravetz}, {Palanque-Delabrouille}, {Pan}, {Prada}, {Ross},
  {Simmons}, {de Simoni}, {Shelden}, {Snedden}, \&
  {Zehavi}}]{2012ApJ...761...13S}
{Seo} H.-J. {et~al.}, 2012, \apj, 761, 13

\bibitem[{{Slosar} {et~al}\mbox{.}(2013){Slosar}, {Ir{\v s}i{\v c}}, {Kirkby},
  {Bailey}, {Busca}, {Delubac}, {Rich}, {Bhardwaj}, {Blomqvist}, {Bolton},
  {Bovy}, {Brownstein}, {Carithers}, {Croft}, {Dawson}, {Font-Ribera}, {Le
  Goff}, {Ho}, {Honscheid}, {Lee}, {Margala}, {McDonald}, {Medolin},
  {Miralda-Escud{\'e}}, {Myers}, {Nichol}, {Noterdaeme}, {P{\^a}ris},
  {Petitjean}, {Pieri}, {Roe}, {Ross}, {Rossi}, {Schlegel}, {Schneider},
  {Sheldon}, {Seljak}, {Viel}, {Weinberg}, \&
  {Y{\`e}che}}]{2013arXiv1301.3459S}
{Slosar} A. {et~al.}, 2013, ArXiv e-prints

\bibitem[{{Smee} {et~al}\mbox{.}(2012){Smee}, {Gunn}, {Uomoto}, {Roe},
  {Schlegel}, {Rockosi}, {Carr}, {Leger}, {Dawson}, {Olmstead}, {Brinkmann},
  {Owen}, {Barkhouser}, {Honscheid}, {Harding}, {Long}, {Lupton}, {Loomis},
  {Anderson}, {Annis}, {Bernardi}, {Bhardwaj}, {Bizyaev}, {Bolton},
  {Brewington}, {Briggs}, {Burles}, {Burns}, {Castander}, {Connolly},
  {Davenport}, {Ebelke}, {Epps}, {Feldman}, {Friedman}, {Frieman}, {Heckman},
  {Hull}, {Knapp}, {Lawrence}, {Loveday}, {Mannery}, {Malanushenko},
  {Malanushenko}, {Merrelli}, {Muna}, {Newman}, {Nichol}, {Oravetz}, {Pan},
  {Pope}, {Ricketts}, {Shelden}, {Sandford}, {Siegmund}, {Simmons}, {Smith},
  {Snedden}, {Schneider}, {Strauss}, {SubbaRao}, {Tremonti}, {Waddell}, \&
  {York}}]{2012arXiv1208.2233S}
{Smee} S. {et~al.}, 2012, ArXiv e-prints

\bibitem[{{Smith} {et~al}\mbox{.}(2002){Smith}, {Tucker}, {Kent}, {Richmond},
  {Fukugita}, {Ichikawa}, {Ichikawa}, {Jorgensen}, {Uomoto}, {Gunn}, {Hamabe},
  {Watanabe}, {Tolea}, {Henden}, {Annis}, {Pier}, {McKay}, {Brinkmann}, {Chen},
  {Holtzman}, {Shimasaku}, \& {York}}]{2002AJ....123.2121S}
{Smith} J.~A. {et~al.}, 2002, \aj, 123, 2121

\bibitem[{{Springel} {et~al}\mbox{.}(2005){Springel}, {White}, {Jenkins},
  {Frenk}, {Yoshida}, {Gao}, {Navarro}, {Thacker}, {Croton}, {Helly},
  {Peacock}, {Cole}, {Thomas}, {Couchman}, {Evrard}, {Colberg}, \&
  {Pearce}}]{2005Natur.435..629S}
{Springel} V. {et~al.}, 2005, \nat, 435, 629

\bibitem[{{Sunyaev} \& {Zeldovich}(1970)}]{1970Ap&SS...7....3S}
{Sunyaev} R.~A., {Zeldovich} Y.~B., 1970, \apss, 7, 3

\bibitem[{{Swanson} {et~al}\mbox{.}(2008){Swanson}, {Tegmark}, {Hamilton}, \&
  {Hill}}]{2008MNRAS.387.1391S}
{Swanson} M.~E.~C., {Tegmark} M., {Hamilton} A.~J.~S., {Hill} J.~C., 2008,
  \mnras, 387, 1391

\bibitem[{{Tegmark} {et~al}\mbox{.}(2006){Tegmark}, {Eisenstein}, {Strauss},
  {Weinberg}, {Blanton}, {Frieman}, {Fukugita}, {Gunn}, {Hamilton}, {Knapp},
  {Nichol}, {Ostriker}, {Padmanabhan}, {Percival}, {Schlegel}, {Schneider},
  {Scoccimarro}, {Seljak}, {Seo}, {Swanson}, {Szalay}, {Vogeley}, {Yoo},
  {Zehavi}, {Abazajian}, {Anderson}, {Annis}, {Bahcall}, {Bassett}, {Berlind},
  {Brinkmann}, {Budavari}, {Castander}, {Connolly}, {Csabai}, {Doi},
  {Finkbeiner}, {Gillespie}, {Glazebrook}, {Hennessy}, {Hogg}, {Ivezi{\'c}},
  {Jain}, {Johnston}, {Kent}, {Lamb}, {Lee}, {Lin}, {Loveday}, {Lupton},
  {Munn}, {Pan}, {Park}, {Peoples}, {Pier}, {Pope}, {Richmond}, {Rockosi},
  {Scranton}, {Sheth}, {Stebbins}, {Stoughton}, {Szapudi}, {Tucker}, {vanden
  Berk}, {Yanny}, \& {York}}]{2006PhRvD..74l3507T}
{Tegmark} M. {et~al.}, 2006, \prd, 74, 123507

\bibitem[{{Tojeiro} {et~al}\mbox{.}(2012){Tojeiro}, {Percival}, {Brinkmann},
  {Brownstein}, {Eisenstein}, {Manera}, {Maraston}, {McBride}, {Muna}, {Reid},
  {Ross}, {Ross}, {Samushia}, {Padmanabhan}, {Schneider}, {Skibba},
  {S{\'a}nchez}, {Swanson}, {Thomas}, {Tinker}, {Verde}, {Wake}, {Weaver}, \&
  {Zhao}}]{2012MNRAS.424.2339T}
{Tojeiro} R. {et~al.}, 2012, \mnras, 424, 2339

\bibitem[{{Tseliakhovich} \& {Hirata}(2010)}]{2010PhRvD..82h3520T}
{Tseliakhovich} D., {Hirata} C., 2010, \prd, 82, 083520

\bibitem[{{Weinberg} {et~al}\mbox{.}(2012){Weinberg}, {Mortonson},
  {Eisenstein}, {Hirata}, {Riess}, \& {Rozo}}]{2012arXiv1201.2434W}
{Weinberg} D.~H., {Mortonson} M.~J., {Eisenstein} D.~J., {Hirata} C., {Riess}
  A.~G., {Rozo} E., 2012, ArXiv e-prints

\bibitem[{{Xu} {et~al}\mbox{.}(2013){Xu}, {Cuesta}, {Padmanabhan},
  {Eisenstein}, \& {McBride}}]{2012arXiv1206.6732X}
{Xu} X., {Cuesta} A.~J., {Padmanabhan} N., {Eisenstein} D.~J., {McBride} C.~K.,
  2013, \mnras\, accepted

\bibitem[{{Xu} {et~al}\mbox{.}(2012){Xu}, {Padmanabhan}, {Eisenstein}, {Mehta},
  \& {Cuesta}}]{2012MNRAS.427.2146X}
{Xu} X., {Padmanabhan} N., {Eisenstein} D.~J., {Mehta} K.~T., {Cuesta} A.~J.,
  2012, \mnras, 427, 2146

\bibitem[{{Yoo} {et~al}\mbox{.}(2011){Yoo}, {Dalal}, \&
  {Seljak}}]{2011JCAP...07..018Y}
{Yoo} J., {Dalal} N., {Seljak} U., 2011, \jcap, 7, 18

\end{thebibliography}

\end{document}